\documentclass[12pt]{article}
\usepackage{amsmath}
\usepackage{amsfonts}
\usepackage{amssymb}
\usepackage{graphicx}
\newlength{\abstractwidth}
\usepackage{epsf}

\renewcommand{\thefootnote}{\fnsymbol{footnote}}
\renewcommand{\thanks}[1]{\footnote{#1}}
\newcommand{\starttext}{
\setcounter{footnote}{0}
\renewcommand{\thefootnote}{\arabic{footnote}}}

\newcommand{\bea}{\begin{eqnarray}}
\newcommand{\eea}{\end{eqnarray}}
\newcommand{\ee}{\end{equation}}
\newcommand{\be}{\begin{equation}}

\newcommand{\ea}{\end{array}}
\newcommand{\bac}{\begin{array}{c}}
\newcommand{\bacc}{\begin{array}{cc}}
\newcommand{\barcl}{\begin{array}{r@{}c@{}l}}
\newcommand{\brcl}{\begin{array}{rcl}}
\newcommand{\bdm}{\begin{displaymath}}
\newcommand{\edm}{\end{displaymath}}
\newcommand{\nn}{\nonumber}

\newcommand{\half}{\frac{1}{2}}
\def \z {\zeta}
\def \tiz {\tilde{\zeta}}
\def \ss {\scriptstyle}
\def \ts {\textstyle} 

\def\half{ {1\over 2}}

\def\a{\alpha}
\def\b{\beta}

\def\g{\gamma}

\def\g{\gamma}
\def\s{\sigma}

\def\no{\nonumber}

\begin{document}
\starttext
\setcounter{footnote}{0}

\begin{flushright}
UCLA/08/TEP/30 
\end{flushright}

\bigskip

\begin{center}

{\Large \bf Instantons and Wormholes in $N=2$ supergravity}

\vskip .7in 

{\large Marco Chiodaroli\footnote{email: mchiodar@ucla.edu} and Michael Gutperle\footnote{email: gutperle@ucla.edu}}

\vskip .2in

 \sl Department of Physics and Astronomy \\
\sl University of California, Los Angeles, CA 90095, USA

\end{center}

\vskip .5in

\begin{abstract}
 
In this paper, we construct Euclidean instanton and wormhole solutions in $d=4$, $N=2$ supergravity theories with hypermultiplets. The analytic continuation of the hypermultiplet action, involving pseudoscalar axions, is discussed using the approach originally developed by Coleman which determines the apparence of boundary terms. In particular, we investigate the conditions obtained by requiring the action   to be positive-definite once the boundary terms are taken into account. The case of  two hypermultiplets parameterizing the coset $G_{2,2}/SU(2)\times SU(2)$ is studied in detail. Orientifold projections which reduce the supersymmetry to $N=1$ are also discussed.

\end{abstract}

\newpage
{\small 
\tableofcontents
}
\newpage

\baselineskip=16pt
\setcounter{equation}{0}
\setcounter{footnote}{0}

\section{Introduction}
\setcounter{equation}{0}
 Instantons and wormholes determine potentially important non-perturbative effects in string theory.
 Both  can be obtained  as saddle-points of the Euclidean action of the corresponding low-energy supergravity \cite{Hawking:1988ae,Lavrelashvili:1988jj,Giddings:1987cg,Coleman:1989zu,Giddings:1989bq}.

The literature discusses how wormholes can lead to several interesting effects. Some examples are the renormalization  of coupling constants, a mechanism setting to zero the cosmological constant, quantum decoherence and creation of baby universes \cite{Strominger:1983ns,Hawking:1987mz,Coleman:1988cy}. In \cite{Rey:1998yx,Gutperle:2002km,Maldacena:2004rf,ArkaniHamed:2007js} wormholes in Anti- de Sitter spaces have been discussed. Recently, in \cite{ArkaniHamed:2007js} it was argued that wormholes  in the AdS bulk can spoil  locality and cluster decomposition  in the context of the AdS/CFT correspondence.\\
 
In contrast to the non-local effects produced by wormholes, instantons produce local non-perturbative contributions to the low-energy effective action. In supergravity theories with extended supersymmetry there are BPS instanton solutions preserving half of the supersymmetries \cite{Gibbons:1995vg}. The broken supersymmetries in the instanton background generate fermionic zero modes which have to be soaked up by instanton-induced interaction terms in the path integral \cite{Green:1997tv}.  Note however that in some theories, such as $N=2$ $d=4$ supergravity theories with $n_{H}>1$ hypermultiplets, extremal non-BPS instanton solutions can exist. 

  Instanton and wormhole solutions have been discussed for various theories and dimensions, in particular the axion/dilaton $SL(2,R)/U(1)$ coset \cite{Gibbons:1995vg,Green:1997tv,Bergshoeff:2004fq,Einhorn:2002sj}, the universal hypermultiplet in $N=2,d=4$ supergravity \cite{Gutperle:2000sb,Davidse:2004gg,Theis:2002er} and general hypermultiplets in $N=2,d=4$ theories \cite{Gutperle:2000ve,deVroome:2006xu, Behrndt:1997ch}. 
 
The structure of the paper is as follows. In section 2 we review important properties of the hypermultplet sector of $N=2$ supergravity theories. 

In section 3 we discuss the general properties of instantons and wormholes  in $N=2$ theories, in particular how the analytic continuation arises using an approach first employed in a paper by Coleman and Lee \cite{Coleman:1989zu} (see also \cite{ArkaniHamed:2007js}). 
In general the instanton and wormhole solutions are constructed by complexifying  the scalar fields and choosing a real section (i.e. the real section defines a particular analytic continuation of the original scalar fields).  We propose a condition which leads to solutions which satisfy reality conditions and lead to a positive definite action. It is an interesting and open problem, whether for other real sections the  instanton solutions are sensible saddle points dominating the path integral. 

In section 4 we cover the universal hypermultiplet which is given by a $SU(2,1)/U(2)$ coset nonlinear sigma model and has been the object of previous work (\cite{Chiodaroli:2008rj}). 

In section 5 we study the case of two hypermultiplets parameterizing a $G_{2,2}/(SU(2)\times SU(2))$ coset. Explicit solutions are obtained using the conserved currents coming from the global symmetries of the coset sigma model. In particular, we study various consistent truncations and we present explicit solutions as well as their actions. We also discuss the existence of extremal non-BPS instanton solutions and the possibility to generate more general solutions using the  $G_{2,2}$ global symmetry. 

 In section 6 the reduction of the supersymmetry due to  orientifold projections is discussed and related to the consistent truncations of section 5 for the $G_{2,2}$ case.
Finally, in section 7, we give a brief discussion of the open problems.

\section{Hypermultiplets in $N=2$ supergravity}
\setcounter{equation}{0}
N=2 supergravity theories are endowed with a very rich structure and stand between phenomenologically viable theories with $N=1$ supersymmetry and theories with more than two supersymmetries which are almost completely fixed by their symmetries. Two recent examples of interest in these theories are the study of the attractor mechanism for extremal black holes \cite{Ferrara:1995ih,Ferrara:1996dd} and the discovery of the role that higher derivative corrections \cite{LopesCardoso:1998wt} and topological string amplitudes  \cite{Ooguri:2004zv} play for the entropy of BPS black holes. In this section we will review the properties of the hypermultiplet sector of $N =2$ theories. 
 
\subsection{Calabi-Yau compactification}
The canonical example of obtaining four-dimensional  $N=2$ supergravity  theories in string theory is the compactification of ten-dimensional type II (A or B) superstring theory on a six-dimensional Calabi-Yau manifold.  The compactification breaks $N=8$ supersymmetry down to $N=2$. For length  scales larger than  the compactification scale (which in turn is larger than the string scale $l_s$) the theory is well approximated by the four-dimensional two-derivative effective supergravity action.   The  moduli space of scalars factorizes into   vector and hypermultiplets, ${\cal M} = {\cal M}_{vector} \times {\cal M}_{hyper}$, where $M_{vector}$ is given by a special K\" ahler manifold \cite{deWit:1984px} and $M_{hyper}$ is given by a quaternionic K\" ahler manifold \cite{Ferrara:1989ik}. The dimensionality of the respective moduli spaces depends on the Hodge numbers $h_{1,1}$ and $h_{2,1}$ of the Calabi-Yau manifold.

\begin{table}[htdp]
\begin{center}\label{tableone}
\begin{tabular}{|c|c |c|} \hline
   &dim(${\cal M}_{vector}$) & dim(${\cal M}_{hyper}$) \\ \hline
   type IIA & $2 h_{1,1}$& $4(h_{2,1}+1) $ \\ \hline 
   type IIB&  $2h_{2,1}$ & $4(h_{1,1}+1) $  \\ \hline 
\end{tabular}
\caption{Dimensionality of moduli spaces}
\end{center}
\end{table}
In terms of the conformal field theory, the compactification is encoded in a $c=9$, $N=2$ superconformal field theory \cite{Cecotti:1988qn}. The massless moduli come from the combination of chiral and anti-chiral primary states of the $N=2$ SCFT.

\subsection{Mirror symmetry and c-map}
For type II string theories compactified on a circle,  T-duality relates type  IIA theory  on a circle of radius $R$ to type IIB an a circle of radius $1/R$ \cite{Dai:1989ua}. There are two analogs of T-duality for  Calabi-Yau compactifications.   

First, Mirror symmetry has a simple realization in terms of the internal SCFT, where one  changes the sign of the chiral $U(1)$ current of the $N=2$ CFT.  This transformation 
relates type IIA on a Calabi-Yau manifold ${\cal M}$ to type IIB on a mirror Calabi-Yau manifold $\tilde{\cal M}$. The two manifolds are topologically different since  the Hodge numbers $h_{1,1}$ and $h_{2,1}$ are interchanged.

Second, the c-map is obtained \cite{Cecotti:1988qn} compactifying one of the four flat  non-compact directions on a circle and performing a T-duality. This T-duality does not act on the internal $N=2$ SCFT and hence the c-map relates string theories compactified on the same Calabi-Yau manifold.  It does however relate the gravity   and  vector multiplets of the type IIA theory to the  hypermultiplets of type IIB and vice versa.

\subsection{Hypermultiplet actions}

The bosonic part of the hypermultiplet action given by a nonlinear sigma model which lives on a    special quaternionic manifold.  In the following,  we will assume that the theory is obtained by compatifying type IIB string theory on a Calabi-Yau manifold.  The quaternionic manifold is $4n_H=4 (h_{1,1}+1)$ dimensional  
\cite{Ferrara:1989ik}.
 It is  parameterized by $n_H-1=h_{1,1}$ complex scalars $z^\alpha, \alpha=1,2,\cdots,n_H-1$ together with $2n_H$ real Ramond-Ramond scalars $\zeta^I,\tilde \zeta_I, I=0,1,\cdots , n_H-1$,  the dilaton $\phi$ and the  NS-NS axion $\sigma$. 
 The explicit form of the action can be obtained by compactification \cite{Bohm:1999uk}
 or applying the c-map on the gravity and vector multiplet action \cite{Ferrara:1989ik}.   The resulting hypermultiplet action can be written as follows
\bea
 S&=& \int d^4x\; \sqrt{-g}\Big\{R - 2 g_{\alpha \bar \beta} \partial_\mu z^{\alpha} \partial \bar z^{\bar \beta}-{1\over 2} (\partial_\mu \phi)^2  -{1\over 2}e^{-2\phi} (\partial_ \mu \sigma + {1\over 2}\zeta^I \partial_\mu \tilde \zeta_I - {1\over 2}\tilde \zeta_I \partial_\mu \zeta^I)^2\nonumber \\
& &-{1\over2} e^{-\phi} I_{IJ} \partial_\mu \zeta^I \partial^\mu \zeta^J -{1\over 2}e^{-\phi} (\partial_\mu\tilde \zeta_I + R_{IK} \partial_\mu \zeta^K) (I^{-1})^{IJ} (\partial^\mu\tilde \zeta_J + R_{JL} \partial^\mu \zeta^L) 
\Big\}\label{hyperone}
 \eea
 The action is completely determined by a prepotential $F(X^I)$, where the projective coordinates $X^I, I=0,1, \cdots,h_{1,1}$ are related to the scalars $z^\alpha$ via $z^\alpha= X^\alpha/X^0$ . 
 The matrices $R_{IJ}$ and $I_{IJ}$ are determined in terms of the prepotential by the relations: 
 \be
 F_{I}= {\partial F\over \partial X^I} , \quad F_{IJ} ={\partial^2 F\over \partial X^I \partial X^J }, \quad N_{IJ} = \bar F_{IJ} + 2 i {Im(F_{IL})  Im(F_{JM}) X^L X^M \over Im(F_{PQ}) X^P X^Q }
 \ee
 and by:
\be  R_{IJ}= Re(N_{IJ}) \quad I_{IJ}=Im(N_{IJ}) \label{R&I} \ee
 The scalars $z^{\alpha}= X^{\alpha} /X^0$ parameterize a  special geometry with K\" ahler potential
\be 
K= -\ln\big( i( \bar X^I F_I- X^I \bar F_I) \big), \quad g_{\alpha \bar \beta} = {\partial^2 K \over \partial z^{\alpha} \partial \bar z^{\beta}}
\ee
In the following we will neglect worldsheet instanton corrections (alternatively one can work with a type IIA compactification where worldsheet instantons modify  the vector multiplet moduli space). The prepotential, in the large volume limit, is then given by
\be
F(X^I)= {1\over 6}C_{\alpha \beta \gamma} {X^\alpha X^\beta X^\gamma \over X^0} \label{prepLCY}
\ee 
where $C_{\alpha \beta \gamma}$ are the intersection numbers of the $H_{1,1}$ cycles on the Calabi-Yau manifold. For the prepotential (\ref{prepLCY}) the matrices $R_{IJ}$ and $I_{IJ}$ are given by

\bea
R_{IJ} &=&
\left(
\begin{array}{cc}
  {1\over3} C_{\alpha \beta  \gamma} x^{\alpha}x^{\beta}x^{\gamma}  & - {1\over 2} C_{\alpha \beta  \gamma} x^{\beta}x^{\gamma}     \\
 - {1\over 2} C_{\alpha \beta  \gamma} x^{\beta}x^{\gamma}  &   C_{\alpha \beta  \gamma} x^{\gamma}
\end{array}
\right) \no \\
I_{IJ} &=&
\left(
\begin{array}{cc}
 {k\over 6}- C_{\alpha \beta  \gamma} x^{\alpha}x^{\beta}y^{\gamma} + {3 C_{\alpha \beta  \gamma} x^{\alpha}y^{\beta}y^{\gamma} C_{\delta \epsilon  \lambda} x^{\delta}y^{\epsilon}y^{\lambda} \over 2 k} &  C_{\alpha \beta  \gamma} x^{\beta}y^{\gamma} -{3 C_{\alpha \beta  \gamma} y^{\beta}y^{\gamma} C_{\delta \epsilon  \lambda} x^{\delta}y^{\epsilon}y^{\lambda} \over 2 k}\\
C_{\alpha \beta  \gamma} x^{\beta}y^{\gamma} -{ 3C_{\alpha \beta  \gamma} y^{\beta}y^{\gamma} C_{\delta \epsilon  \lambda} x^{\delta}y^{\epsilon}y^{\lambda}\over 2 k} &  -  C_{\alpha \beta  \gamma} y^{\gamma} + { 3 C_{\alpha \ \gamma \delta}  y^{\gamma} y^{\delta} C_{ \beta  \epsilon \lambda} y^{\epsilon} y^{\lambda}  \over 2 k} 
\end{array}
\right) \qquad
 \eea
where $k= C_{\alpha \beta \gamma}y^{\alpha}y^{\beta}y^{\gamma}$ and the  complex scalars $z^{\alpha}
$ have been split into real and imaginary part $z^{\a}=x^{\a}+ i y^{\a}, \a=1,2,\cdots n_H-1$. Note that with our conventions the matrix $I_{IJ}$ is positive-definite if $y^{a}>0$ for $a=1,2,\cdots n_H-1$ .

\subsection{Supersymmetry  variations}
The supersymmetry variation parameters $\epsilon^i, \; i=1,2$ and the hyperinos $\xi_a, \; a=1,2,\cdots 2n_H$ are complex Weyl spinors.
The fermionic supersymmetry variations for the gravitino is given by:
\begin{equation}\label{delpsi}
\delta \psi^i_\mu = D_\mu \epsilon^i +(Q_\mu)_{\;\;j}^{i}\epsilon^j
\end{equation} 
where $D_{\mu}$ is the standard covariant derivative which  includes the spin connection and 
$Q_{\;j}^{i}$ is a composite $SU(2)$ gauge connection 
defined by:
\begin{equation}\label{sutwoc}
Q_{\; \;j}^{i}=
\left(
\begin{array}{cc}
 {\textstyle v-\bar v \over \textstyle 4} -{  \bar X Im(F) dX-X Im(F)d\bar X \over  4 \bar X Im(F) X} &  - u    \\
 \bar u  &   { \textstyle \bar v- v \over \textstyle 4} + { \bar X Im(F) 
d X-X Im(F) d \bar X \over 4 \bar X (ImF) X}
\end{array}
\right)
\end{equation}
The hyperino variation is defined as:
\begin{equation}\label{hypervar}
\delta \xi_a= -i C_{ab} V_\mu^{b i} \gamma^\mu \epsilon_i
\end{equation}
Where $C_{ab}$ is the $Sp(2n_H)$ invariant tensor and $ \epsilon_{ab}$ is the two-dimensional antisymmetric tensor. The quaternionic vielbein $V$
 is a $2n_H \times 2 $ dimensional matrix 
 \begin{equation}\label{quadvb}
 V_\mu^{a i} =
\left(
\begin{array}{cc}
  u_\mu &    v_\mu   \\
  &\\
   e^A_\mu &  E_\mu^A   \\
   &\\
  - \bar E_\mu^A  & \bar e^A_\mu \\
&\\
 -\bar v_\mu  & \bar  u_\mu    
\end{array}
\right)
\end{equation}
 where the components appearing in (\ref{sutwoc}) and (\ref{quadvb}) are given by
 \begin{eqnarray} \label{vbein}
e^{A}_{\mu}&=& e^{A}_{\alpha} \partial_{\mu} z^{\alpha} \nonumber \\
E^{A}_{\mu}&=&-{i \over \sqrt{2}} e^{-{ \phi \over 2}}e^{A \alpha} \bar f_{\alpha}^{I}\Big ({N}_{IJ} \partial_{\mu} \zeta^{J}+ \partial_{\mu}\tilde \zeta_{I}\Big)\nonumber \\
u_{\mu}&=& {i \over \sqrt{2} } e^{ K - \phi \over 2 } X^{I}\Big( N_{IJ}\partial_{\mu} \zeta^{J}+ \partial_{\mu} \tilde \zeta_{I} \Big)\nonumber\\
v_{\mu}&=& {1 \over 2 } \partial_{\mu} \phi +{i\over 2}e^{- \phi}\Big(\partial_{\mu}
\sigma + {1 \over 2} \zeta^{I}\partial_{\mu }\tilde \zeta - {1 \over 2 } \tilde \zeta_{I}\partial_{\mu}\zeta^{I}\Big)
\end{eqnarray}
where
\be X^I = \left( \begin{array}{c} 1 \\ z^\alpha \end{array} \right), \qquad f^I_\alpha = D_\alpha e^{K \over 2} X^I =\Big( \partial_\alpha +{\partial_\alpha K \over 2} \Big) e^{K \over 2} X^I  \ee

\subsection{Shift symmetries}
The hypermultiplet action (\ref{hyperone}) is invariant under $2n_{H}+1$  shift symmetries. From the ten-dimensional point of view these symmetries arise because the scalars $\zeta^I,\tilde \zeta_I, I=0,1,\cdots , n_H-1$ are descending  from RR tensors which only have derivative couplings. Similarly the axion $\sigma$ comes from the dualized NS-NS two-tensor field. The non-trivial Wess-Zumino term in the ten-dimensional IIB action leads to a mixing of $\sigma$ and $\zeta^I,\tilde \zeta_I$ shifts.
\be
\delta \zeta^I = \gamma^I , \quad \delta \tilde \zeta_I = \tilde \gamma_I  \, \quad \delta \sigma= \alpha+{1\over 2} \tilde \gamma_I \zeta^I -{1\over 2} \gamma^I \tilde \zeta_I \label{hypertwo}
\ee
where $\gamma^I, \tilde \gamma_I, \alpha$ parameterize the $2 n_H +1$ shift symmetries.
The shift symmetries (\ref{hypertwo}) have generators $\Gamma^I, \tilde \Gamma_I$ and $E$,  which satisfy a $n_H$ dimensional Heisenberg algebra with central element $E$:
\be
[\Gamma_I,\tilde \Gamma^J]=\delta^J_I E, \quad [E, \Gamma_I]=0, \quad [E, \tilde \Gamma^I]=0 \label{hyperthree}
\ee
The action  (\ref{hyperone}) also has a scaling symmetry
\be
\delta \phi = 2 \epsilon ,\quad \delta \sigma = 2 \sigma \epsilon, \quad \delta \zeta^I = \zeta^I \epsilon, \quad \delta  \tilde \zeta_I = \tilde \zeta_I \epsilon, 
\ee
which is generated by $H$ and satisfies the following commutation relations with the generators of the shifts (\ref{hyperthree}):
\be
[H, \Gamma_I]= {1\over 2}  \; \Gamma_I, \quad [H,\tilde \Gamma^I]={1\over 2} \; \tilde \Gamma^I, \quad [H, E]= E 
\ee
Finally, for  a prepotential of the form (\ref{prepLCY}), there are additional shift symmetries \cite{deWit:1990na,deWit:1992wf} which involve the  real parts of the NS-NS scalars $z^\alpha=x^{\alpha}+ i y^{\alpha}$. We denote the generators of these shift symmetries $B_\alpha$: 
\be
\delta x^\alpha = \beta^\alpha, \quad \delta \zeta^\alpha= \beta^{\alpha} \zeta^0, \quad \delta \tilde{\zeta}_0=-\beta^\alpha \tilde{\zeta}_\alpha, \quad \delta \tilde{\zeta}_\alpha= -  C_{\alpha \beta \gamma} \beta^\beta \zeta^\gamma, \quad \delta \sigma=0 \label{xshift}
\ee
These extra axionic symmetries satisfy the commutation relations:
\be [\Gamma_0, B_\alpha]=-\Gamma_\alpha, \quad [\Gamma_\alpha, B_\beta]= C_{\alpha \beta \gamma} \tilde \Gamma^\gamma, \quad [\tilde \Gamma^0, B_\beta ]=0, \quad [\tilde \Gamma^\alpha, B_\beta ]=\delta^{\alpha}_{\beta} \tilde \Gamma^0, \quad [E,B_\beta]=0 \ee

\subsection{Quaternionic coset actions}
The scalars in extended supergravities with more than eight supersymmetries   are always described by sigma models with target spaces which are coset manifolds ${\cal M}_{scalar}=G/H$. Here $G$ is a noncompact group and $H$ is a maximal compact subgroup. The simplest  example is a $SL(2,R)/U(1)$ coset sigma-model with Lagrangian
\be 
 {\cal L}= {1\over 2} \Big( \partial_\mu \phi \partial^\mu \phi + e^{a\phi } \partial_\mu\chi \partial^\mu \chi\Big) \label{hyperthree}
 \ee
 
 The value of the constant $a$ depends on the theory. 
For example, in ten dimensions for $a=2$ one gets the action of the dilaton/axion scalars of IIB supergravity \cite{Schwarz:1983qr,Howe:1983sr}.  In the case of $N=2$ theories with hypermultiplets, the action (\ref{hyperthree}) will appear as a subsector of the full hypermultiplet action.  Instanton and wormhole solution  for the action (\ref{hyperthree}), in various dimensions and for various values of the parameter $a$, have been discussed in the literature \cite{Gutperle:2002km,ArkaniHamed:2007js,Bergshoeff:2004fq,Einhorn:2002sj}

\begin{table}[htdp]
\begin{center}\label{tabletwo}
\begin{tabular}{|c|c|c |c|} \hline
 coset  manifold $G/H$& dim($G$) & dim($H$) &  $n_H$   \\ \hline
  ${SU(n,2) / U(n)\times SU(2)}$ &$(n+2)^2-1$& $n^2+3$ & $ n$\\ \hline 
 ${SO(n,4)/ SO(n) \times SU(2)\times SU(2)}$ & $(n+4)(n+3)/2$ & $n(n-1)/2+6$& $n$  \\ \hline 
 ${Sp(n,1) /  Sp(n)\times SU(2)}$ & $(n+1)(2n+3) $&$ n(2n+1) +3$&$n$ \\  \hline
\end{tabular}
\caption{Infinite series of quaternionic coset spaces}
\end{center}
\end{table}

There are cases, where the full hypemultiplet action (\ref{hyperone}) parameterizes a coset manifold. First,  there are three infinite sets of cosets, which are non-compact versions of Wolf spaces \cite{wolfa,alexa}.
The first row  of table 2 with  $n=1$ is the  $S(2,1)/SU(2)\times U(1)$ coset which parameterizes the universal hypermultiplet and will be discussed briefly in section 4.

Second,  there are exceptional cosets which are given in table 3.  The coset ${G_{2,2}\over SU(2)\times SU(2) }$  is an eight dimensional quaternionic manifold which will be discussed in detail in section 5. Note that an  hypermultiplet sigma model coming from a generic  Calabi-Yau compactification will in general not be a coset manifold.  The increased symmetry of the coset manifolds makes it easier to find explicit instanton and wormhole solutions.


\begin{table}[htdp]
\begin{center}\label{tablethree}
\begin{tabular}{|c|c|c  |c|} \hline
 coset  manifold $G/H$& dim($G$) & dim($H$) & $n_H$\\ \hline
  ${G_{2,2}/ SU(2)\times SU(2) }$ & 14&  6&  $2$\\ \hline 
   ${F_{4,4}/ Sp(3)\times SU(2) }$ &52 &  24&   7\\ \hline 
    $E_{6(+2)} / SU(6)\times SU(2) $ &78 & 38 &  10\\ \hline 
        $E_{7(-5)} / SO(12)\times SU(2) $ &133 & 69&  16\\ \hline 
            $E_{8(-24)} / E_7 \times SU(2) $ &248 &136 &  28\\ \hline 
 \end{tabular}
\caption{Exceptional quaternionic coset spaces}
\end{center}
\end{table}

\section{Euclidean instantons and wormholes}
\setcounter{equation}{0}
In a semiclassical approximation,  instantons and wormholes are viewed as saddle-points of the Euclidean action, i.e. they are solutions to the classical Euclidean equations of motion.  Instantons with finite action  can provide an important contribution in the path integral calculation of   some processes. 


In a theory with pseudo-scalars fields that posses shift symmetries (the so-called axionic scalars), the analytic continuation from Minkowskian to Euclidean signature is non-trivial. In particular, regular instanton and wormhole solutions which carry charges associated with axionic scalars only exist if the sign of the kinetic terms for the axionic scalars are flipped  as the theory is continued from Minkowskian to Euclidean signature. Since axionic scalars are ubiquitous in supergravity and string theory, it is important to have a sensible prescription for the analytic continuatin in order to study non-perturbative effects in string theory.

A first approach  is to dualize (in four dimensions) axions to rank-three  antisymmetric tensor fields  \cite{Giddings:1987cg,Green:1997tv} and rewrite the hypermultiplet action as a tensor multiplet action
 \cite {Theis:2002er}. For this theory the analytic continuation to Euclidean signature poses no problems and one obtains a positive-definite action.  Dualization and analytic continuation, however, do not commute and one has to pick the order above to obtain a sensible result.

 A more formal approach is to replace the Minkowskian quaternionic geometry by a para-quaternionic geometry in Euclidean space.  
  \cite{Gibbons:1995vg,Cortes:2005uq} and to define the theory in Euclidean spacetime 
 from the beginning. 
  
   In the following we discuss  a third approach, originally formulated in a paper by Coleman and Lee \cite{Coleman:1989zu}. This method  was applied  to   the axion of the $SL(2,R)/U(1)$ coset in \cite{ArkaniHamed:2007js}    and to the universal hypermultiplet in  \cite{Chiodaroli:2008rj}. Here we want to apply the method to a general hypermultiplet action.

\subsection{The Coleman approach}

In this section, we consider imaginary-time transition amplitudes between initial and final states with constant values of the hypermultiplet fields:
\bea
| I \rangle &=& |z^\alpha_0, \phi_0, \zeta^I_0, \tilde{\zeta}_{I0},\sigma_0 \rangle \nonumber \\ 
| F \rangle &=& |z^\alpha_F, \phi_F, \zeta^I_F, \tilde{\zeta}_{IF},\sigma_F \rangle
\eea
Following the approach of \cite{Coleman:1989zu}, we can project initial and final states into eigenspaces of the shift symmetry charge densities inserting delta function projectors of the form:
\be
\delta (\rho_{0,F}-j^\tau_{S})|I,F\rangle \propto \int{\mathcal{D}\alpha \exp{\left\{ -i \int{d^3\vec{x} \; \alpha(\vec{x}) [\rho_{0,F}(\vec{x})-j^\tau_S(\vec{x})]} \right\} }}|I,F\rangle
\ee
Here $j^\tau_S(\vec{x})$ is the Noether charge density corresponding to the shift of some field $\chi$.
\be
j^\tau_S(\vec{x})= \frac{\delta S}{\delta \partial_\tau \chi (\vec{x})}  
\ee
$\rho_0(\vec{x})$ and $\rho_F(\vec{x})$ are the charge density eigenvalues.
The overall amplitude can be obtained summing over the charge density eigenspaces. As we shall see, each term of this sum can be expressed through a path integral dominated by a single  saddle-point of the Euclidean action. These saddle-points are exactly the instantons and wormholes which we will analyze in this paper.

The hypermultiplet action shift charge densities obey to commutation relations of the form:
\be
[j^\tau_{I}(\vec{x},\tau),j^\tau_{n_H+J}(\vec{y},\tau)]= \delta_{IJ} \delta^3(\vec{x}-\vec{y}) j^\tau_E (\vec{x},\tau) \label{cdensities}
\ee 
where $j^\tau_I$, $j^\tau_{n_H+J}$ and $j^\tau_E$ are the charge densities associated to the symmetries $\Gamma_I$, $\tilde \Gamma^J$ and $E$. Because of the non-trivial commutation relation (\ref{cdensities}), initial and final states cannot be projected into eigenspaces of all the $2n_H+1$ shift densities. Instead, $|I\rangle$ and $|F\rangle$ can be decomposed into irreducible representations of the $n_{H}$-dimensional Heisenberg group $H_n$ generated by the shift symmetries. 
Elements of such representations can be labeled by the charge density eigenvalues of a properly chosen set of commuting generators.

According to the Stone-Von Neumann theorem, there is a unique unitary irreducible representation of the Heisenberg group $H_n$ for each  value the central element $E$. 
There are two qualitatively different cases:

\medskip


 \noindent $\bullet$ { Vanishing $E$ charge density:}  The central element in the algebra (\ref{cdensities}) is zero and it follows from the Stone-Von Neumann theorem that we can project initial and final states into eigenspaces of all the shift densities $j^\tau_a$ , $a=0,1, \cdots, 2n_H-1$. Saddle points of this kind correspond to instantons and wormholes charged only under the RR scalars shift symmetries (pure RR-charged instantons and wormholes).

\medskip

  \noindent $\bullet$  {Nonzero $E$ charge density:}  If we project initial and final states into eigenspaces of non-zero $E$ charge density, elements within the same eigenspace belong to a unique irreducible representation of $H_n$ with non-zero value of the central element. In the analogy with standard quantum mechanics, $j^\tau_{I},I=0,1,\cdots n_H -1$ play the role of position operators and $j^\tau_{n_H+J}, J=0,1,\cdots n_H-1$ play the role of momentum operators. $j^\tau_E $ is a central element and can be identified with $\hbar$.
These saddle-points correspond to instantons and wormholes charged under the NS-NS shift symmetry (NS-charged instantons and wormholes). 

\subsection{Classification of analytic continuations \label{ac}}

In this  section we will classify the possible analytic continuations corresponding to the two cases discussed above. The analytic continuation for the pure RR-charged and the NS-charged cases are different. Morover, in case of a prepotential of the form (\ref{prepLCY}), we need to take into account the extra shift symmetries $B_\alpha$. We will see in the next section that extra conditions need to be satisfied in order for the action to have a real saddle-point after anayltic continuation.

\subsubsection{Pure RR-charged case}

For analyzing pure RR-charged instanton and wormhole solutions it is convenient to define:
\be
\chi^I=\zeta^I, \qquad \chi^{n_H+I}=\tilde{\zeta}_I, \qquad \tilde{\sigma}=\sigma + \frac{1}{2} \chi^I \chi^{n_H+I}
\ee
After analytic continuation to Euclidean time $t \rightarrow -i \tau $, the action (\ref{hyperone}) can be rewritten as follows:
\be S_E = \int d^4x \sqrt{g} \left\{ -R + 2 g_{\alpha \bar{\beta}} \partial_{\mu} z^{\alpha} \partial^{\mu} \bar{z}^{\bar{\beta}}  + \frac{1}{2} (\partial_{\mu} \phi)^2 +\frac{1}{2} e^{2\phi} (j_{E \; \mu})^2 + \frac{1}{2} e^{- \phi} M_{a b} \partial_{\mu} \chi^a \partial^{\mu} \chi^b \right\} \label{S-pureD}
\ee
With $a=0 \dots 2 n_H-1$. The matrix $M$ is positive-definite and given by: 
\be
M=\left( \begin{array}{cc} I^{-1} & I^{-1} R \\ R I^{-1} & R I^{-1} R + I  \end{array} \right) \label{matrixM}
\ee
with the matrices $R$ and $I$ from equation (\ref{R&I}). We project initial and final state into charge density eigenspaces of the commuting $2 n_H$ shift charges:
 \bea P_I|I \rangle &=& \delta(j_E^\tau) \prod_{a} \delta(\rho_{a0}-j^\tau_a) |I\rangle \nn \\
 &\propto &\int{ \mathcal{D}\alpha \prod_{a} \mathcal{D}\gamma^a e^{i \int{d^3 \vec{x} \alpha j_E^\tau}} e^{-i \int{d^3\vec{x} \gamma^a (\rho_{a0}-j^\tau_a)}}}|z^\alpha_0,\phi_0,\chi^a_0, \tilde{\sigma}_0 \rangle \nonumber \\
 &=& \int \mathcal{D} \alpha \prod_{a} \mathcal{D}\gamma^a e^{-i \int{d^3\vec{x} \gamma^a \rho_{a0} }}|z^\alpha_0, \phi_0, \chi_0^a+\gamma^a, \tilde{\sigma}_0 + \gamma^{n_H+I} \chi^I + \alpha \rangle \eea 
Redefining:
\be \chi^a \rightarrow \chi^a-\gamma^a, \qquad \tilde{\sigma} \rightarrow \tilde{\sigma} -\gamma^{n_H+I} \chi^I- \alpha \label{redefD} \ee
The transition amplitude becomes:

\be 
\langle F |P_F e^{-H (\tau_F- \tau_0)} \delta( j_E^0) P_I |I \rangle=
e^{i \int (\rho_{a0}\chi_0^a -\rho_{aF} \chi^a_F) } \int \mathcal{D}\Phi e^{-(S_E+\Sigma)}
\ee 
Where $\Sigma$ is a surface term given by:
\be
\Sigma=i \int{d^3\vec{x} [\rho_{a0}(\vec{x}) \chi^a (\vec{x},\tau_0)-\rho_{aF}(\vec{x}) \chi^a(\vec{x},\tau_F)]}
\label{Sigma-D} \ee
As an effect of the redefinition (\ref{redefD}), the functional integration of the fields $\chi^a$ and $\sigma$ goes over configurations without fixed initial and final values. In particular $\chi^a(\vec{x},\tau_{0,F})$ and $\tilde{\sigma}(\vec{x},\tau_{0,F})$ do not equal $\chi^a_{0,F}(\vec{x})$ and $\tilde{\sigma}_{0,F}(\vec{x})$. 
Varying the action with respect to $\chi^a$ and $\tilde \sigma$ on the boundary leads to:
\bea
 \left. \frac{\delta S_E}{\delta \partial_\tau \chi^a } \right|_{\tau_{0,F}} &=&  j^\tau_a = i \rho_{a0,F} \nonumber \\
\left. \frac{\delta S_E}{\delta \partial_\tau \tilde \sigma} \right|_{\tau_{0,F}} &=& j_E^\tau = 0 \label{EOM-pureD}
\eea
We can see that because of the surface term $\Sigma$, the path integral is dominated by a complex saddle-point. In order to evaluate the path integral with the semiclassical approximation, we have to analytically continue:
\be
\chi^a \rightarrow  i \chi'^a, \quad \tilde{\sigma} \rightarrow \tilde{\sigma}'  \label{AC-pureD} 
\ee
while the currents are continued as:
\be j^\tau_a \rightarrow i j'^\tau_a \ee
After this analytic continuation, the action has a real saddle-point and the analytically continued currents obey the following relation
\bea
j'^\tau_a \nonumber &=&  \rho_{a0,F} \\
j'^\tau_E &=& 0 \eea 

\subsubsection{NS-charged case}
In case of NS-charged instanton and wormholes solutions, the choice of projectors is not unique. For every $Sp(2n_H,R)$ matrix $S$, we can define: 
\be \left( \begin{array}{c} \chi^I \\ \chi^{n_H + I} \end{array} \right) = S \left( \begin{array}{c} \zeta^I \\ \tilde{\zeta}_I  \end{array} \right) \qquad \tilde{\sigma}=\sigma+ \frac{1}{2} \chi^I \chi^{n_H +I} 
\ee
and use the shift symmetries of the $\chi^I$ ($I=0 \dots n_H-1$) as the commuting generators. The Euclidean hypermultiplet action can then be rewritten as:
\bea  S_E &=& \int{d^4x \sqrt{g} \left\{ -R + 2 g_{\alpha \bar{\beta}} \partial_{\mu} z^{\alpha} \partial^{\mu} \bar{z}^{\bar{\beta}}  + \right. } \frac{1}{2} (\partial_{\mu} \phi)^2 +\frac{1}{2} e^{-2\phi} (\partial_{\mu}\tilde{\sigma}-\chi^{n_H+I} \partial_{\mu} \chi^I )^2 \nonumber \\ && \left. +
 \frac{1}{2} e^{- \phi} \tilde{M}_{a b} \partial_{\mu} \chi^a \partial^{\mu} \chi^b \right\}  \label{S-NScharged}
\eea
With this notation $\tilde{M}=S^T M S$, $M$ given by (\ref{matrixM}) and $\tilde M$ is positive-definite. Now we can project initial and final states into charge density eigenspaces corresponding to the shifts of $\tilde{\sigma}$ and $\chi^I$:  
\bea P_I |I\rangle &=&\delta(\rho_E - j_E^\tau) \prod_{I} \delta(\rho_{I0}-j_I^\tau) |I\rangle \nn \\
 &\propto& \int{\mathcal{D}\alpha \prod_{I} \mathcal{D}\gamma^I e^{-i \int{d^3\vec{x} (\alpha \rho_{E 0} + \gamma^I \rho_{I0} )}}|z^\alpha_0, \phi_0, \chi_0^I+\gamma^I, \chi_0^{n_H+I}, \tilde{\sigma}+\alpha \rangle }
\eea 
The transition amplitude becomes: 
\be
\langle F | P_F e^{-H (\tau_F- \tau_0)} P_I |I \rangle = \quad e^{i \int ( \rho_{I0} \chi_0^I+ \rho_{E \; 0} \tilde{\sigma}_0 - \rho_{IF} \chi^I_F - \rho_{E \; F} \tilde{\sigma}_F ) } \int \mathcal{D}\Phi e^{-(S_E+\Sigma)}
 \ee 
The surface term $\Sigma$ is given by:
\be
\Sigma=i \int{d^3\vec{x} [\rho_{I0} (\vec{x}) \chi^I (\vec{x},\tau_0) + \rho_{E \; 0} (\vec{x}) \tilde{\sigma}(\vec{x},\tau_0)- \rho_{IF}(\vec{x}) \chi^I(\vec{x},\tau_F) -\rho_{E \; F}(\vec{x})\tilde{\sigma}(\vec{x},\tau_F) ]}
\ee
As in the pure RR-charged case, the functional integration of the fields $\chi^I$ and $\tilde{\sigma}$ goes over configurations which do not have fixed initial and final values. In order to evaluate the path integral with the semiclassical approximation, we have to use a different kind of analytic continuation:
\be  \begin{array}{rcl} \chi^I &\rightarrow& i \chi'^I \\
\chi^{n_H+I} &\rightarrow& \chi'^{n_H+I} \\
\tilde{\sigma} &\rightarrow& i \tilde{\sigma}'  \end{array}  \label{AC-NS} \ee
Varying the action with respect to $\chi^I$ and $\tilde{\sigma}$ on the boundary leads to:
\bea
j'^{\tau}_I(\vec{x},\tau_{0,F}) &=& \rho_{I0,F}  \nonumber \\
j'^\tau_E(\vec{x},\tau_{0,F}) &=& \rho_{E \; 0,F} \label{EOM-NS} \eea

\subsubsection{Large Calabi-Yau Manifolds}

In case of a prepotential of the form (\ref{prepLCY}), there are extra shift symmetries corresponding to the shift of the $n_H-1$ NS-NS axions $x^{\alpha}$. In the general case, the number of commuting generators is still $n_H$, but there are extra possible choices for the analytic continuation. \\
In particular, the generators $\tilde \Gamma^{\alpha}$, $B_{\alpha}$ and $\tilde \Gamma^0$ form a $n_H -1$ dimensional Heisenbeg albegra with central element $\tilde \Gamma^0$. 
We can re-define some of the axions so that $n_H$ of above simmetries become simple shift symmetries:
\bea \chi^{0} &=& \tilde \zeta_0 + {1 \over 2} x^\alpha \tilde \zeta_\alpha + {1 \over 2} \chi^\alpha \chi^{n_H + \alpha} + {1 \over 12} C_{\alpha \beta \gamma } x^\alpha x^\beta x^\gamma \zeta^0 \nonumber \\
\left(  \begin{array}{c} \chi^\alpha \\ \chi^{n_H + \alpha} \end{array} \right) &=& S \left( \begin{array}{c}  x^\alpha \\ \tilde \zeta_\alpha +  C_{\alpha \beta \gamma} x^\beta \zeta^\gamma - {1 \over 2} C_{\alpha \beta \gamma}    x^\beta x^\gamma \zeta^0 \end{array} \right) \nonumber \\
\chi^{n_H} &=& \sigma - {1 \over 2} \z^I \tilde \z_I -{1 \over 2} C_{\alpha \beta \gamma} x^\alpha \z^\beta \z ^\gamma + {1 \over 2} C_{\alpha \beta \gamma} x^\alpha x^\beta \z^\gamma \z^0 - {1 \over 6} C_{\alpha \beta \gamma} x^\alpha x^\beta x^\gamma {\zeta^0}^2   \nonumber \\
\chi^{2n_H} &=& \zeta^0  \nonumber \\
\chi^{2n_H+ \alpha } &=& \zeta^\alpha - x^\alpha \zeta^0 \eea
Here $S$ is a $Sp(2n_H-2, R)$ matrix. We can then project initial and final states into charge density eigenstates corresponding to the shifts of $\chi^I$ with $I = 0 \dots n_H$ as done in the NS-charged case. We obtain a third class of analytic continuations:
\bea  \chi^I & \rightarrow & i \chi'^I \nn \\ 
\chi^{n_H+\alpha} & \rightarrow & \chi'^{n_H+\alpha} \nn \\  
\chi^{2n_H+I} &\rightarrow& \chi'^{2n_H+I}  \label{AC-CY1} \eea
Similarly, the case of vanishing $\tilde \Gamma^0$ leads to $2 n_H -2 $ commuting generators and is analogous to the pure RR-charged case. 

\subsection{Positive definite action for instantons and wormholes}

In this section we will study the conditins which need to be satisfied in order for the actions (\ref{S-pureD}) and (\ref{S-NScharged}) to be  positive-definite  for instanton and wormhole solutions. 
As we shall see, the boundary term introduced in section 3.2  is essential to obtain a positive-definite action. As we shall see, this condition restricts the possible analytic continuations in some cases.
\subsubsection{Pure RR-charged case}
In case of pure RR-charged  instanton and wormhole solutions, the RR scalars shift symmetries have the following Noether currents:
\be 
j'_{a\mu} =  e^{-\phi'} M_{ab} \partial_\mu \chi'^b \\
\ee
Using the equations of motion (\ref{EOM-pureD}) and the analytic continuation (\ref{AC-pureD}), the surface term can be rewritten:
\be \Sigma = \int{d^4x \partial_{\mu}(\sqrt{g} j'^{\mu}_a \chi'^a})= \int{d^4x \sqrt{g}\partial_\mu \chi'^a j'^{ \mu}_a } =\int{d^4x \sqrt{g} (e^{-\phi'} M_{ab} \partial_\mu \chi'^b \partial^\mu \chi'^b) } \label{boundprr}
\ee
The analytic continuation (\ref{AC-pureD}) flips the sign of the kinetic term for the $\chi^a$ fields in the  bulk part of the action (\ref{S-pureD}). As discussed before this flip of the sign in the action is essential for the existence of regular instanton and wormhole solutions. After the flip of the sign, the bulk part of the action is not positive-definite. However, adding the boundary term  (\ref{boundprr}) makes $S_E+\Sigma$ manifestly positive-definite:
\be
S_E+\Sigma = \int{d^4x \sqrt{g} \left\{ -R + 2 g_{\alpha \bar{\beta}} \partial_{\mu} z'^{\alpha} \partial^{\mu} \bar{z}'^{\bar{\beta}}  + \frac{1}{2}(\partial_\mu \phi')^2  + \frac{1}{2} e^{- \phi'} M_{a b} \partial_{\mu} \chi'^b \partial^{\mu} \chi'^b \right\} } \ee

\subsubsection{NS-charged case}
For the NS-charged case we perform the analytic continuation (\ref{AC-NS}). After analytic continuation the relevant shift symmetries have Noether currents:
\bea
j'_{I\mu} &=&  e^{-\phi'} \tilde M_{IJ} \partial_\mu \chi'^I - i e^{-\phi'} \tilde M_{I \; n_H+J} \partial_\mu \chi'^{n_H+J} - \delta_{IJ} \chi'^{n_H+J} j'_{E \mu} \label{jmu-NS}  \\
j'_{E \mu} &=&e^{-2 \phi'} (\partial_\mu \tilde{\sigma}' - \chi'^{n_H +I} \partial_\mu \chi'^{I}) 
\eea 
Because of the second term of (\ref{jmu-NS}), the analytically continued action will not have a real saddle-point in the general case. Using the equations of motion (\ref{EOM-NS}) and the analytic continuation (\ref{AC-NS}), the surface term can be rewritten as follows:
\bea \Sigma &=& \int{d^4x \partial_{\mu}[\sqrt{g} (j'^{\mu}_I \chi'^I+ j'^\mu_E \tilde{\sigma}'})]=  \int{d^4x \sqrt{g}(\partial_\mu \chi'^I j'^{ \mu}_I + \partial_\mu \tilde{\sigma}' j'^\mu_E)} \nonumber \\
&=& \int{d^4x \sqrt{g} [e^{-\phi'} \tilde M_{IJ} \partial_\mu \chi'^I \partial^\mu \chi'^J -i e^{-\phi'} \tilde M_{I \; n_H+J} \partial_\mu \chi'^I \partial^\mu \chi'^{n_H+J} + (j'_{E \;\mu})^2  ] } \label{Sigma-NS}
\eea
The action (\ref{S-NScharged}) after the analytic continuation (\ref{AC-NS}) can be written as:
\bea  S_E &=& \int{d^4x \sqrt{g} \left\{ -R + 2 g_{\alpha \bar{\beta}} \partial_{\mu} z'^{\alpha} \partial^{\mu} \bar{z}'^{\bar{\beta}}  + \right. } \frac{1}{2} (\partial_{\mu} \phi')^2 -\frac{1}{2} e^{-2\phi'} (j'_{E \;\mu} )^2 - \frac{1}{2} e^{- \phi'} \tilde{M}_{I J} \partial_{\mu} \chi'^I \partial^{\mu} \chi'^J \nonumber \\ && \left. + i
e^{- \phi'} \tilde{M}_{I \; n_H+J} \partial_{\mu} \chi'^I \partial^{\mu} \chi'^{n_H+J} + \frac{1}{2} e^{- \phi'} \tilde{M}_{n_H+I \; n_H+J} \partial_{\mu} \chi'^{n_H+I} \partial^{\mu} \chi'^{n_H+J} \right\} \label{S-NScharged2}
\eea
$S_E+\Sigma$ becomes:
\bea
S_E+\Sigma &=& \int{d^4x \sqrt{g} \left\{ -R + 2 g_{\alpha \bar{\beta}} \partial_{\mu} z'^{\alpha} \partial^{\mu} \bar{z}'^{\bar{\beta}}  + \right. } \frac{1}{2} e^{2 \phi'} (j'_{E \; \mu})^2 + \frac{1}{2} e^{- \phi'} \tilde{M}_{I J} \partial_{\mu} \chi'^I \partial^{\mu} \chi'^J \nonumber \\ && \left. +\frac{1}{2} e^{-\phi'} \tilde{M}_{n_H+I \; n_H+J} \partial_\mu \chi'^{n_H + I} \partial^\mu \chi'^{n_H + J}  \right\}  \label{instactthree}
\eea
Note the first term in the second line of the equation for the euclidean action  (\ref{S-NScharged2})  is imaginary. The equations of motion derived from (\ref{S-NScharged2}) imply that the solution is complex. A further analytic continuation $\chi^{n_{H}+I} \to i \chi^{n_{H}+I}, \sigma\to i \sigma$  can be used to obtain real equations of motion. Note, however, that in this case the total action $S_{E}+\Sigma$ is not positive definite anymore.   Consequently, it is not guaranteed that after the further analytic continuation  saddle point solution will give a dominant contribution to the path integral. For example, it might be possible that for some assignments of  charges the instanton action could be negative and the multi instanton contribution in the dilute gas approximation would diverge. 

One way to obtain real positive definite saddle point solutions,  is to impose the following condition on the solution
\be 
\tilde M_{I \; n_H+J} \partial_\mu \chi'^I \partial^\mu \chi'^{n_H+J}  = 0 \label{realcond}
\ee 
This eliminates the imaginary part in the equations of motion derived from (\ref{S-NScharged2}). Consequently, the solution are real and the saddle point action including the boundary term (\ref{instactthree}) is positive definite. 

Note, however, that the condition is not a constant of motion and it will lead to a truncation of the space of all solutions of the equations of motion.
In the following we will impose the condition (\ref{realcond}) and discuss the resulting truncations   in section 5 using the concrete example of the $G_{2,2}/SU(2)\times SU(2)$ coset. 

It is an important open problem whether the more general solutions, where (\ref{realcond}) is not imposed, make physical sense and contribute to the instanton induced effective action. In this paper we will not discuss the complexified solutions further.

\subsection{SO(4)-invariant solutions}

We will now focus on $SO(4)$ invariant solutions of the equations of motion obtained by varying the actions (\ref{S-pureD}) and (\ref{S-NScharged}). It is convenient to start with the following  ansatz for the Euclidean metric:
\be
ds_E^2={e^{3U}\over 4 \tau^3}d\tau^2+ {e^U\over \tau}d\Omega^2
\ee
Here $d \Omega^2_3$ is the metric of the unit  three-sphere and $\tau $ is a radial coordinate. The ansatz reduces to the flat metric in case $U(\tau)\equiv 0$, with the identification $\tau=1/r^2$. Moreover, as a consequence of the $SO(4)$ invariance, the scalar fields depend only on $\tau$. 
With this choice for the metric, Einstein equation gives:
\bea
   G_{\tau \tau}&=&  \frac{3}{4 \tau^2} (1- e^{2U}-2 \tau \dot U +\tau^2 \dot{U}^2)= T_{\tau \tau} \nonumber \\
G_{\alpha \beta }&=& e^{-2U} (1-e^{2U}+6 \tau \dot U - 3 \tau^2 { \dot U}^2 + 4 \tau^2 {\ddot  U}) \eta_{\alpha \beta}=T_{\alpha \beta} \label{EinsteinEq1}
\eea
where the dot indicates a derivative with respect to $\tau$ and $\eta_{\alpha \beta}$ is the metric on the unit three-sphere. The radial and angular components of the energy-momentum tensor can be obtained from  (\ref{S-pureD}) 
for the pure RR-charged case:
\bea
T_{\tau \tau}&=& 2 g_{\alpha \bar{\beta}} \partial_\tau z^{ \prime \alpha } \partial_\tau \bar{z}^{ \prime \bar{\beta} } +\frac{1}{2} (\partial_\tau \phi^\prime )^2 - \frac{1}{2}e^{-\phi^\prime} M_{ab} \partial_\tau \chi^{ \prime a}  \partial_\tau \chi^{ \prime b }\nonumber\\
T_{\alpha \beta} &=& -4 \tau^2 e^{-2U} T_{\tau \tau} \eta_{\alpha \beta} \label{EinsteinEq2}
\eea
For the NS-charged case one gets from (\ref{S-NScharged})
\bea
T_{\tau \tau}&=& 2 g_{\alpha \bar{\beta}} \partial_\tau z^{\prime \alpha } \partial_\tau \bar{z}^{\prime \bar{\beta}} +\frac{1}{2} ( \partial_\tau \phi^\prime )^2 -\frac{1}{2} e^{-2 \phi^\prime} (\partial_\tau \tilde{\sigma}^{\prime} - \chi^{\prime n_H+I} \partial_\tau \chi^{\prime I })^2 - \frac{1}{2}e^{-\phi^\prime} \tilde{M}_{IJ} \partial_\tau \chi^{ \prime I} \partial_\tau \chi^{\prime J} \nonumber \\
&&\;
 + i e^{-\phi^\prime} \tilde{M}_{I\; n_H+J } \partial_\tau \chi^{\prime I} \partial_\tau \chi^{\prime n_H+ J} + \frac{1}{2}e^{-\phi^\prime} \tilde{M}_{n_H+I \; n_H+ J} \partial_\tau \chi^{\prime n_H+I } \partial_\tau \chi^{\prime n_H+J }  \nonumber\\
\eea
A linear combination of the (\ref{EinsteinEq2}) depends only from the function $U(\tau)$  and not on the energy momentum tensor.  This gives a   second order ordinary differential equation for the metric factor $U(\tau)$.
\be \partial^2_\tau U=\frac{e^{2U}-1}{\tau^2} \label{Uode} \ee
All solutions can be brought in a form where $U(\tau)\rightarrow 0$ as $\tau \rightarrow 0$ with a simple rescaling of the radial coordinate $\tau$. The equation (\ref{Uode}) has two linearly independent solution. \\
The first  solution  has the form:
\be e^{U(\tau)}=\frac{4 \sqrt{c} \tau}{\sinh{4 \sqrt{c} \tau}} \label{SingSol}\ee
where $c$ is a positive constant and $\tau$ can assume any value from $0$ to $+ \infty$. These solutions are always singular for $\tau \rightarrow \infty$ and will not be studied in this paper. \\
The second solution  has the form:
\be e^{U(\tau)}=\frac{4 \sqrt{c} \tau}{\sin{4 \sqrt{c} \tau}} \label{WHSol} \ee  
The radial coordinate $\tau$ can assume any value from $0$ to $\pi / 4 \sqrt{c}$. These solutions are regular and exhibit two flat asymptotic regions ($\tau \rightarrow 0$ and $\tau \rightarrow \pi/4 \sqrt{c}$) connected by a wormhole. 
The neck of the wormhole is located at $\tau = \pi / 8 \sqrt{c}$. The area of the three sphere at the neck is given by
\be
Area(S^{3}_{\rm neck})=  2 \pi^{2} R_{\rm neck}^{3}= 16 \pi^{2 } c^{\;3/ 4}
\ee
 
\medskip

 Using equation (\ref{WHSol}) we can rewrite the first equation of (\ref{EinsteinEq1}) as follows for the pure RR-charged case 
\bea
-24c &=& 2 g_{\alpha \bar{\beta}} \partial_\tau z^{\prime \alpha} \partial_\tau \bar{z}^{\prime \bar{\beta}} +\frac{1}{2} (\partial_\tau \phi^\prime )^2 - \frac{1}{2}e^{-\phi^\prime} M_{ab} \partial_\tau \chi^{\prime a} \chi^{ \prime b} \label{Constraint1}
\eea
For the NS-charged case (\ref{EinsteinEq2}) one obtains
\bea
-24c &=& g_{\alpha \bar{\beta}} \partial_\tau z^{\prime \alpha}  \partial_\tau \bar{z}^{\prime \bar{\beta}} +\frac{1}{2} (\partial_\tau \phi^\prime )^2 - \frac{1}{2} e^{-2 \phi^\prime} ( \partial_\tau \tilde{\sigma}^{\prime} -\chi^{\prime n_H+I} \partial_\tau \chi^{\prime I })^2 - \frac{1}{2}e^{-\phi^\prime} \tilde{M}_{IJ} \partial_\tau \chi^{ \prime I}  \partial_\tau \chi^{\prime J} \nonumber  \\ 
& &\; + i e^{-\phi^\prime} \tilde{M}_{I\; n_H+J } \partial_\tau \chi^{\prime I } \partial_\tau \chi^{\prime n_H+ J} + \frac{1}{2}e^{-\phi^\prime} \tilde{M}_{n_H+I \; n_H+ J} \partial_\tau \chi^{\prime n_H+I } \partial_\tau \chi^{\prime n_H+J }\label{Constraint2}
\eea

These equations contain only the first derivatives of the fields. The limit $c\rightarrow 0$ of (\ref{WHSol}) gives the flat metric. Solutions of this kind are the extremal instantons.
Note that for the $SO(4)$ symmetric NS-charged solution the reality condition (\ref{realcond})  becomes
\be
e^{-\phi^\prime} \tilde{M}_{I\; n_H+J } \partial_\tau \chi^{\prime I} \partial_\tau \chi^{\prime n_H+ J }=0 \label{realcondb}
\ee
If this condition is satisfied, the constraint (\ref {Constraint2}) can be satisfied by a real solution. As mentioned earlier,  in general (\ref{realcondb}) is not a conserved quantity, i.e.  its time derivative does not vanish  if the equations of motion are satisfied. This implies that  (\ref{realcondb}) imposes severe constraints on the solution since it has to be obeyed for all values of $\tau$. As well shall see for the explicit example of $G_{2,2}/SU(2)\times SU(2)$ coset only a few  consistent truncations 
satisfy  (\ref{realcondb}).

\subsection{BPS-condition, Extremality, non-extremality and attractors} 
In this section  we limit ourselves to $SO(4)$ invariant solutions. A bosonic solution preserves half of the supersymmetries if there exists a linear combination of the supersymmetry parameters $\epsilon^{1}$ and $\epsilon^{2}$ for which the gravitino (\ref{delpsi}) and hyperino variation (\ref{hypervar}) vanish. The gravitino variation determine the radial dependence  of the unbroken supersymmetry, while the hyperino variation will determine which linear combination of the supersymmetries is unbroken.
The condition that the hyperino variation  $\delta_{\epsilon}\xi_{a}=0$ vanishes for the unbroken supersymmetry variation 
 is equivalent to the statement that the the quaternionic vielbein $V$ defined in (\ref{quadvb})
 is a $2n_H \times 2 $ dimensional matrix has non-maximal rank. This is the case if the $2n_{H}$ dimensional columns are proportional.
\be
\left(
\begin{array}{c}
  u_\mu   \\
  \\
   e^A_\mu     \\
   \\
  - \bar E_\mu^A    \\
  \\
 -\bar v_\mu      
\end{array}
\right)=c\left(
\begin{array}{cc}
    v_\mu   \\
    \\
    E_\mu^A   \\
    \\
   \bar e^A_\mu \\
   \\
  \bar  u_\mu    
\end{array}
\right)\label{bpscon}
\ee
where the complex constant $c$ determines the linear combination of the unbroken susy:

\be
\left(
\begin{array}{c}
  \epsilon^{1}  \\
   \epsilon^{2}  \\   
\end{array}
\right)= \epsilon \left(
\begin{array}{c}
  1\\
   c  \\   
\end{array}
\right)
\ee
Note that it follows from (\ref{bpscon}) that
\be\label{bpscontwo}
u_{\tau}\bar u_{\tau}+ v_{\tau}\bar v_{\tau}+ e^{A}_{\tau}\bar e^{A}_{\tau}+  E^{A}_{\tau}\bar E^{A}_{\tau}=0
\ee
Using the explicit form of the vielbein components (\ref{vbein})  and the following identity of special geometry
\be
f_{\alpha}^{I} \bar f_{\beta}^{J} g^{\alpha\bar\beta}+ e^{K}\bar Z^{I}Z^{J}= {1\over 2} (I^{-1}) ^{ IJ}
\ee
One can show that the left hand side of (\ref{bpscontwo}) is proportional to $T_{\tau\tau}$. It follows that all half-BPS  solutions have $c=0$ and are therefore extremal instanton solutions. Note that the extremality condition (\ref{bpscontwo}) is single equation whereas the BPS conditions (\ref{bpscon}) are $2n_{H}$ equations. It is therefore possible if $n_{H}>1$ to have an extremal solutions which break all supersymmetries.   We will discuss such a case for the $G_{2,2}/SU(2)\times SU(2)$ coset in section \ref{sectionfive}.

For a very similar system of hypermultiplets in five-dimensional supergravity it was shown in \cite{Gutperle:2000ve} (see also \cite{ Behrndt:1997ch,Gunaydin:2005mx,Gunaydin:2007bg} for  discussion  for four-dimensional N=2 supergravity) that the BPS equations for pure RR-charged solutions are equivalent to the BPS attractor equations. The purely RR-charged  instanton solution is related via the c-map to extremal BPS black holes.  Similarly the extremal non-BPS instanton solutions are related to extremal non-BPS black hole solutions. For recent work on the attractor mechanism for extremal non-BPS black holes see e.g.  \cite{Goldstein:2005hq,Kallosh:2006bt,Gaiotto:2007ag,Saraikin:2007jc,Bergshoeff:2008be}.

Note also, that the $SO(4)$ symmetric BPS-instanton solutions that can be mapped by the c-map to BPS black holes  correspond to single center black holes. For black holes in $N=2$ supergravity there are however multicenter black hole solutions which are BPS \cite{Shmakova:1996nz,Behrndt:1997ny,Denef:2000nb,LopesCardoso:2000qm}. 
If the c-map relates these solutions to instantons they would not be $SO(4)$ invariant. Since multicenter black hole solutions are stationary instead of static they would necessarily be NS-charged. It is an interesting question whether such instantons exist and contribute to the path integral in the semiclassical approximation.

The extremal BPS instanton solutions break four of the eight real supersymmetries. The broken supersymmetries induce fermionic zero-modes and correlation function are non-zero only when the zero modes are soaked up by appropriate operator insertions \cite{Green:1997tv,Becker:1995kb}.

Four fermionic zero modes induce non-perturbative four-fermion terms  which couple to the curvature tensor of the $N=2$ sigma model. By supersymmetry such terms are related to kinetic terms for the scalars in the sigma model. Hence instantons provide non-perturbative corrections to the geometry of the quaternionic hypermultiplet moduli space. For a recent discussion of such issues see \cite{deVroome:2006xu,Alexandrov:2006hx,Alexandrov:2008ds,Alexandrov:2008gh}

\section{The $SU(2,1)/U(2)$ coset model}
\setcounter{equation}{0}
The universal hypermultiplet action  can be derived from the general formulae given in section  by setting $n_{H}=1$ and $F=i X_{0} ^{2}/2$. The action  is given by:

\begin{equation}\label{unione}
S= \int d^{4}x \sqrt{g} \;\Big\{ {1\over 2} (\partial_{\mu} \phi)^{2} +{1\over 2}e^{2\phi}(\partial_{\mu} \sigma +\tilde \zeta \partial_{\mu} \zeta)^{2}+{1\over 2}e^{\phi}\big( (\partial_{\mu} \zeta)^{2}+(\partial_{\mu }\tilde\zeta)^{2}\big)\Big\}
\end{equation}
 The possible analytic continuations and associated instanton and wormhole solutions for the universal hypermultiplet were discussed by the authors in a previous paper \cite{Chiodaroli:2008rj}. In this section we briefly review the results in the interest of completeness. 
 
\begin{figure}[htbp]
  \label{figone}
\begin{center}
\centering
\includegraphics[angle=270,scale=0.4]{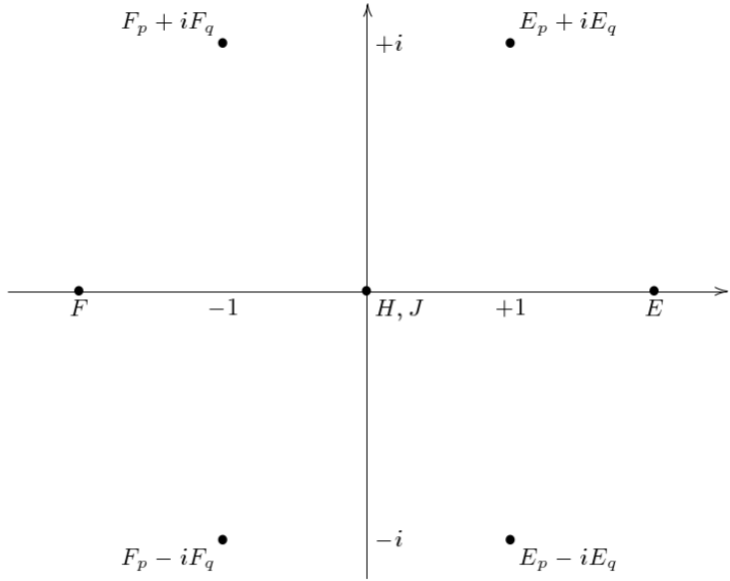}
 \caption{Root diagram of $SU(2,1)$ with identification of the symmetry generators.}
\end{center}
\end{figure}

 The action (\ref{unione}) is the  sigma model action for the $ SU(2,1)/SU(2)\times U(1) $ coset. The coset  has eight  global $SU(2,1)$ symmetries. The eight infinitesimal generators are given by

\bea
 &&  E  \left\{  \barcl \delta \phi  &=&  0\\
		\delta \sigma &=&1\\
		\delta \zeta &=& 0 \\
		\delta \tilde \zeta &=& 0    \ea \right.
		\quad 
		E_{q}\left\{  \barcl \delta \phi  &=&  0\\
		\delta \sigma &=&0\\
		\delta \zeta &=& -\sqrt{2}\\
		\delta \tilde \zeta &=& 0    \ea \right.
\quad
E_{p} \left\{  \barcl \delta \phi  &=&  0\\
		\delta \sigma &=&\sqrt{2}\zeta\\
		\delta \zeta &=& 0 \\
		\delta \tilde \zeta &=& -\sqrt{2}    \ea \right.
		\quad
H \left\{  \barcl \delta \phi  &=&  2\\
		\delta \sigma &=&2\sigma\\
		\delta \zeta &=& \zeta \\
		\delta \tilde \zeta &=& \tilde\zeta    \ea \right.
\no\\
&&  F_{p}  \left\{  \barcl \delta \phi  &=&  \sqrt{2} \tilde \zeta,\\
		\delta \sigma &=& {\sqrt{2}\over 4}( \zeta^{3}- 3\zeta \tilde \zeta^{2})\\
		\delta \zeta &=& \sqrt{2} \left( \sigma +{3\over 2} \zeta \tilde \zeta \right) \\
		\delta \tilde \zeta &=&  - \sqrt{2} \left( e^{\phi}+{3\over 4}  \zeta^{2}-{1\over 4}\tilde \zeta^{2} \right)  \ea \right.
		\quad 
		 F_{q} \left\{  \barcl \delta \phi  &=&  \sqrt{2} \zeta\\
		\delta \sigma &=&  \sqrt{2} \left( \sigma\zeta + e^{\phi}\tilde\zeta +{1\over 2}\tilde \zeta^{3} \right)\\
		\delta \zeta &=& \sqrt{2}\left(- e^{\phi}+{1\over 4} \zeta^{2}-{3\over 4} \tilde\zeta^{2} \right) \\
		\delta \tilde \zeta &=&   \sqrt{2} \left(- \sigma +{1\over 2} \zeta\tilde\zeta \right)    \ea \right. \no\\
	&&J \left\{  \barcl \delta \phi  &=&  0\\
		\delta \sigma &=& {1\over 2}(\tilde\zeta^{2}-\zeta^{2})\\
		\delta \zeta &=& - \tilde \zeta \\
		\delta \tilde \zeta &=& \zeta  \ea \right.  
		\quad	
  F  \left\{  \barcl \delta \phi  &=&  - (2\sigma+\zeta\tilde \zeta)\\
		\delta \sigma &=&  e^{2\phi}-\sigma^{2 } + e^{\phi }\tilde \zeta^{2}-{1\over 16} \zeta^{4}+{3\over 16} \tilde\zeta^{4}+{3\over 8} \zeta^{2}\tilde \zeta^{2}\\
		\delta \zeta &=&\sigma \zeta-e^{\phi}\tilde \zeta -{3\over 4}\zeta^{2}\tilde\zeta-{1\over 4} \tilde\zeta^{3}  \\
		\delta \tilde \zeta &=&    - \sigma \tilde \zeta+e^{\phi}\zeta +{1\over 4}\zeta^{3}-{1\over 4} \zeta \tilde\zeta^{2} \ea \right.
		\quad 
\no\\
		\eea 
The relation of the symmetry generators and the roots of $SU(2,1)$ are given in Figure  \ref{figone}.
The eight global symmetries lead to eight  Noether currents given by:
\begin{equation}
j^{\mu}_{a} = \sum_{i=1}^{4}{\delta  L\over \delta(\partial_{\mu } \Phi_{i})} \delta_{{a}} \Phi_{i}
\end{equation}
The shift symmetries of the NS-NS axion $\sigma$ and the RR axions $\zeta$ and $\tilde \zeta$ are generated by $E,E_{q}$ and $E_{p}$ respectively and form a Heisenberg algebra. We have to distinguish two cases depending on whether the central element vanishes or not.

\begin{itemize}
\item For a zero value of the charge density $j_{E}$, the initial and final states can be projected onto eigenstates of $j_{E_{p}}$ and $j_{E_{q}}$. This case is called "pure RR charged".   Applying the Coleman approach  the path integral is dominated by a complex saddle-point where both $\zeta$ and $\tilde \zeta$ are pure imaginary. We can make the saddle-point real by the analytic continuation 
 \be
\zeta \to  i \zeta' ,\quad  \tilde \zeta \to  i \tilde \zeta' \label{anaconrr}
\ee
\item  For a non-zero value of the charge density $\rho_{E}$, the initial and final states are projected onto eigenstates of fixed $\rho_{E}$ and $\rho_{E_q}$.  This case is called "mixed  NS-R charged". Applying the Coleman approach the path integral is dominated by a complex saddle-point where both $\zeta'$ and $\sigma'$ are pure imaginary. We can make the saddle-point real by the analytic continuation
\be
\zeta \to i \zeta' ,\quad \sigma \to i \sigma' \label{anaconnsr}
\ee
\end{itemize}
In both cases the instanton and wormhole solutions can be expressed in terms of the conserved charges. See  our  previous paper \cite{Chiodaroli:2008rj} for more  details.

\section{The ${ \ts G_{2,2}/ SU(2)\times SU(2)}$ coset model}\label{sectionfive}
\setcounter{equation}{0}


The next simplest example is the quaternionic symmetric space  $G_{2,2}/SU(2)\times SU(2)$. This model has $n_H=2$ and corresponds to the prepotential $F=(X^1)^3/X^0$.  The complex scalar $z$ can be split into real and imaginary part $z=x+i y$.
The Euclidean action  (before any analytic continuations are performed)  is:
\be S_E = \int \sqrt{g} \left\{ \frac{3}{2} \frac{(\partial_\mu x)^2 + (\partial_\mu y)^2}{y^2} + \frac{1}{2}(\partial_\mu \phi)^2 + \frac{1}{2} e^{-2 \phi} (\partial_\mu \sigma - \zeta^I \partial_\mu \tilde{\zeta}_I)^2 + \frac{1}{2} e^{-\phi} \partial_\mu \zeta^T  M \partial^\mu \zeta \right\} \label{action-G22} 
\ee
where the matrix $M_{ab}$ is given by
\be M=\frac{1}{y^3} 
\left(\begin{array}{cccc} 1 & x & -x^3 & 3 x^2 \\
x & y^2/3+x^2 & -x^4-x^2y^2 & 3x^3+2 x y^2  \\
-x^3 & -x^4-x^2 y^2 & (x^2+y^2)^3  & -3(x^2+y^2)^2 x \\
3x^2 & 3x^3+2xy^2 & -3(x^2+y^2)^2 x & (3x^2 +2 y^2)^2-y^4 \ea \right) \label{def-M} \ee
It is convenient to collect the four RR axions into a vector:
\be\label{def-zeta}
\zeta=(\tilde{\zeta}_0, \tilde{\zeta}_1, \zeta^0, \zeta^1) 
\ee
The model has a period matrix:
\be N=\frac{1}{2}\left( \bacc z^3 + 3 z^2 \bar{z} &  - 3 z^2 - 3 z \bar{z} \\ - 3 z^2 - 3 z \bar{z} & 9 z +3 \bar{z}  \ea \right) \ee
Hence  the $I$ and $R$ matrices are given by
\be I=\left( \bacc 3x^2y+y^3 & -3xy \\ -3xy & 3 y  \ea \right) \quad R= \left( \bacc 2x^3 & -3 x^2 \\ -3 x^2 & 6x   \ea \right) \ee

With our conventions, $y$ is positive and $I$ is positive-definite. The one forms $u,v,E,e$ are defined as:
\bea u &=&  \frac{i}{4} y^{-\frac{3}{2}} e^{-\frac{\phi}{2}} \big(- z^3 d \zeta^0 +3  z^2 d\zeta^1 +  d\tilde{\zeta}_0+ z d\tilde{\zeta}_1 \big) \label{viel-G22-u}\\
		v &=&  \frac{1}{2} d\phi + \frac{i}{2} e^{-\phi}(d \sigma - \zeta^I d \tilde{\zeta}_I)  \\
		E &=& \frac{\sqrt{3}}{4} y^{-\frac{3}{2}} e^{-\frac{\phi}{2}} \Big(z\bar{z}^2 d\zeta^0 -\bar{z}(2z+\bar{z}) d\zeta^1 - d\tilde{\zeta}_0 -\frac{2 \bar{z}+ z}{3} d\tilde{\zeta}_1 \Big)  \\ 
		e &=& \frac{\sqrt{3}}{2} \frac{ dz}{y} \label{viel-G22-e}	 \eea

\subsection{Global symmetries and conserved charges}
Since the action he action (\ref{action-G22}) is a sigma model for the $G_{2,2}/SU(2)\times SU(2)$ coset manifold, there are 14 global symmetries which form a  $G_{2,2}$ group. 
\def\latticebody{\drop{}}
\begin{figure}
\centering
\includegraphics[angle=270,scale=0.45]{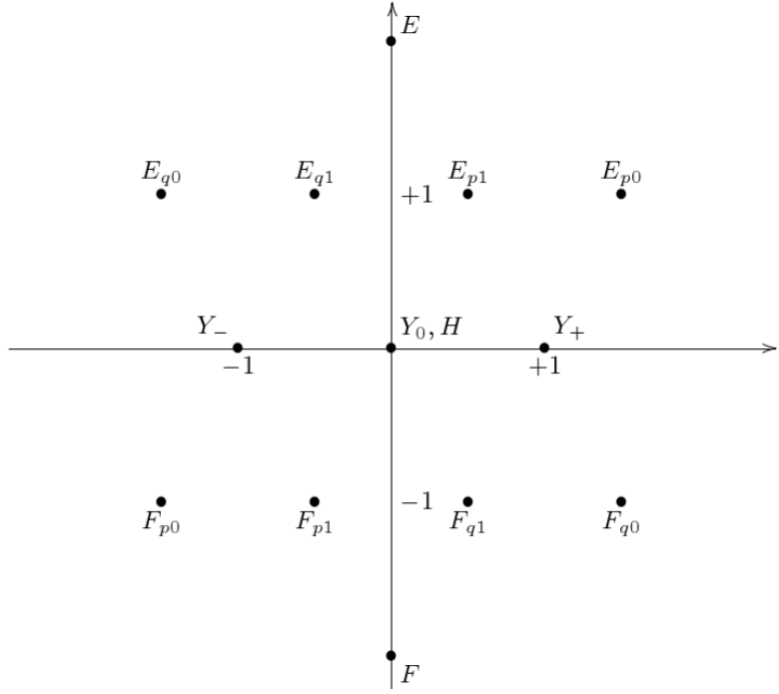}
\caption{ The $G_{2,2}$ root diagram. $Y_0$ and $H$ are the Cartan generators.}
\end{figure}
Our definition of the generators agrees with \cite{Gunaydin:2007qq,Berkooz:2008rj}.:
For completeness we give the explicit form of all generators of the group in appendix A.
Here we display the root diagram in Figure 2.
There are five generators which produce shift symmetries

\bea   E_{p0} \left\{ \barcl \delta x &=& 0 \\
		\delta y &=& 0\\
		\delta \phi &=& 0 \\
		\delta \z^0 &=& \frac{1}{3} \\
		\delta \z^1 &=& 0 \\
		\delta \tiz_0 &=& 0 \\
		\delta \tiz_1 &=& 0 \\
		\delta \s &=& \frac{1}{3} \tiz_0 \ea\right. \;
E_{p1}\left\{ \barcl \delta x &=& 0 \\
		\delta y &=& 0\\
		\delta \phi &=& 0 \\
		\delta \z^0 &=& 0\\
		\delta \z^1 &=& \frac{1}{3} \\
		\delta \tiz_0 &=& 0 \\
		\delta \tiz_1 &=& 0 \\
		\delta \s &=& \frac{1}{3} \tiz_1 \ea\right.\; 
 E_{q0} \left\{ \barcl \delta x &=& 0 \\
		\delta y &=& 0\\
		\delta \phi &=& 0 \\
		\delta \z^0 &=& 0\\
		\delta \z^1 &=& 0 \\
		\delta \tiz_0 &=& \sqrt{3} \\
		\delta \tiz_1 &=& 0 \\
		\delta \s &=& 0 \ea \right. \;
		E_{q1}  \left\{ \barcl \delta x &=& 0 \\
		\delta y &=& 0\\
		\delta \phi &=& 0 \\
		\delta \z^0 &=& 0\\
		\delta \z^1 &=& 0 \\
		\delta \tiz_0 &=& 0 \\
		\delta \tiz_1 &=& \sqrt{3} \\
		\delta \s &=& 0 \ea \right.  \;
		E \left\{ \barcl \delta x &=& 0 \\
		\delta y &=& 0\\
		\delta \phi &=& 0 \\
		\delta \z^0 &=& 0\\
		\delta \z^1 &=& 0\\
		\delta \tiz_0 &=& 0 \\
		\delta \tiz_1 &=& 0 \\
		\delta \s &=& -\frac{1}{2\sqrt{3}} \ea\right.   \label{gen-G22-3a}
		\eea
		These five generators obey a two-dimensional Heisenberg algebra where $E$ is the central element.
\be
[E_{q_{I}}, E_{p_{J}}]= 2 \delta_{IJ} E, \quad \quad I,J=1,2
\ee
It is convenient to express the conserved Noether currents associated to symmetries other than the shift generators (\ref{gen-G22-3a}) in terms of the shift currents (\ref{gen-G22-3a}) using the following relations,
\bea \frac{\delta L}{\delta \partial_\mu \z^I } &=&  3 j^\mu_{E_{pI}} + 2 \sqrt{3} \tiz_I j^\mu_{E} \no\\
 \frac{\delta L}{\delta \partial_\mu \tiz_I} &=& \frac{1}{\sqrt{3}} j^\mu_{E_{qI}} \no\\
 \frac{\delta L}{\delta \partial_\mu \sigma} &=& - 2 \sqrt{3} j^\mu_E \label{def-jE} \eea
For example we obtain the following expressions for the currents associated to the generators $H$ and $Y_0$:
\bea j_H^{\mu}  &=&   3 \z^0 j^{\mu}_{E_{p0}} + 3 \z^1 j^{\mu}_{E_{p1}} + \frac{ \tiz_0 }{ \sqrt{3}} j^{\mu}_{E_{q0}} + \frac{ \tiz_1}{ \sqrt{3}} j^{\mu}_{E_{q1}} -2 \sqrt{3} 
(2\sigma - \z^I \tiz_I)j^{\mu}_E +2 \partial^{\mu} \phi \nn \\
 j^{\mu}_{Y_0}  &=&  \frac{ 3}{ 2} \big( 3 \z^0 j^{\mu}_{E_{p0}} + \z^1 j^{\mu}_{E_{p1}} - \frac{ \tiz_0 j^{\mu}_{E_{q0}}}{ \sqrt{3}} -\frac{ \tiz_1 j^{\mu}_{E_{q1}} }{ 3 \sqrt{3}}\big)  + \sqrt{3} (3 \z^0 \tiz_0 + \z^1 \tiz_1) j^{\mu}_E  -3 \frac{   \partial^\mu |z|^2 }{ 2 y^2} \nn
 \eea  
We have similar expressions for all the symmetry generators in the appendix. Like in the case of the universal hypermultiplet, we focus our analysis on $SO(4)$ invariant solutions. With the spherically symmetric ansatz, the angular components of the Noether currents vanish and the $\tau$ components of the Noether charges are constants.  Since the action (\ref{action-G22}) contains eight scalar fields the equation of motion are eight second order differential equations. 
The expressions for the fourteen conserved charges would allow us, in principle, to reduce the problem to a system of two ODEs and to express the solutions in terms of fourteen charges and two integration constants.\\
However, the problem is algebraically too complex for obtaining the general solution in this way. In the general discussion about the analytic continuation in section \ref{ac}, it became clear that the instanton and wormhole action will be in real only in particular cases. For these reasons we will consider truncations for which the conserved charges can be used to completely solve the equations of motion and for which the action is real. The general solutions can be obtained from the truncated solutions by acting with a group transformation, as outlined in section \ref{general-sol}.\\
Since $G_{2,2}$ has rank two, there are two linearly independent Casimir operators of degree two and six. Like in the case of the $SU(2,1)/SU(2)\times U(1)$ coset, the quadratic Casimir is proportional to the non-extremality parameter $c$. However, the $G_{2,2}/SU(2) \times SU(2) $ solutions do not obey a constraint in terms of the degree six Casimir, which is not in general equal to zero. \\
 
 \subsection{Consistent truncations and instanton solutions \label{cond-trunc}}

The problem greatly simplifies if we consider solutions where only a subset of the eight scalar fields have a non-trivial profile. In order to identically set some fields to constants, we need the truncation to be consistent with the equations of motion. It is easy to express the first derivatives of the RR scalars in terms of the conserved charges:
\be \left( \begin{array}{c} \dot \z^I  \\ \dot \tilde \z_I \end{array}  \right) =e^\phi M^{-1} \left( \begin{array}{c} 3 q_{E_{PI}} + 2 \sqrt{3} \tilde \zeta_I q_E \\ {\ts q_{E_{qI}} \over \ts \sqrt{3}} - 2 \sqrt{3} \zeta^I q_E \ea \right) \ee 
To be able to set to constants some of the RR axions without setting to zero all the four corresponding shift charges, we need the matrix  $M$ which is defined in (\ref{def-M}) to have vanishing nondiagonal terms. This requires the NS axion $x$ to vanish identically as all the non-diagonal terms are linear, quadratic or cubic in $x$. The equations on motion for $x$ give an extra condition for consistency: two RR axions cannot both have non-trivial profile if the corresponding off-diagonal term of (\ref{def-M}) is linear in $x$. Moreover, if we want a solution with $q_E \neq 0$ we need both $\tilde \z_I$ and $\z^I$ to be constant for some $I=0,1$. This leves us with several possible consistent truncations, which we will list in the following.
%
\subsubsection{The RR truncation I}
To get a consistent truncation involving non-trivial RR fields we need to set the modulus $x$ to zero. We also set the shift charge corresponding to the NS axion $\sigma$ to zero obtaining a pure RR truncation.
\be x \equiv 0, \quad \z^1 \equiv \text{const}, \quad \tiz_0 \equiv \text{const}, \quad q_E \equiv 0  \ee 
This truncation is consistent as the $\partial_\tau \z^0 \partial_\tau \tiz_1 $ term in the action is quadratic in $x$. It is easy to see that the charges $q_{E_{p1}}$ and $q_{E_{q0}}$ are equal to zero as well. Moreover, the equations for the conserved charges simplify. It is possible to solve the equations for $q_Y$, $q_{Y_-}$, $q_{Y_+}$ and $q_{H}$ (\ref{eq-qY0}-\ref{eq-qY+}) for the fields $\z^0$, $\z^1$, $\tiz_0$ and $\tiz_1$. We get:
\bea \z^0 &=& \frac{q_H}{12 q_{E_{p0}}}+\frac{q_{Y0}}{6 q_{E_{p0}}}-\frac{1}{6 q_{E_{p0}}} \left(  \partial_\tau \phi - 3 \frac{\partial_\tau y}{y} \right) \\
\z^1 &=& \frac{q_{Y_-}}{3 \sqrt{2} q_{E_{q1}}} \\
\tiz_0 &=& \sqrt{\frac{3}{2}} \frac{q_{E_{p0}} q_{Y_-}}{q^2_{E_{q1}}} - \sqrt{2} \frac{q_{Y_+}}{q_{E_{q1}}} \\
\tiz_1 &=& \frac{3 \sqrt{3} q_{H}}{4 q_{E_{q1}}} - \frac{\sqrt{3} q_{Y_0}}{2 q_{E_{q1}}} - \frac{3 \sqrt{3}}{2 q_{E_{q1}}}\left(\partial_{\tau} \phi + \frac{\partial_\tau y}{y} \right) \label{Sol-RR1} \eea
Substituting these expressions into equation (\ref{eq-qFp0}) and equation (\ref{eq-qFq1}) we get two decoupled ODEs:
\be \Big( \frac{q_H^2}{4} +q_H q_{Y_0} + q_{Y_0}^2+ 2 q_{F_{p0}} q_{E_{p0}} - 2 \frac{q_{E_{p0}}q_{Y_-}^2}{\sqrt{3} q_{E_{q1}}} \Big) - 36 q_{E_{p0}}^2 \frac{e^{\phi}}{y^3} - \Big( 3 \frac{\partial_\tau y}{y} -  \partial_\tau \phi \Big)^2=0 \ee
And: 
\be \Big( \frac{ q_H^2}{4} -\frac{q_H q_{Y_0}}{3}+\frac{q_{Y_0}^2}{9}+{2 \over 3} q_{E_{q1}}q_{F_{q1}} + \frac{2 q_{E_{p0}} q_{Y_-}^2}{3 \sqrt{3} q_{E_{q1}}} -\frac{8 q_{Y_-}q_{Y_+}}{9} \Big) - \frac{4}{3} q_{E_{q1}}^2 e^{\phi} y -  \left(\frac{\partial_\tau y}{y} + \partial_\tau \phi \right)^2 = 0 \ee
The appropriate analytic continuation for this case is:
\be \z^I \rightarrow i \z'^I, \quad \tiz_I \rightarrow i \tiz'_I  \ee
It is convenient to define:
\bea \gamma_1 &=& {q'_{F_{p0}}q'_{E_{p0}} \over 2 } + {q'_{E_{p0}} q'^2_{Y_-} \over 2 \sqrt{3} q'_{E_{q1}}} -{q'^2_H \over 16 } -{q'_H q'_{Y_0} \over 4} - {q^2_{Y_0} \over 4} \nn \\
\gamma_2 &=& {q'_{F_{q1}}q'_{E_{q1}} \over 6 } - {q'_{E_{p0}} q'^2_{Y_-} \over 6 \sqrt{3} q'_{E_{q1}} } + {2 q'_{Y_+} q'_{Y_-} \over 9 } -{q'^2_H \over 16 } + {q'_H q'_{Y_0} \over 12} - {q'^2_{Y_0} \over 36}   \eea
The solution is then given by:
\bea e^{- \eta'_1} &=& { 9 q'^2_{E_{p0}} \over \gamma_1 }  \cos^2 \Big[ \sqrt{ \gamma_1 } (\tau + c_1) \Big] \nn  \\
e^{- \eta'_2} &=& {  q'^2_{E_{q1}} \over 3 \gamma_2 } \cos^2 \Big[ \sqrt{\gamma_2} (\tau + c_2) \Big]  \label{Sol-RR2}  \eea 
And:
\be \phi' = {\eta'_1 + 3 \eta'_2 \over 4}, \qquad y'=e^{\eta'_2-\eta'_1 \over 4} \ee
Equation (\ref{eq-qFq0}) and the equation for $q_F$ simplify and reduce to two constraints on the charges:  
\bea q'_{E_{p0}} q'_{F_{p1}}+q'_{F_{q0}} q'_{E_{q1}} - \sqrt{2} \big( q'_H - {2 \over 3} q'_{Y_0}  \big) { q'_{E_{p0}} q'_{Y_-} \over q'_{E_{q1}} } - \sqrt{ 2 \over 3} (2 q'_{Y_0}- q'_H) q'_{Y_+} &=& 0 \\
q'_{F_{p1}} q'_H - {1 \over \sqrt{2}} q'_{F_{q1}} q'_{Y_-} - \sqrt{2 \over 3} q'_{F_{p0}} q'_{Y_+} - q'_F q'_{E_{q1}} + {2 \sqrt{2} q'_{E_{p0}} q'^3_{Y_-} \over 3 \sqrt{3} q'^2_{E_{q1}} } + && \nn \\
 \big( q'_{E_{p0}} q'_{F_{q0}} - q'^2_H - {2 \over 3} q'_H q'_{Y_0} - {4 \over 3} q'_{Y_-} q'_{Y_+}  \big) {q'_{Y_-} \over \sqrt{2} q'_{E_{q1}} } &=& 0 \label{RR-I-constraints}  \eea
It is instructive to consider the particular case in which the fields $\zeta^1$, $\tilde \zeta_0$ and $\sigma$ are equal to zero. In this case, the charges $q_{Y_\pm}, q_{F_{p1}}, {q_{F_{q0}}}$ and $q_F$ vanish while the constraints (\ref{RR-I-constraints}) are authomatically satisfied and the non-vanishing charges correspond to two perpendicular $SL(2,R)$ subalgebra in root space. Moreover, $\gamma_{1,2}$ are proportional to the quadratic Casimirs of these subalgebras.\\
 
The action of the instanton solution is given by the surface term (\ref{Sigma-D}). Plugging in (\ref{Sol-RR1}) and (\ref{Sol-RR2}) we get: 
\bea S & = & 3 q'_{E_{p0}} \Delta \zeta'^0 + { q'_{E_{q1}} \over \sqrt{3} } \Delta \tilde \zeta'_1 =\left.  -{1 \over 2} \partial_\tau \eta_1 - {3 \over 2} \partial_\tau \eta_2 \right|^{\tau_F}_{\tau_I} \nn  \\
& = & \left. \sqrt{\gamma_1} \tan{\Big[\sqrt{\gamma_1}(\tau + c_1) \Big]} \right|^{\pi \over 4 \sqrt{c}}_0 + 3 \left. \sqrt{\gamma_2} \tan{\Big[\sqrt{\gamma_2}(\tau+c_2) \Big]} \right|^{\pi \over 4 \sqrt{c}}_0  \eea
The non-extremality parameter $c$ can be expressed as:
\be c= {\gamma_1 + 3 \gamma_2 \over 48} \ee
Imposing regularity leads to the conditions:
\be \gamma_{1,2} \ge 0, \quad \sqrt{\gamma_{1,2}} c_{1,2} \ge - {\pi \over 2}, \quad \sqrt{\gamma_{1,2}}\big( {\sqrt{3} \pi \over \sqrt{\gamma_1 + 3 \gamma_2} } + c_{1,2} \big) \le {\pi \over 2}    \ee
In particular, there are non-singular solutions for some values of the integration constants provided that:
\be \gamma_2 \ge {2 \over 3} \gamma_1 \ee
\begin{figure}
\centering
\includegraphics[angle=90,scale=0.38]{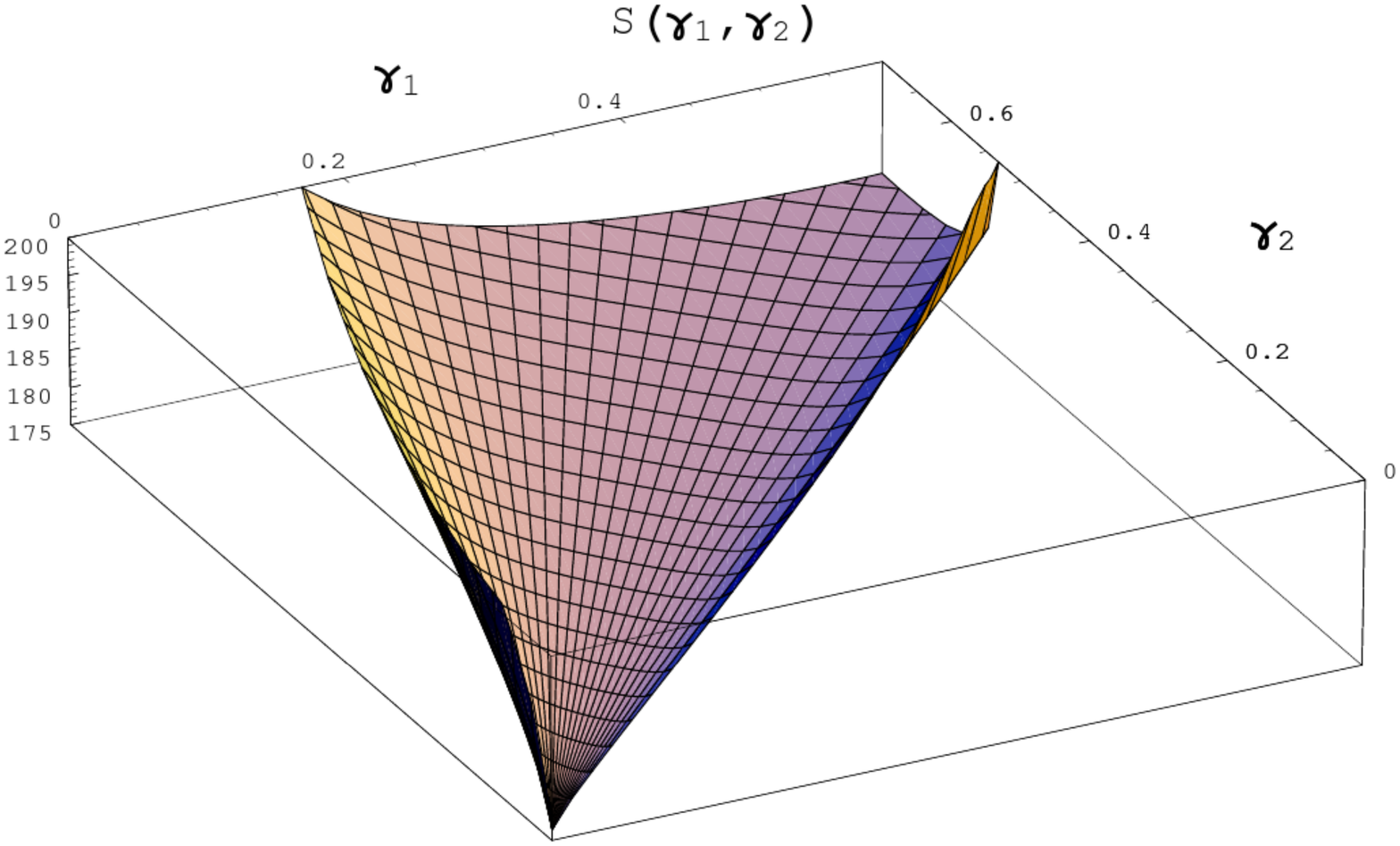}
\caption{ Action of regular RR wormhole solutions as a function of $\gamma_1$ and $\gamma_2$ for fixed values of $\phi$ and $y$ at $\tau=0 $. The plot is obtained setting $e^{-\phi_0}=0.001$, $y_0=10$ and $q'_{E_{q1}}=q'_{E_{p0}}=1$. Extremal instantons correspond to an absolute minimum of the action for $\gamma_1=\gamma_2=0$. }
\end{figure}
It is particularly interesting to consider the limit $g_S \ll 1/y_{0} \ll 1$ with $g_S = e^{- \phi_{0}/ 2}$. In this case the action reduces to:
\be S={ 3 { \ts |q'_{E_{p0}}| \over \ts  g_S  \sqrt{y^3_0} }  \over 1 + { \ts \sqrt{\gamma_1 y^3_{0} }g_S  \over \ts 3 |q'_{E_{p0}}|   } \cot \Big(\sqrt{ \ts 3 \gamma_1 \over \ts \gamma_1 + 3 \gamma_2} \pi \Big) } +{ \sqrt{3} {\ts |q'_{E_{q1}}| \sqrt{y_0} \over \ts g_S}  \over 1 + { \ts \sqrt{3 \gamma_2 } g_S \over \ts  \sqrt{y_0} |q'_{E_{q1}}|   } \cot \Big(\ \sqrt{ \ts 3 \gamma_2 \over \ts  \gamma_1 + 3 \gamma_2} \pi \Big) } \ee 
This expression further simplifies if the value for $\gamma_2$ is not close to ${2 \over 3} \gamma_1$. Note that the action is proportional to $1/g_S$ as expected for D-brane instantons.

\subsubsection{The RR truncation II}
A second consistent truncation has a different pair of non-trivial RR fields.
\be x \equiv 0, \quad \z^0 \equiv \text{const}, \quad \tiz_1 \equiv \text{const}, \quad q_E \equiv 0  \ee 
In this case the solution is given by:
\bea \z^0 &=& -\sqrt{\frac{2}{27}} \frac{q_{Y_-}}{q_{E_{p1}}} - \frac{q_{E_{q0} }q_{Y_+}}{ 3 \sqrt{2} q^2_{E_{p1}}}  \label{RR2-in} \\
\z^1 &=& \frac{ q_{H}}{4 q_{E_{p1}}} + \frac{ q_{Y_0}}{6 q_{E_{p1}}} - \frac{1}{2 q_{E_{p1}}}\left(\partial_{\tau} \phi - \frac{\partial_\tau y}{y} \right) \\
\tiz_0 &=& \frac{\sqrt{3} q_H}{4 q_{E_{q0}}}-\frac{\sqrt{3} q_{Y_0}}{2 q_{E_{q0}}}-\frac{\sqrt{3}}{2 q_{E_{q0}}} \left(  \partial_\tau \phi + 3 \frac{\partial_\tau y}{y} \right)\\
\tiz_1 &=& - \sqrt{3 \over 2} \frac{q_{Y_+}}{q_{E_{p1}}}  \eea
And by:
\bea e^{- \eta'_1} &=& {  q'^2_{E_{q0}} \over 3 \gamma_1} \cos^2 \Big[ \sqrt{\gamma_1} (\tau + c_1) \Big] \nn  \\
e^{- \eta'_2} &=& {  q'^2_{E_{p1}} \over \gamma_2} \cos^2 \Big[ \sqrt{ \gamma_2} (\tau + c_2) \Big] \eea 
with:
\bea  \phi' &=& {3 \eta'_2 + \eta'_1 \over 4}, \qquad y' \; = \;e^{\eta'_1-\eta'_2} \nn  \\
 \gamma_1 &=& {q'_{F_{q0}}q'_{E_{q0}} \over 2 } - {q'_{E_{q0}} q'^2_{Y_+} \over 2 \sqrt{3}  q'_{E_{p1}}} -{q'^2_H \over 16 } +{q'_H q'_{Y_0} \over 4} - {q^2_{Y_0} \over 4} \nn \\
\gamma_2 &=& {q'_{F_{p1}}q'_{E_{p1}} \over 6 } + {q'_{E_{q0}} q'^2_{Y_+} \over 6 \sqrt{3} q'_{E_{p1}} } + {2 q'_{Y_+} q'_{Y_-} \over 9 } -{q'^2_H \over 16 } - {q'_H q'_{Y_0} \over 12} - {q'^2_{Y_0} \over 36}   \label{RR2-fin} \eea
The charges obey to the constraints:
\bea q'_{E_{q0}} q'_{F_{q1}}+q'_{F_{p0}} q'_{E_{p1}} + \sqrt{2} \big( q'_H + {2 \over 3}  q'_{Y_0}  \big) { q'_{E_{q0}} q'_{Y_+} \over q'_{E_{p1}} } + \sqrt{ 2 \over 3} (2 q'_{Y_0}+ q'_H) q'_{Y_-} &=& 0 \label{RR2-con1} \\
q'_{F_{q1}} q'_H + {1 \over \sqrt{2}} q'_{F_{p1}} q'_{Y_+} - \sqrt{2 \over 3} q'_{F_{q0}} q'_{Y_-} + q'_F q'_{E_{p1}} + {2 \sqrt{2} q'_{E_{q0}} q'^3_{Y_+} \over 3 \sqrt{3} q'^2_{E_{p1}} } - && \nn \\
 \big( q'_{E_{q0}} q'_{F_{q0}} - q'^2_H + {2 \over 3}  q'_H q'_{Y_0} - {4 \over 3} q'_{Y_-} q'_{Y_+}  \big) {q'_{Y_+} \over \sqrt{2} q'_{E_{p1}} } &=& 0  \label{RR2-con2} \eea
Like in the case of the RR truncation I, these solutions are regular for some values of the integration constants if $\gamma_2 \geq  2/3 \gamma_1$ and have actions proportional to $1/g_S$ as expected for D-instantons.

\subsubsection{The NS-NS truncation}
A different kind of truncation can be obtained by setting all RR axions to be constant.
\be \z^I = \text{const}, \qquad \tiz_I = \text{const} , \qquad I=0,1\ee
The definitions of the shift currents (\ref{def-jE}) give the following expressions for the RR axions in terms of the charges:
\bea \zeta^0 = { q_{E_{q0}} \over 6 q_E}, & \qquad & \tilde \zeta_0 = - { \sqrt{3} q_{E_{p0}} \over 2 q_E} \\
\zeta^1 = { q_{E_{q1}} \over 6 q_E }, & \qquad & \tilde \zeta_1 = - { \sqrt{3} q_{E_{p1}} \over 2 q_E}    \eea
Substituting equations (\ref{eq-qH}-\ref{eq-qY+}) into the expression for $q_F$ and equations (\ref{eq-qY0}) and (\ref{eq-qY-}) into equation (\ref{eq-qY+}) we are left with two ODE:  
\bea -{(3 q_{E_{p0}} q_{E_{q0}} + q_{E_{p1}} q_{E_{q1}} - 4 q_E q_{Y_0}) ^2 \over 144 q^2_E} + {2 \tilde q_{Y_-} \tilde q_{Y_+}\over 9} + {2 \over 27}  \tilde q^2_{Y_-} y^2  + {(\partial_\tau y)^2  \over y^2} &=& 0   \\
{(q_{E_{p0}} q_{E_{q0}} + q_{E_{p1}} q_{E_{q1}}) ^2 \over 16 q^2_E} - q_E  \tilde q_F  - {q^2_H \over 4} 
 + 12  q^2_E e^{2 \phi}  + { (\partial_\tau \phi)^2  } &=& 0   \eea
In the above equations, we have re-defined the charges $\tilde q_{Y_\pm}$ as:
\be  \tilde q_{Y_-}  =  q_{Y_-} + {\sqrt{6} q_{E_{p1}} q_{E_{q0}} - \sqrt{2} q^2_{E_{q1}} \over 4 q_E }  \qquad \qquad \quad \tilde q_{Y_+}  =  q_{Y_+}- {\sqrt{2} q^2_{E_{p1}} + \sqrt{6} q_{E_{p0}} q_{E_{q1}} \over 4 q_E} \ee
and the charge $\tilde q_F$ as:
\bea  \tilde q_F  = q_F + \Big( {q_{E_{p0}} q_{E_{q0}} \over 2 q^2_E} +{ q_{E_{p1}} q_{E_{q1}}\over 6 q^2_E} \Big) q_{Y_0} - {q^3_{E_{p1}} q_{E_{q0}} \over 12 \sqrt{3} q^3_E} + { 3 q^2_{E_{p0}} q^2_{E_{q0}} \over 8 q^3_E } + { q_{E_{p0}} q_{E_{p1}} q_{E_{q0}} q_{E_{q1}} \over 4 q^3_E }  & &\nn \\
\quad + { q^2_{E_{p1}} q^2_{E_{q1}} \over 8 q^3_E} + { q_{E_{p0}} q^3_{E_{q1}} \over 12 \sqrt{3} q^3_E } + \Big( { q^2_{E_{p1}} \over 3 \sqrt{2} q^2_E} +{ q_{E_{p0}} q_{E_{q1}} \over  \sqrt{6} q^2_E } \Big) \tilde q_{Y_-} - \Big( {q_{E_{p1}} q_{E_{q0}} \over  \sqrt{6} q^2_E } -{ q^2_{E_{q1}}  \over 3 \sqrt{2} q^2_E } \Big) \tilde q_{Y_+}   &&  \eea
The equations (\ref{eq-qFp0}-\ref{eq-qFq1}) lead to four constraints in the charges:
\bea
 {q_{E_{p0}} q^2_{E_{q0}} \over 2 q_E}-  { q^3_{E_q1}\over 3 \sqrt{3} q_E } -    q_{E_{q0}} \Big( {q_H \over 2}+  q_{Y_0} \Big) +  q_{E_{q1}} \Big( { q_{E_{p1}} q_{E_{q0}} \over 2 q_E} +  \sqrt{2 \over 3}  q_{Y_-} \Big) =&  q_E q_{F_{p0}} & \\
  {q_{E_{p0}} q_{E_{q0}}  \over 2 q_E } + { q^2_{E_{p1}} q_{E_{q0}}\over \sqrt{3} q_{E_{q1}} q_E} - q_{E_{p1}} \Big( { q_{E_{q1}} \over 6 q_E} - { \sqrt{8} q_{Y_-} \over 3 q_{E_{q1}} } \Big) - { q_H \over 2} -  {q_{Y_0} \over 3} -\sqrt{2 \over 3} {q_{E_{q0}} q_{Y_+}\over q_{E_{q1}}} =&  {\ts q_E q_{F_{p1}} \over \ts q_{E_{q1}}}  & \\
{ q^3_{E_{p1}} \over 3 \sqrt{3} q_E}+ q_{E_{p0}} \Big({q_{E_{p0}} q_{E_{q0}} \over 2 q_E} + {q_H \over 2}- q_{Y_0}\Big) + q_{E_{p1}} \Big( { q_{E_{p0}} q_{E_{q1}} \over 2 q_E} -  \sqrt{2 \over 3}   q_{Y_+} \Big) =& q_E q_{F_{q0}} & \\
   {q_H \over 2} - {q_{Y_0} \over 3 }- {q_{E_{p1}} q_{E_{q1}} \over 6 q_E }- {q_{E_{p0}}q^2_{E_{q1}}\over \sqrt{3} q_E q_{E_{p1}}}+ q_{E_{p0}} \Big( { q_{E_{q0}} \over 2 q_E} + { \sqrt{2} q_{Y_-} \over \sqrt{3} q_{E_{p1}}} \Big) +{\sqrt{8}  q_{E_{q1}} q_{Y_+}\over 3 q_{E_{p1}}} =& {\ts q_E q_{F_{q1}}\over \ts q_{E_{p1}}}  &  \eea
Finally, the fields $x$ and $\sigma$ can be expressed as follows:
\bea x &=& - \sqrt{3 \over 32} { 3 q_{E_{p0}} q_{E_{q0}} + q_{E_{p1}} q_{E_{q1}} \over q_E \tilde q_{Y_-}} + \sqrt{3 \over 2} {q_{Y_0} + 3 {\partial_\tau y \over y} \over  \tilde q_{Y_-}} \\
\sigma &=& - { q_{E_{p0}} q_{E_{q0}} + q_{E_{p1}} q_{E_{q1}} \over 8 \sqrt{3} q^2_E } - { q_H - 2 \partial_\tau \phi  \over 4 \sqrt{3} q_E} \eea
These solutions have non-zero $q_E$ charge and require an analytic continuation of the kind (\ref{AC-NS}). As the condition (\ref{realcond}) is satisfied, the action after analytic continuation has a real saddle-point and is definite-positive after the surface term is taken into account. 
If we choose the analytic continuation:
\be \zeta^0 \rightarrow i \zeta'^0, \qquad \zeta^1 \rightarrow i \zeta'^1, \qquad \sigma \rightarrow i \sigma'  \ee
the charges are continued as follows:
\bea q_{E_{pI}} \rightarrow i q'_{E_{pI}}, & q_E \rightarrow i q'_E ,& \tilde q_{Y_\pm} \rightarrow \tilde q'_{Y_\pm} \nn \\
q_{E_{qI}} \rightarrow q'_{E_{qI}}, & \tilde q_F \rightarrow i \tilde q'_F, & q_{Y_0} \rightarrow  q'_{Y_0}   \eea 
Note that the charges $q'_{Y_\pm}$ and $q'_F$ are not real nor purely imaginary. In the particular case of the NS-NS truncated solution, the real and imaginary parts of these charges are separately conserved while the analytically continued fields are real. We then obtain the solution: 
\bea y^{-2} &=& { 2 \tilde q'^2_{Y_-}   \over 27 \gamma_1 }  \cosh^2 \Big[ \sqrt{ \gamma_1 } (\tau + c_1) \Big] \nn \\
e^{-2 \phi} &=& { 12 q'^2_{E} \over \gamma_2 } \cos^2 \Big[ \sqrt{\gamma_2} (\tau + c_2) \Big]   \eea 
with:
\bea \gamma_1 &=& {(3 q'_{E_{p0}} q'_{E_{q0}} + q'_{E_{p1}} q'_{E_{q1}} - 4 q'_E q'_{Y_0}) ^2 \over 144 q'^2_E} - {2 \tilde q'_{Y_-} \tilde q'_{Y_+}\over 9} \nn \\
\gamma_2 &=&  {(q'_{E_{p0}} q'_{E_{q0}} + q'_{E_{p1}} q'_{E_{q1}}) ^2 \over 16 q'^2_E} + {q'_E \tilde q'_F } - {q'^2_H \over 4}  \eea
Note that these instantons are charged under the shifts of the NS axions and correspond to worldsheet instantons. The non-extremality parameter can be expressed as:
\be c= {\gamma_2 - 3 \gamma_1 \over 48}  \ee
These solutions are always singular. \\
\begin{figure}
\centering
\includegraphics[angle=90,scale=0.38]{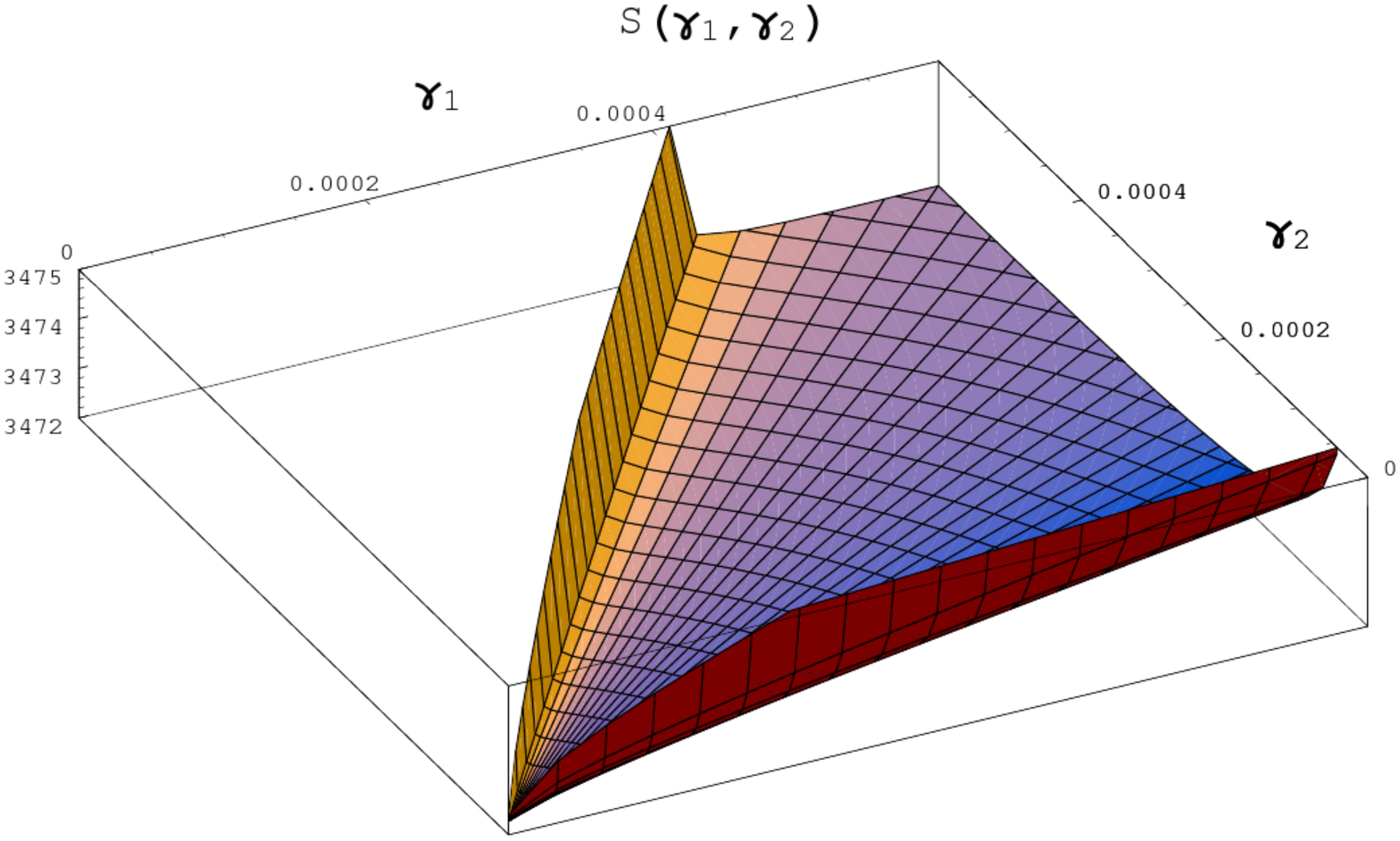}
\caption{ Action of regular NS-NS wormhole solutions as a function of $\gamma_1$ and $\gamma_2$ for fixed values of $\phi$ and $y$ at $\tau=0 $. The plot is obtained setting $e^{-\phi_0}=0.001$, $y_0=10$ and $q'_{E}=\tilde q'_{Y_-}=1$. Extremal instantons correspond to an absolute minimum of the action. }
\end{figure} NS-NS truncated solutions also admit analytic continuations of the kind (\ref{AC-CY1}) \footnote{This analytic continuation is slightly different from (\ref{AC-CY1}) since the field $\tilde \zeta_1$ is continued as well. However, the lack of a surface term for the field $\tilde \zeta_1$ poses no problem in this case because $\tilde \zeta_1$ has costant profile.}:
\be \tilde \zeta_0 \rightarrow i \tilde \zeta'_0, \qquad \tilde \zeta_1 \rightarrow i \tilde \zeta'_1, \qquad x \rightarrow i x', \qquad \sigma \rightarrow i \sigma'  \ee
In this case, the charges are continued as:
\bea q_{E_{qI}} \rightarrow  i q'_{E_{qI}}, & q_E \rightarrow i q'_E,  & \tilde q_{Y_\pm} \rightarrow i \tilde q'_{Y_\pm}    \nn \\
 q_{E_{pI}} \rightarrow q'_{E_{pI}}, & \tilde q_F \rightarrow i \tilde q'_F &  \eea
With this analytic continuation we obtain the non-extremality parameter:
\be c = {3 \gamma_1 + \gamma_2 \over 48} \ee
With $\gamma_1$ and $\gamma_2$ given by:
\bea \gamma_1 &=& -{(q'_{E_{p1}} q'_{E_{q1}} - 4 q'_E q'_{Y_0}) ^2 \over 144 q'^2_E} - {2 \tilde q'_{Y_-} \tilde q'_{Y_+}\over 9} \nn \\
\gamma_2 &=&  {(q'_{E_{p0}} q'_{E_{q0}} + q'_{E_{p1}} q'_{E_{q1}}) ^2 \over 16 q'^2_E} + {q'_E \tilde q'_F } - {q'^2_H \over 4}  \eea
The solution for $\phi$ and $y$ is given by:
\bea y^{-2} &=& { 2 \tilde q'^2_{Y_-}   \over 27 \gamma_1 }  \cos^2 \Big[ \sqrt{ \gamma_1 } (\tau + c_1) \Big] \nn  \\
e^{-2 \phi} &=& { 12 q'^2_{E} \over \gamma_2 } \cos^2 \Big[ \sqrt{\gamma_2} (\tau + c_2) \Big]   \eea
In this case, there exist non-singular solutions for some value of the integration constants provided that $\gamma_1 \ge {2 \over 3} \gamma_2$. The action for these solutions is given by:
\be S= \left. 3 \sqrt{\gamma_1} \tan \Big[\sqrt{\gamma_1} (\tau+ c_1)\Big] + \sqrt{\gamma_2} \tan \Big[ \sqrt{\gamma_2} (\tau + c_2) \Big] \right|^{\pi \over 4 \sqrt{c}}_0  \ee
As done for the RR truncation, it is instructive to consider the $g_S \ll 1/y_0 \ll 1$ limit (corresponding to the weak coupling limit for a large Calabi-Yau manifold). In this limit, we get:
\be S= { \sqrt{2 \over 3} \ts | \tilde q'_{Y_-}| y_0  \over 1 + { \ts \sqrt{3 \gamma_1 } \over \ts \sqrt{2}  y_0   |\tilde q'_{Y_-}|   } \cot \Big(\ \sqrt{ \ts 27 \gamma_1 \over \ts 3 \gamma_1 +  \gamma_2} \pi \Big) }+{ \sqrt{12} { \ts |q'_E| \over \ts  g^2_S  }  \over 1 + { \ts \sqrt{\gamma_2} g^2_S  \over \ts \sqrt{12} |q'_E|   } \cot \Big(\sqrt{ \ts 3 \gamma_2 \over \ts 3 \gamma_1 + \gamma_2} \pi \Big) }  \ee
The second term presents a $1/g^2_S$ dependence characteristic of a fivebrane instanton, while the first term is proportional to the volume of a two-cycle of the manifold as expected for a worldsheet instanton.

\subsubsection{NS-R truncations}

A fourth truncation in which the off-diagonal terms of (\ref{def-M}) are at least quadratic in x is given by setting $\zeta^{1} = \tilde \zeta_{1}={\rm const}$ and $x \equiv 0$. The resulting truncation is however more complicated, since the $y$ and $\phi$ equations do not decouple. This is related to the fact that the non-zero global symmetry charges are not associated with commuting subalgebras. \\

It is possible to generate NS-R truncated solutions acting on a particular solution with a global $G_{2,2}$ transformation. We first consider the case in which the modulus $y$ is constant. It is convenient to set: 
\be q_E=q_{E_{p1}}=q_{E_{q1}}=q_{F_{p1}}=q_{F_{q1}}=0, \qquad y \equiv y_0= \sqrt{3} \left( { q_{E_{p0}}\over q_{E_{q0}} } \right)^{1 \over 3} \label{NS-R-trunc} \ee
If $q_E=0$, it is easy to see from the equations of motion of the truncated lagangian that $y_0$ corresponds to a minimum of the potential for the field $y$.  
In this case, we can substitute (\ref{eq-qH}), (\ref{eq-qY0}) amd (\ref{eq-qFp0}) into (\ref{eq-qFq0}) obtaining:
\be - {q_{E_{p0}} q_{F_{p0}} \over 2} - {q_{E_{q0}} q_{F_{q0}} \over 2} - {q_H^2 \over 4} -  {q_{Y_0}^2 \over 3}  +  4 \sqrt{3} |q_{E_{p0}} q_{E_{q0}}| e^{\phi} + \left( \partial_\tau \phi \right)^2 = 0  \ee
This ODE is solved by:
\be e^{-\phi} = \frac{2 \sqrt{3} |q_{E_{p0}} q_{E_{q0}}|} {C_2} \cosh^2 \left({ \sqrt{C_2} \over 2 } \tau + A_1 \right) \label{NS-R-sol1}\ee
With this notation, $C_2$ is the quadratic Casimir operator:
\be  C_2 = {q_H^2 \over 4} + {q_{Y_0}^2 \over 3}+ {q_{E_{p0}} q_{F_{p0}} \over 2} + {q_{E_{q0}} q_{F_{q0}} \over 2} = -48 c   \label{c-NS-R} \ee
where $c$ is the non-extremality parameter. The expression for $q_F$ reduces to a constraint in the charges:
\bea 54 q_{E_{p0}} q_{E_{q0}} q_F - 9 q_{E_{p0}} q_{F_{p0}} \left(3 q_H-2 q_{Y_0}\right)+\left(3 q_H+2 q_{Y_0}\right) \left(4 q_{Y_0}^2-6 q_H q_{Y_0}+9 q_{E_{q0}} q_{F_{q0}} \right) =0  \quad&& \label{NS-R-cons}\eea 
The non-trivial RR fields can be obtained from (\ref{eq-qH}) and (\ref{eq-qY0}):
\bea 
\zeta^0 &=& \frac{3 q_H+2 q_{Y_0}+6 \sqrt{C_2} \tanh \left( { \sqrt{C_2} \over 2} \tau + A_1 \right)}{18 q_{E_{p0}}} \label{NS-R-sol2} \\
\tilde \zeta_0 &=& \frac{3 q_H-2 
   q_{Y_0}+6 \sqrt{C_2} \tanh \left({\sqrt{C_2} \over 2} \tau + A_1 \right)}{2 \sqrt{3} q_{E_{q0}}}  \label{NS-R-sol3}\eea
\\
If we consider a set of charges such that: 
\be q_{Y_\pm}=q_{E_{p1}}=q_{E_{q1}}=q_{F_{p1}}=q_{F_{q1}}=0 \label{NS-R-charges} \ee
together with a set of arbitrary values for the fields $\phi$ and $y$ at $\tau=0$ and act with a group transformation of the form:
\be g_{\alpha, \beta, \gamma}=e^{-\alpha F_{p0}} e^{-\beta E_{q0}} e^{-\gamma E_{p0}}  \label{NS-R-trans} \ee
we can solve numerically for the values of the transformation parameters $\a, \b, \g$ such that the transformed solution obeys to (\ref{NS-R-trunc}) and (\ref{NS-R-cons}). The equations we obtain admit solutions only for some of the random values of charges and fields at $\tau=0$. The interpreatation of this fact is that the solutions obtained applying the inverse of (\ref{NS-R-trans}) on the solution (\ref{NS-R-sol1}-\ref{NS-R-sol3}) constitute a patch of non-zero measure of the set of solutions obeying to (\ref{NS-R-charges}).\\

To obtain an explicit form for these solutions we note that a finite transformation generated by $F_{p0}$ acts on the truncated fields as:
\bea {e^{\phi / 2} \over y^{3 /2}} &\rightarrow& { {e^{\phi / 2} \over y^{3 /2}} \over (1 + 6 \alpha \zeta^0 )^2 + 36 \alpha^2  {e^{\phi} \over y^{3}}} \nn \\
e^{\phi} y &\rightarrow& e^{\phi} y \nn \\
\zeta^0 & \rightarrow & { \zeta^0 + 6 \alpha {\zeta^0}^2 + 6 \alpha { {e^{\phi } \over y^{3}}} \over (1+  6 \alpha \zeta^0)^2 + 36 \alpha^2  {e^{\phi} \over y^{3}} }  \eea
The finite transformations generated by $F_{p0}$ leave the fields $x$, $\z^1$ and $\tiz_1$ equal to zero and allows to obtain an expression for the fields $\phi$ and $y$ of a truncated solution obeying to (\ref{NS-R-charges}). \\
The appropriate analytic continuation is the one corresponding to the NS-charged case:
\bea \zeta^0 \rightarrow i \zeta'^0 & \tiz_0 \rightarrow \tiz'_0 & \sigma \rightarrow i \sigma'  \\
q_{E_{p0}} \rightarrow i q'_{E_{p0}} & q_{E_{q0}} \rightarrow q'_{E_{q0}} & q_E \rightarrow i q'_E \\
q_{F_{p0}} \rightarrow i q'_{F_{p0}} & q_{F_{q0}} \rightarrow q'_{F_{q0}} & q_F \rightarrow i q'_F \eea
we need to continue two of the transformation parameters as well:
\be \alpha \rightarrow i \alpha \qquad \gamma \rightarrow i \gamma \ee
The solution is:
\bea e^{\phi'} &=& {12 \sqrt{ 2 c } |\alpha| \over |q'^2_{E_{p0}} q'_{E_{q0}}|^{1 \over 3} } {\sqrt{b_1 \sin^2  \left( 2 \sqrt{3 c} \tau + A_2 \right) - 1 } \over |\cos \left( 2 \sqrt{3 c} \tau + A_1  \right)| } \\ 
y' &=& { \sqrt{ 2 c } \over  |\alpha  q'_{E_{q0}}| } {|\sec \left( 2 \sqrt{3 c} \tau + A_1  \right)|  \over \sqrt{b_1 \sin^2  \left( 2 \sqrt{3 c} \tau + A_2 \right) - 1 }  } \label{NS-R-sol-gen} \eea
The constants $b_1$ and $A_2$ are given by:
\bea b_1 &=& 2 + {(q'_{E_{p0}}+\alpha q'_H + {2 \over 3} \alpha q'_{Y_0})^2 \over 96 \alpha^2 c } \nn \\
A_2 &=& A_1 -\tan^{-1} \left(\frac{q'_{E_{p0}}+\alpha  q'_H+\frac{2 \alpha  q'_{Y_0}}{3}}{8 \alpha  \sqrt{3 c}}\right) \eea
This solution is parametrized by eight charges, three transformation parameters and one integration constant. The actual Noether charges of the solution (\ref{NS-R-sol-gen}) can be obtained applying the inverse of (\ref{NS-R-trans}) to the charges $q'_H, q'_{Y_0} \dots q'_F$ which transform in the adjoint representation.
The expression for the two RR axions and the NS-NS axion can be obtained solving the equations (\ref{eq-qH}-\ref{eq-qY0}) and (\ref{eq-qFp0}) for the fields $\zeta^0, \tiz_0$ and $\s$. Finally, more general NS-R truncated solutions can be obtained by acting with the shifts of the the fields $\s$, $\z^1$ and $\tiz_1$ on the above solution. \\

One may wonder whether it is possible to recover the universal hypermultiplet as a truncation of the $G_{2,2} / (SU(2) \times SU(2))$ model. From a ten-dimensional perspective, the NS-R truncation has the same non-trivial axionic fields of the universal hypermultiplet. However, NS-R truncated solutions have a non-trivial profile of the modulus $y$ which corresponds to the volume of the compactfication manifold and has no analogue in the $SU(2,1)/(SU(2)\times U(1))$ sigma model covered in section 4.

\subsection{Extremal limit and supersymmetry analysis}

In case of the RR truncated solutions, we have regular (extremal) instanton solutions for $\gamma_1 \rightarrow 0$ and $\gamma_2 \rightarrow 0$. For the RR truncation I, we get:
\be
e^{- \phi'} = \sqrt{|q'_{E_{p0}} \tau + h_1 ||q'_{E_{p1}} \tau + h_2|^3 \over 3}, \qquad y' =  \sqrt{27 \left|{q'_{E_{p0}} \tau + h_1 \over q'_{E_{q1}} \tau + h_2}  \right| } 
\ee
Direct computation of the quaternionic vielbein (\ref{viel-G22-u}-\ref{viel-G22-e}) leads to:
\bea
u'_{\tau}&=& -{3 \over 4} i \Big( q'_{E_{p0}} e^{\eta'_1 \over 2} +{q'_{E_{q1}} \over \sqrt{3}} e^{\eta'_2 \over 2} \Big) \\
v'_{\tau} &=& -{3 \over 4}  \Big( |q'_{E_{p0}}| e^{\eta'_1 \over 2} +{|q'_{E_{q1}}| \over \sqrt{3}} e^{\eta'_2 \over 2} \Big) \\
E'_{\tau}&=& {1 \over 4} \Big( 3 \sqrt{3} q'_{E_{p0}} e^{\eta'_1 \over 2} -q'_{E_{q1}} e^{\eta'_2 \over 2} \Big) \\
e'_{\tau} &=& {i \over 4}  \Big( 3 \sqrt{3} |q'_{E_{p0}}| e^{\eta'_1 \over 2} -|q'_{E_{q1}}|e^{\eta'_2 \over 2} \Big) \eea
with $\eta'_1$ and $\eta'_2$ defined in (\ref{Sol-RR1}). It is easy to see that the BPS condition (\ref{bpscon}) is satisfied only if $q'_{E_{p0}} q'_{E_{q1}} \ge 0 $. In contrast, solutions with $q'_{E_{p0}} q'_{E_{q0}} < 0$ are extremal non-BPS instantons.
Note that the extremal non-BPS instantons will be related to extremal non-BPS black holes by the c-map. The existence and properties of such black hole solutions and their relation to the attractor mechanism was discussed in several papers recently \cite{Goldstein:2005hq,Kallosh:2006bt,Gaiotto:2007ag,Saraikin:2007jc,Gimon:2007mh}. 
Note that an extremal non-BPS solutions is still an instanton, i.e. it has only one asymptotic region and induces a local operator insertion. However the fact that it breaks all supersymmetries implies that there are twice as many fermionic zero-modes and consequently the instanton will only contribute to higher derivative corrections to the $N=2$ effective action. \\
Similarly, direct computation of the vielbein for the RR truncation II leads to:
\bea
u'_{\tau}&=& -{1 \over 4}  \Big( { q'_{E_{q0}} \over \sqrt{3}} e^{\eta'_1 \over 2} - 3 q'_{E_{p1}} e^{\eta'_2 \over 2} \Big) \\
v'_{\tau} &=& -{1 \over 4}  \Big({ |q'_{E_{q0}}| \over \sqrt{3}} e^{\eta'_1 \over 2} + 3 |q'_{E_{p1}}|  e^{\eta'_2 \over 2} \Big) \\
E'_{\tau}&=& -{i \over 4} \Big(  q'_{E_{q0}} e^{\eta'_1 \over 2} + \sqrt{3} q'_{E_{p1}} e^{\eta'_2 \over 2} \Big) \\
e'_{\tau} &=& -{i \over 4}   \Big(  |q'_{E_{q0}}| e^{\eta'_1 \over 2} - \sqrt{3} |q'_{E_{p1}}|e^{\eta'_2 \over 2} \Big) \eea
These solutions are supersymmetric only if $q'_{E_{q0}} q'_{E_{p1}} \leq 0 $. We will have non-supersymmetric extremal solutions if $q'_{E_{q0}} q'_{E_{p1}} \geq 0 $. \\
Finally, taking the extremal limit for the NS-NS truncation leads to:
\be e^{-\phi'} = 2 \sqrt{3}  \Big| q'_E \tau + h_2 \Big| , \qquad y'= \sqrt{2 \over 27} \Big|\tilde q'_{Y_-}  \tau + h_1 \Big|^{-1} 
\ee
It is easy to see  the NS-NS truncated extremal solutions are BPS since $u_{\tau}=E_{\tau}=0$.

\subsection{General solution \label{general-sol}}
The truncations allowed for the exact solution of the equations of motion using the conserved charges to solve for the radial dependence of all fields.  Applying this method in the most general case leads to algebraic equations which cannot be solved explicitly. In the following we will describe using the method of solution generating transformations to obtain the general solution \footnote{The dimensional reduction to coset sigma models and the use of global symmetries to generate solutions has a long history for black holes in (super)gravity, see e.g. \cite{Ehlers,Geroch:1970nt,Kinnersley:1977pg,Breitenlohner:1987dg}.}.

The RR truncated solutions obey to five constraints in the charges. It is in principle possible to obtain the general solution using a five parameter group transformation. The general solution will then be characterized by fourteen parameters: two integration constants, seven charges and five transformation parameters. We first integrate the  $F_{p1}$ generator to obtain a finite group transformation. The result for the fields $x$, $y$, $\phi$ and $\zeta^0$ is:
\bea x &\rightarrow& x + 6a(\zeta^0 x -\zeta^1 ) \nn \\
y &\rightarrow& y \sqrt{(1 + 6a \zeta^0 )^2 + 36 a^2 {e^\phi \over y^3}} \nn \\
e^{\phi} & \rightarrow & { e^{\phi} \over \sqrt{(1 + 6 a \zeta^0 )^2 + 36 a^2 {e^\phi \over y^3}} } \nn \\
\zeta^0 & \rightarrow & { \zeta^0 + 6 a {\zeta^0}^2 + 6 a {e^{\phi} \over y^3 } \over (1+  6 a \zeta^0)^2 + 36 a^2 {e^\phi \over y^3}  } \eea
It is easy to show that the other fields transform so that:
\bea \zeta^1 - \zeta^0 x &=& \text{const} \nn \\
3 \zeta^1 x + \tilde \zeta_1 &=& \text{const} \nn \\
\sigma + 2 {\zeta^1}^2 x - \zeta^0 \zeta^1 x^2 &=& \text{const}  \nn \\
\delta_{F_{p0}} \delta_{F_{p0}} (\tilde \zeta_0 - \zeta^1 x^2 ) &=& 0 \eea
The finite transformation generated by $Y_+$ will be simpler as the $Y_+$ action on the RR axions is nilpotent:
\bea x &\rightarrow& \sqrt{6} { (\sqrt{6} + a x )x + a y^2  \over (\sqrt{6} + a x)^2 + a^2 y^2 } \nn \\
y &\rightarrow& {6 y \over (\sqrt{6} + a x)^2 + a^2 y^2 } \nn \\ 
\zeta^0 &\rightarrow& \zeta^0 + \sqrt{3 \over 2} a \zeta^1 + {a^2 \over 6} \tilde \zeta_1 - {a^3 \over 6 \sqrt{6}} \tilde \zeta_0 \nn \\
\zeta^1 &\rightarrow& \zeta_1 - \sqrt{2 \over 27} a \tilde \zeta_1 + {a^2 \over 6} \tilde \zeta_0 \nn\\
\tilde \zeta_1 &\rightarrow& \tilde \zeta_1 - \sqrt{3 \over 2} a \tilde \zeta_0  \eea
It is slightly more involved to obtain the finite transformation generated by $F_{q0}$. It is convenient to consifer the transformation rules for the combination $e^{\phi \over 2} \Big( {x^2 + y^2 \over y} \Big)^{3 \over 2}$. We get:
\bea e^{\phi \over 2} \Big( {x^2 + y^2 \over y} \Big)^{3 \over 2} & \rightarrow & { e^{\phi \over 2} \Big( {x^2 + y^2 \over y} \Big)^{3 \over 2} \over \Big( 1 + {2 \over \sqrt{3}} a \tilde \zeta_0 \Big)^2  + {4 \over 3} a^2  e^{\phi } \Big( {x^2 + y^2 \over y} \Big)^{3} } \\
\tilde \zeta_0 & \rightarrow &   { \tilde \zeta_0 + {2 \over \sqrt{3}} a \Big( \tilde \zeta^2_0 + e^{\phi} \Big( {x^2+ y^2 \over y} \Big)^3  \Big) \over \Big( 1 + {2 \over \sqrt{3}} a \tilde \zeta_0 \Big)^2  + {4 \over 3} a^2  e^{\phi } \Big( {x^2 + y^2 \over y} \Big)^{3} } \eea
The transfromations for the fields $x$, $y$ and $\phi$ can be obtained observing that:
\bea \delta_{F_{q0}} \Big( e^{- \phi} { x^2 + y^2 \over y }   \Big) & =& 0 \\
\delta_{F_{q0}} \Big( { x \over x^2 + y^2 } \tilde \zeta_0 + { \tilde \zeta_1 \over 3} \Big) & = & 0  \eea
and that:
\be { x \over x^2 + y^2 } \rightarrow {x \over x^2 + y^2} + {2 a \over \sqrt{3}} \Big (  { x \over x^2 + y^2 } \tilde \zeta_0 + { \tilde \zeta_1 \over 3} \Big) \ee
The general solution can be obtained by acting with a five parameter group transformation on one of the truncated solutions. In particular, we can act with the $G_{2,2}$ element:
\be g = e^{\alpha_5 F_{q0}} e^{\alpha_4 Y_+} e^{\alpha_3 E_{p1}} e^{\alpha_2 E_{q1}} e^{\alpha_1 F_{p0}} \label{5-trans} \ee
on the RR truncation II (\ref{RR2-in}- \ref{RR2-fin}).
It can be checked numerically that the inverse of the above transformation can map generic values of the fourteen $G_{2,2}$ charges into values obeying to the constraints characterizing the RR truncation (equations \ref{RR2-con1}-\ref{RR2-con2} and the vanishing of three shift charges). On the other side, some random assignments for the values of the fourteen charges cannot be mapped into a truncated solution. As in the case of the NS-R truncation, the interpretation of this fact is that the set of solutions obtained by acting with a transformation of the form (\ref{5-trans}) on a truncated solution represent a patch of non-zero measure in the space of general solutions. \\

Solutions with $q'_E=0$ admit an analytic continuation of the form (\ref{AC-pureD}) and lead to a real positive-definite action. On the other hand, solutions with $q_E \neq 0$ need to satisfy the condition (\ref{realcond}) in order to have a real positive-definite action. This condition poses a strong constraint on the solutions. Indeed, we suspect that a real positive-definite action can be obtained only with the truncations studied in section 5.2 \footnote{Solutions obtained by applying a finite transformation generated by $Y_-$ to these truncations will have positive-definite action as well.}.

\section{Orientifolding and $N=1$ supergravities}
\setcounter{equation}{0}

Orientifolding of $N=2$ supergravity theories can be used to obtain consistent truncations which reduces the supersymmetry of the theory to $N=1$.  The orientifolding can be understood purely from the perspective of the four-dimensional supergravity \cite{Andrianopoli:2001gm,DAuria:2004kx} or microscopically from the Calabi-Yau compactification of type II string theory \cite{Grimm:2004uq,Grimm:2004ua}, where the orientation reversal on the worldsheet is accompanied by an involution acting on the Calabi-Yau manifold.

\subsection{Orientifolding $N=2$ theories}
 The simplest orientifold ${\cal O}_{1}$ projection (corresponding to an orientation reversal on the worldsheet with a trivial involution on the Calabi-Yau manifold)
 \bea
&& 
{\cal O}_{1}\phi = \phi, \quad  {\cal O}_{1}\sigma= -\sigma, \quad   {\cal O}_{1}\zeta^{0}= \zeta^{0} , \quad {\cal O}_{1}\tilde\zeta_{0}= -\tilde \zeta_{0}\no\\
 &&{\cal O}_{1} x^{a}= - x^{a}, \quad {\cal O}_{1} y^{a}= y^{a}, \quad   {\cal O}_{1}\zeta^{a}= -\zeta^{a} , \quad {\cal O}_{1}\tilde\zeta_{a}= \tilde \zeta_{a}, \quad a=1,2,\cdots h_{1,1} 
 \eea
 Projecting out the odd fields one obtains  the action:
\be
 S= \int d^4x\; \sqrt{-g}\Big\{R - 2 g_{a b } \partial_\mu y^a \partial y^{b }-{1\over 2} (\partial_\mu \phi)^2  -{1\over 2}e^{\phi} \;\partial_{\mu} \tilde\zeta_{a} (I^{-1})^{ab} \partial^{\mu }\tilde\zeta_{b}  -{1\over 2}e^{\phi} \;\partial_{\mu}  \zeta^{0}  I_{00} \partial^{\mu }\zeta^{0} \Big\}
\ee
The microscopic projection in this case is:
\be
O_{1} =  \Omega_{p} \sigma_{i}, \quad \sigma_{i} \Omega_{3,0}= \Omega_{3,0}, \quad \sigma_{i}\omega_{a} = \omega_{a},\quad  a=1,2, \cdots ,h_{1,1}
\ee
and  corresponds to the insertion of spacetime filling orientifold O5/O9 planes.  Here $\Omega_{p}$ corresponds to orientation reversal on the worldsheet. The map $\sigma_{i}$ is an involution on the Calabi-Yau and  is not to be confused with the spacetime field $\sigma$, which is the NS-NS axion.

For special  quaternionic geometries a  more general version of the orientifold projection ${\cal O}_{1}$ can be constructed. Defining the following split of the indices into a positive and  a negative set 
\be\label{splitin}
A_{+}=\{ 1,2,\cdots n_{+}\}, \quad A_{-}=\{n_{+}+1, n_{+}+2,\cdots h_{1,1} \}
\ee
We can define the following projection
\bea\label{origen}
 && {\cal O}_{1}' y_{A_{+}} = y_{A_{+}}, \quad {\cal O}_{1}' x_{A_{+}} = -x_{A_{+}},\quad 
{\cal O}_{1}' x_{A_{-}} = x_{A_{-}}, \quad  {\cal O}_{1}' y_{A_{-}} = -y_{A_{-}} \nonumber \\
&&
  {\cal O}_{1}' \zeta^{A_{+}} = -\zeta^{A_{+}} \;\quad  {\cal O}_{1}' \tilde \zeta_{A_{+}}= \tilde\zeta_{A_{+}}, \quad 
 {\cal O}_{1}' \zeta^{A_{-}} = \zeta^{A_{-}}, \quad  {\cal O}_{1}'\tilde \zeta_{A_{-}}= -\tilde\zeta_{A_{-}}\nonumber \\
 &&{\cal O}_{1}'\phi = \phi, \quad  {\cal O}_{1}'\sigma= -\sigma, \quad {\cal O}_{1}' \zeta^{0}=\zeta^{0}, \quad {\cal O}_{1}' \tilde\zeta_{0}=-\tilde\zeta_{0}
   \eea
 
 It can be shown that  all terms in the  hypermultiplet  action (\ref{hyperone}) linear in the odd fields under ${\cal O}_{1}'$ vanish if the intersection numbers $C_{{abc}}$ satisfy the following condition
\be
C_{A_{+}A_{+}A_{-}}= C_{A_{-}A_{-}A_{-}}=0
\ee
Hence projecting out the odd fields is a consistent truncation and the projected action is given by
\bea
 S&=& \int d^4x\; \sqrt{-g}\Big\{R - 2 g_{+ + } \partial_\mu y^{+} \partial y^{+ }- 2 g_{- - } \partial_\mu x^{+} \partial x^{- } -{1\over 2} (\partial_\mu \phi)^2 \no\\
 &&   -{1\over 2}e^{\phi} \;\partial_{\mu}  \zeta^{-}  I_{--} \partial^{\mu }
 \zeta^{-} -{1\over 2}e^{\phi} \;\partial_{\mu}  \zeta^{-} R_{- I} (I^{-1})^{IJ}R_{J -} \partial^{\mu }
 \zeta^{-} \no\\
 &&-{1\over 2}e^{\phi} \;\partial_{\mu} \tilde\zeta_{+} (I^{-1})^{+ +}  \partial^{\mu }\tilde\zeta_{+}  -   e^{\phi} \;\partial_{\mu} \tilde\zeta_{+} (I^{-1})^{+ I} R_{I -}  \partial^{\mu } \zeta^{-}   \Big\}
\eea
Where schematically the idex $+$ runs over $A_{+}$ and the index $-$ runs over $\{0\} \cup A_{-}$.
Microscopically this orientifold projections  is the same as $O_{1}$ where in addition the involution $\sigma_{i}$  of the Calabi-Yau manifold acts non-trivially on the $(1,1)$ forms splitting them into even and odd forms.
\bea
O_{1}' &=&  \Omega_{p} \sigma_{i}, \quad\quad \quad \quad\sigma_{i}\Omega_{3,0}= \Omega_{3,0}\\
\sigma_{i} \omega_{a}&=& + \omega_{a}, \quad a\in A_{+}, \quad \quad  \sigma_{i}\omega_{a}= - \omega_{a}, \quad a\in A_{-}
\eea
Like the first orientifold projection $O_{1}'$ corresponds to the insertion of space filling O5/O9 planes, the non-trivial action of the involution $\sigma_{i}$ arises from the way the $O5$ plane is embedded in the Calabi-Yau manifold.

A second  projection $O_{2}$ exchanges the role of $\zeta^{I}$ and $\tilde \zeta_{I}$.  The same argument as above shows that the orientifold projection is consistent.
\bea
&& 
{\cal O}_{2}\phi = \phi, \quad  {\cal O}_{2}\sigma= -\sigma \quad   {\cal O}_{2}\zeta^{0}= -\zeta^{0} , \quad {\cal O}_{2}\tilde\zeta_{0}= \tilde \zeta_{0}\no\\
 &&{\cal O}_{2} x^{a}= - x^{a}, \quad {\cal O}_{2} y^{a}= y^{a}, \quad   {\cal O}_{2}\zeta^{a}= \zeta^{a} , \quad {\cal O}_{2}\tilde\zeta_{a}= -\tilde \zeta_{a}, \quad a=1,2,\cdots h_{1,1} 
 \eea
 Since the action is even under the orientifold projection  there are no linear terms involving odd fields.  Hence setting to zero the odd fields under ${\cal O}_{2}$  is a consistent truncation, The projected action is given by
\be
S= \int d^4x\; \sqrt{-g}\Big\{R - 2 g_{a b } \partial_\mu y^a \partial y^{b }-{1\over 2} (\partial_\mu \phi)^2  -{1\over 2}e^{\phi} \;\partial_{\mu} \zeta^{a} I_{ab} \partial^{\mu }\zeta^{b}  -{1\over 2}e^{\phi} \;\partial_{\mu} \tilde \zeta_{0}  (I^{-1})^{00} \partial^{\mu }\tilde \zeta_{0} \Big\}
\ee
Microscopically, the orientifold projections  on a IIB Calabi-Yau compactification can be understood as combining a worldsheet parity reversal $\Omega_{p}$ with an involution $\sigma_{i} $ of the Calabi-Yau manifold. 
\be
O_{2} = (-1)^{F_{L}} \Omega_{p} \sigma_{i}, \quad \sigma_{i} \Omega_{3,0}= -\Omega_{3,0}, \quad \sigma_{i} \omega_{a} = \omega_{a},\quad  a=1,2, \cdots ,h_{1,1}
\ee
where $F_{L}$ is the left-moving spacetime fermion number. This projection corresponds to the insertion of spacetime filling orientifold O3/O7 planes.

There there is a generalized  orientifold projection $O_{2}'$ associated with ${\cal O}_{2}$  which can be obtained from $O_{1}'$ by exchanging the roles of $\zeta^{I}$ and $\tilde \zeta_{I}$. Microscopically this projection is given by
\bea
O_{2}' &=&  \Omega_{p} (-1)^{F_{L}}\sigma_{I}, \quad\quad \quad \quad\sigma_{i}\Omega_{3,0}= -\Omega_{3,0}\\
\sigma_{i} \omega_{a}&=& + \omega_{a}, \quad a\in A_{+}, \quad \quad  \sigma_{i} \omega_{a}= - \omega_{a}, \quad a\in A_{-}
\eea
and corresponds to the insertion of space filling O3/O7 planes. The form of the projected action can be easily worked out. 

It is useful to compare the orientifold projections to the truncations of the $N=2$ action    for the $G_{2,2}/(SU(2)\times SU(2))$ coset space,  which was explictely constructed. The orientifold projections $O_{1}$ and $O_{2}$ set to zero  the NS-NS fields $\sigma$ and $x$.  Hence they correspond to the RR truncations discussed in section 5.2.   For a general $N=2$ action all four orientifold projections set to zero $\sigma$ and correspond to a pure RR-charged case. Whether the solution exists depends on whether the behavior of the $x^{a}$ and $\xi^{a}, \tilde \xi_{a}$ $a=1,2,\cdots ,h_{1,1} $ in the full $N=2$ instanton solution  is consistent with the particular orientifold projection. This means that the fields 
which are to be projected out in the orientifold projection are zero for all values of Euclidean time in the full $N=2$ solution.

In section 5  the  NS-NS truncation  was  also discussed. One may wonder whether there is an orientifold projection associated with this truncation as in the case of the RR truncation above. From the perspective of four-dimensional $N=2$ supergravity such a projection indeed exists and  was called  ``heterotic'' in \cite{D'Auria:2005yg}, where all RR axions are projected out. Such a projection does  not have an interpretation as an action on the ten-dimensional type II string theory and will not be discussed further here.  

\subsection{Supersymmetry}
For all  orientifold projections given above  the supersymmetry is reduced from $N=2$ to $N=1$. 
  The projection on the fermions can also be worked out either from the consistency of the supersymmetry transformations or microscopically from the Calabi-Yau compactification.

The reduction of the supersymmetry can be achieved by choosing  a linear combination of the two $N=2$  gravitinos $\psi_{\mu}^{\alpha}$  as the  single $N=1$  gravitino.
Also the  half the degrees of freedom of the hyperino $\xi^{a}$ is set to zero producing the fermionic components of the $N=1$ chiral multiplet.

The supersymmetry is reduced by setting to zero a linear combination of the two infinitesimal supersymmetry transformation parameters $\epsilon^{1},\epsilon^{2}$ which are Weyl spinors of positive chirality $\gamma_{5}\epsilon^{1,2}=+\epsilon^{1,2}$. The supersymmetry transformation parameters of negative chirality are labelled $\epsilon_{1,2}$ related to the positive chirality by $\epsilon_{1}=(\epsilon^{1})^{*},  \epsilon_{2}=(\epsilon^{2})^{*}$. 
The supersymmetries which are preserved by the O3/O7 and the O5/O9 orientifold can be derived by the consistency of the orientifold projection on the bosonic fields and the supersymmetry variations of the fermionic fields \cite{D'Auria:2005yg}. 
We remind the reader of the $N=2$ gravitino variation 
\begin{equation}\label{delpsiII}
\delta \psi^i_\mu = D_\mu \epsilon^i +(Q_\mu)_{\;\;j}^{i}\epsilon^j
\end{equation} 
where $D_{\mu}$ is the standard covariant derivative which  includes the spin connection, and 
$Q_{\;j}^{i}$ is the composite $SU(2)$ gauge connection, which for both orientifold projections $O_{1}$ and $O_{2}$ reduces to
\begin{equation}\label{sutwocII}
Q_{\; \;j}^{i}=
\left(
\begin{array}{cc}
  0 &  - u    \\
 \bar u  &  0
\end{array}
\right)
\end{equation}
The components of the vielbein  $u_{\mu}$ and $v_{\mu}$  after the orientifold projection are given by
  \bea
O_{1}:&&v_{\mu}={1\over 2} \partial_{\mu}\phi, \quad  u_{\mu}= -{1\over \sqrt{2}} e^{K-\phi\over 2}\Big({1\over 6} C_{abc} y^{a}y^{b}y^{c} \partial_{\mu} \zeta^{0}+y^{a}\partial_{\mu}\tilde \zeta_{a}\Big)\no\\
&& \quad \quad \quad\quad\quad\quad \bar u_{\mu}= -{1\over \sqrt{2}} e^{K-\phi\over 2}\Big({1\over 6} C_{abc} y^{a}y^{b}y^{c} \partial_{\mu} \zeta^{0}+y^{a}\partial_{\mu}\tilde \zeta_{a}\Big)\label{udeforia}\\
O_{2}:&& v_{\mu}={1\over 2} \partial_{\mu}\phi, \quad u_{\mu }= {i\over \sqrt{2}} e^{K-\phi\over 2}\Big( \partial_{\mu }\tilde \zeta_{0}-{1\over 2} C_{abc}y^{a}y^{b}\partial_{\mu} \zeta^{a}\Big)\no\\
&&\quad \quad \quad \quad \quad\quad \bar u_{\mu }=- {i\over \sqrt{2}} e^{K-\phi\over 2}\Big( \partial_{\mu }\tilde \zeta_{0}-{1\over 2} C_{abc}y^{a}y^{b}\partial_{\mu} \zeta^{a}\Big)\label{udeforib}
\eea 

The supersymmetry is reduced by setting to zero a linear combination of the two infinitesimal supersymmetry transformation parameters $\epsilon^{1},\epsilon^{2}$ which are Weyl spinors of positive chirality $\gamma_{5}\epsilon^{1,2}=+\epsilon^{1,2}$. The supersymmetry transformation parameters of negative chirality are labelled $\epsilon_{1,2}$ related to the positive chirality by $\epsilon_{1}=(\epsilon^{1})^{*},  \epsilon_{2}=(\epsilon^{2})^{*}$. 
The supersymmetries which are preserved by the O3/O7 and the O5/O9 orientifold can be derived by the consistency of the orientifold projection on the bosonic fields and the supersymmetry variations of the gravitino \cite{D'Auria:2005yg}. The following combination of supercharges is consistent with the orientifold projection
 \bea\label{ofoldsusy}
{ O_{1}: }&&  \epsilon_{\alpha}=\epsilon^{1}_{\alpha}+ i  \epsilon^{2}_{\alpha} , \quad \epsilon_{\dot\alpha}=\epsilon_{1\dot\alpha }-i  \epsilon_{2\dot\alpha}\no\\
{ O_{2}: } &&\epsilon_{\alpha}=\epsilon^{1}_{\alpha}-  \epsilon^{2}_{\alpha}, \;\quad \epsilon_{\dot \alpha}=\epsilon_{1 \dot \alpha}-  \epsilon_{2\dot \alpha}
\eea
Were for clarity we have written the un-dotted (positive chirality) and dotted (negative chirality) spinor indices.  A second approach uses the microscopic definition of the orientifold projection and the world sheet definition of the  supersymmetry generators \cite{Brunner:2003zm,Jockers:2005pn} and leads to the same conditions.

Since the chirality of the surviving supersymmetries for the instanton solutions  is very  important, we repeat the hyperino variations for both chiralities for the special case of $SO(4)$ symmetric solutions
\be
\delta \xi^{a}= -i  V_\tau^{a \alpha} \gamma^\tau \epsilon_\alpha, \quad \delta \xi_{a}= i  V_{\tau a \alpha} \gamma^\tau \epsilon^\alpha
\ee
Where $(V_\tau^{a \alpha} )^{*}=V_{\tau a \alpha} $.  After multiplication by $\gamma^{\tau}$ the condition that $N=1$ supersymmetry is preserved for the hyperino variation  for the negative chirality supersymmetry becomes
\be\label{bpscona}
u_{\tau} \epsilon_{1}+v_{\tau}\epsilon_2=0, \quad -\bar v_{\tau} \epsilon_{1}+ \bar u_{\tau} \epsilon_{2}=0, \quad e^{A}_{\tau }\epsilon_{1}+ E^{A}_{\tau}\epsilon_{2}=0, \quad  -\bar E^{A}_{\tau}\epsilon_{1}+\bar e^{A}_{\tau }\epsilon_{2}=0
\ee
and for the positive chirality supersymmetry one gets
\be\label{bpsconb}
-v_{\tau} \epsilon^{1}+u_{\tau}\epsilon^{2}=0, \quad \bar u_{\tau} \epsilon^{1}+ \bar v_{\tau} \epsilon^{2}=0, \quad -e^{A}_{\tau }\epsilon^{1}+ E^{A}_{\tau}\epsilon^{2}=0, \quad  \bar E^{A}_{\tau}\epsilon^{1}+\bar e^{A}_{\tau }\epsilon^{2}=0
\ee

 Note that for all orientifold projections the NS-NS axion field $\sigma$ is projected out and the shift isometries of the remaining RR fields all commute.  The analytic continuation \`{a} la Coleman always leads to a saddlepoint with a real action since  (\ref{realcond}) is projected out. The analytic continuation of the RR axion fields for the RR-charged instanton solution is
\bea
O_{1}: &&  \zeta^{0}\to i \zeta^{'0}\quad  \tilde \zeta_{a}\to i \tilde \zeta_{a}' \no \\
O_{2}: && \tilde\zeta_{0}\to i \tilde\zeta_{0} ', \quad  \zeta^{a}\to i \zeta^{'a }
\eea
 and it follows that the vielbein component $u$  after analytic continuation is imaginary for $O_{1}$ and real  for $O_{2}$.\footnote{Note however that it follows from (\ref{udeforia}) that after analytic continuation one still has the relations  $u_{\mu}=\bar u_{\mu}$ for the $O_{1}$ truncation  and  $u_{\mu}=-\bar u_{\mu}$  for the $O_{2}$
 truncation.}
The same is true for the components $e^{A}$ and $E^{A}$ 
 The consistency of the first two equations in (\ref{bpscona}) and (\ref{bpsconb}) implies that the following linear combinations $\epsilon'$ parameterize the  unbroken supersymmeties  for an extremal BPS instanton solution.
 \bea\label{instsusy}
  O_{1}: && \epsilon'_{\alpha}=\epsilon^{1}_{\alpha}\pm   i \epsilon^{2}_{\alpha}, \;\quad \epsilon'_{\dot \alpha}=\epsilon_{1 \dot \alpha}\pm i \epsilon_{2\dot \alpha}  \no\\
O_{2} :&& \epsilon'_{\alpha}=\epsilon^{1}_{\alpha}\pm  \epsilon^{2}_{\alpha} , \quad \;\;\epsilon'_{\dot\alpha}=\epsilon_{1\dot\alpha }\mp  \epsilon_{2\dot\alpha}
 \eea
 The choice of sign corresponds to a choice of sign for  the RR charge and gives an instanton or anti instanton solution. The comparison of (\ref{ofoldsusy}) and (\ref{instsusy}) shows of the four real supersymmetries which survive the orientifold projection, two are identical to unbroken  supersymmetries of the (anti)-instanton solution.  Two of the four broken supersymmetries of the $N=2$ (anti)-instanton survive the orientifold projection  (\ref{ofoldsusy}). The two broken supersymmetries generate fermionic zero modes in the instanton background. In order to obtain  non-zero correlation function the fermionic zero modes have to be soaked up by the appropriate operator insertions. The resulting terms are instanton induced F-terms 
 \be
 \int d^{4}x \int d^{2}\theta F(\Phi) e^{-S_{inst}} + \int d^{4}x \int d^{2}\bar \theta F(\bar \Phi) e^{-S_{inst}} 
 \ee
 Such are potentially important in constructing phenomenologically viable superstring models as they can lift moduli and be responsible for supersymmetry breaking.
 
  Such terms have been analyzed in several models such as intersecting D-branes in orientifolds \cite{Blumenhagen:2006xt,Akerblom:2007uc,Cvetic:2007sj}. We will leave the evaluation of such terms for the theories obtained by orientifold projections discussed in this section for future work.


\section{Discussion}
\setcounter{equation}{0}

In this paper instanton and wormhole solutions in $d=4$ $N=2$ supergravity theories coming from large volume Calabi-Yau compactification of type II string theories were discussed using a method due to Coleman. It is an interesting question whether other prescriptions (e.g. the dualization of axions to tensor fields) give the same results for solutions, boundary term and saddlepoint action. For the case of the 
$SU(2,1)/SU(2)\times U(1)$ coset  this is indeed the case as shown in a previous paper 
\cite{Chiodaroli:2008rj}). It would be interesting to find a general proof for the equivalence of the prescriptions in order to show that there is no arbitrariness in the analytic continuation procedure.

The Coleman method allows for a classification of possible analytic continuation depending on the charge the Euclidean solution is carrying. Furthermore this prescription produces the boundary terms which are necessary to get a non-zero  action for the instanton.  
The positive definiteness  of the saddlepoint action is however not guaranteed. We proposed an additional condition which guarantees the reality of the solution as well as the positive definiteness of the action.
This condition can only be satisfied for truncated solutions More general real solutions exist after an additional analytic continuation is performed and the action is not positive definite anymore.  Wether these solutions give physical sensible saddle point contributions  is an open problem.

We discussed two cases: the $SU(2,1)/SU(2)\times U(1)$ coset (which was discussed in \cite{Chiodaroli:2008rj}) and  the $G_{2,2}/ SU(2)\times SU(2)$ coset. Instanton and wormhole solutions were constructed using the conserved Noether charges associated with all  the global symmetries of the coset.  The solutions are then explicitly obtained 
for some truncations which give a real saddle-point action.  The method of using the global symmetries to generate the most general solution was discussed for the $G_{2,2}/ SU(2)\times SU(2)$ coset. 
 
For higher dimensional cosets the Noether-method can also be applied to reduce  number of independent equations of motion by using the conservation equations to replace fields and their derivatives by conserved charges.  The usefulness of this approach for more complicated cases is however limited by the  fact that
  for an exact solution it would be necessary to solve algebraic equations in terms of the charges of high degree which can not be done analytically  in general. Generic Calabi-Yau compactifications which are not cosets, have fewer symmetries and the Noether-method is less useful.

The various orientifold projections   which reduce the four dimensional supersymmetry from $N=2$ to $N=1$   provide truncations of the $N=2$ theory. $N=2$ instanton solutions will lead to solutions of the truncated theory, as long as in the solutions all the fields which are projected out are trivial. Such $N=1$ instanton solution can lead to interesting contributions,  since the orientifold projection reduces the number of fermionic   
zero modes. Hence such  instantons can contribute to F-terms in the effective action.

In this paper we focussed on solutions which are $SO(4)$ invariant since the equations of motion reduce to ordinary differential equations, it would be interesting to generalize the solutions to situation with less symmetry. Via the c-map such solutions could be related to rotating or multi-center black holes.

\bigskip

\noindent{\large \bf Acknowledgements}

 \medskip

 This work was supported in
part by NSF grants PHY-04-56200 and PHY-07-57702. M.G. gratefully acknowledges a useful conversation with  Inaki Garcia-Etxebarria. We thank Boris Pioline for a clarifying correspondence. 
M.G. gratefully acknowledges the hospitality of the
Department of Physics and Astronomy, Johns Hopkins University
during the course of this work.

\newpage 

\appendix
\section{Some details on $G_{2,2}$}
In this appendix we give the explict form of the fourteen infinitesimal transformations of the action which generate the Lie algebra of  $G_{2,2}$.  
For completeness we recall the definition of  the matrix $M_{ab}$ is 
\be M=\frac{1}{y^3} 
\left(\begin{array}{cccc} 1 & x & -x^3 & 3 x^2 \\
x & y^2/3+x^2 & -x^4-x^2y^2 & 3x^3+2 x y^2  \\
-x^3 & -x^4-x^2 y^2 & (x^2+y^2)^3  & -3(x^2+y^2)^2 x \\
3x^2 & 3x^3+2xy^2 & -3(x^2+y^2)^2 x & (3x^2 +2 y^2)^2-y^4 \ea \right) \label{def-Mapp} \ee

First, we repeat the infinitesimal shift generators, which produce a Heisenberg algebra.

\bea   E_{p0} \left\{ \barcl \delta x &=& 0 \\
		\delta y &=& 0\\
		\delta \phi &=& 0 \\
		\delta \z^0 &=& \frac{1}{3} \\
		\delta \z^1 &=& 0 \\
		\delta \tiz_0 &=& 0 \\
		\delta \tiz_1 &=& 0 \\
		\delta \s &=& \frac{1}{3} \tiz_0 \ea\right. \;
E_{p1}\left\{ \barcl \delta x &=& 0 \\
		\delta y &=& 0\\
		\delta \phi &=& 0 \\
		\delta \z^0 &=& 0\\
		\delta \z^1 &=& \frac{1}{3} \\
		\delta \tiz_0 &=& 0 \\
		\delta \tiz_1 &=& 0 \\
		\delta \s &=& \frac{1}{3} \tiz_1 \ea\right.\; 
 E_{q0} \left\{ \barcl \delta x &=& 0 \\
		\delta y &=& 0\\
		\delta \phi &=& 0 \\
		\delta \z^0 &=& 0\\
		\delta \z^1 &=& 0 \\
		\delta \tiz_0 &=& \sqrt{3} \\
		\delta \tiz_1 &=& 0 \\
		\delta \s &=& 0 \ea \right. \;
		E_{q1}  \left\{ \barcl \delta x &=& 0 \\
		\delta y &=& 0\\
		\delta \phi &=& 0 \\
		\delta \z^0 &=& 0\\
		\delta \z^1 &=& 0 \\
		\delta \tiz_0 &=& 0 \\
		\delta \tiz_1 &=& \sqrt{3} \\
		\delta \s &=& 0 \ea \right.  \;
		E \left\{ \barcl \delta x &=& 0 \\
		\delta y &=& 0\\
		\delta \phi &=& 0 \\
		\delta \z^0 &=& 0\\
		\delta \z^1 &=& 0\\
		\delta \tiz_0 &=& 0 \\
		\delta \tiz_1 &=& 0 \\
		\delta \s &=& -\frac{1}{2\sqrt{3}} \ea\right.  \label{gen-G22-3}
		\eea
		In the next set of generator  $H$ generates a scale transformation, whereas $Y_{0,+,-}$ generators a $SL(2,R)$ action on the moduli $x,y$.
 \bea   H  \left\{  \barcl \delta x &=&  0\\
		\delta y &=& 0\\
		\delta \phi &=& 2 \\
		\delta \z^0 &=& \z^0 \\
		\delta \z^1 &=& \z^1 \\
		\delta \tiz_0 &=& \tiz_0 \\
		\delta \tiz_1 &=& \tiz_1 \\
		\delta \s &=&  2 \s  \ea \right. 
\quad Y_0 \left\{ \barcl \delta x &=& - x \\
		\delta y &=& -  y\\
		\delta \phi &=& 0 \\
		\delta \z^0 &=& \frac{3}{2}  \z^0 \\
		\delta \z^1 &=& \half  \z^1 \\
		\delta \tiz_0 &=& - \frac{3}{2}  \tiz_0 \\
		\delta \tiz_1 &=& - \half  \tiz_1 \\
		\delta \sigma &=& 0  \ea \right. 
\quad Y_+ \left\{ \barcl \delta x &=& \frac{1}{\sqrt{6}} (y^2-x^2) \\
		\delta y &=&  - \sqrt{\frac{2}{3}}  xy \\
		\delta \phi &=& 0 \\
		\delta \z^0 &=& \sqrt{\frac{3}{2}} \z^1 \\
		\delta \z^1 &=&  - \frac{\sqrt{2}}{3 \sqrt{3}} \tiz_1 \\
		\delta \tiz_0 &=& 0 \\
		\delta \tiz_1 &=& -\sqrt{\frac{3}{2}} \tiz_0  \\
		\delta \s &=& - \frac{1}{3 \sqrt{6}} \tiz_1^2 \ea \right. 
		\quad Y_-  \left\{ \barcl  \delta x &=&  - \sqrt{\frac{3}{2}} \\
		\delta y &=& 0 \\
		\delta \phi &=& 0 \\
		\delta \z^0 &=& 0 \\
		\delta \z^1 &=& - \sqrt{\frac{3}{2}} \z^0 \\
		\delta \tiz_0 &=&  \sqrt{\frac{3}{2}} \tiz_1 \\
		\delta \tiz_1 &=& 3 \sqrt{6} \z^1 \\
		\delta \sigma &=& \frac{3 \sqrt{3}}{\sqrt{2}} {\z^1}^2 \ea \right. 
		\eea
		The rest of the generators are quite complicated and complete the $G_{2,2}$ algebra. $F_{p0}$ and $F_{p1}$ are given by:
		\bea 
F_{p0} \left\{ \barcl \delta x &=& 6 (x \z^0 - \z^1)\\
		\delta y &=&  6 y \z^0 \\
		\delta \phi &=& -6 \z^0 \\
		\delta \z^0 &=& 6 \left( \frac{e^{\phi}}{y^3} -{\z^0}^2  \right) \\
		\delta \z^1 &=& 6  \left( \frac{e^{\phi} x}{y^3} - \z^0 \z^1 \right) \\
		\delta \tiz_0 &=& 6 \left( \frac{e^{\phi}x^3}{y^3} + \s \right) \\
		\delta \tiz_1 &=&  18 \left( {\z^1}^2 - \frac{e^{ \phi}x^2}{y^3} \right) \\
		\delta \s &=& 6 \left( \frac{e^{\phi}x^3}{y^3} \z^0 -\frac{3 e^{\phi}x^2}{y^3} \z^1 + 2 {\z^1}^3 \right) \ea \right. 
		F_{p1}  \left\{ \barcl \delta x &=&  2 (x^2-y^2) \z^0 + 2 x \z^1 + \frac{4}{3} \tiz_1\\
		\delta y &=& 2  y (\z^1+ 2x\z^0) \\
		\delta \phi &=& -6 \z^1 \\
		\delta \z^0 &=& 6 \left( \frac{e^{\phi}x}{y^3} - \z^0 \z^1  \right) \\
		\delta \z^1 &=& 2 \left[ \frac{e^{\phi} (3 x^2 +y^2)}{y^3} - {\z^1}^2 + \frac{2}{3} \z^0 \tiz_1 \right] \\
		\delta \tiz_0 &=& 6 \left[\frac{e^{\phi}(x^4+x^2y^2)}{y^3} - \frac{1}{9} \tiz^2_1 \right] \\
		\delta \tiz_1 &=&  -6 \left[ \frac{e^{\phi}(3x^3+2 x y^2)}{y^3} - \s +\frac{4}{3}\z^1 \tiz_1 \right] \\
		\delta \sigma &=& 6 \left[ \ss  \frac{e^{\phi}x^2 |z|^2}{y^3} \z^0 -\frac{e^{\phi}(3x^3+2 xy^2)}{y^3} \z^1  + \frac{2}{3} {\z^1}^2 \tiz_1 \right] \ea \right. \qquad
		 \eea
with $|z|^2=x^2+y^2$. The $F_{q0}$ generator is: 
\bea &&F_{q0} \left\{ \barcl \delta x &=& \frac{2}{\sqrt{3}} [\frac{1}{3}(y^2-x^2) \tiz_1 - x \tiz_0 ]\\
		\delta y &=& -\frac{2}{\sqrt{3}} y ( \tiz_0+ \frac{2}{3} x \tiz_1) \\
		\delta \phi &=& -\frac{2}{\sqrt{3}} \tiz_0 \\
		\delta \z^0 &=& \frac{2}{\sqrt{3}} \left( \frac{e^{\phi}x^3}{y^3} +\z^0 \tiz_0 + \z^1 \tiz_1 - \s \right) \\
		\delta \z^1 &=&  \frac{2}{\sqrt{3}} \left[ \frac{e^{\phi}  x^2 |z|^2}{y^3} - \frac{1}{9} {\tiz_1}^2 \right] \\
		\delta \tiz_0 &=& \frac{2}{\sqrt{3}} \left[ \frac{e^{\phi}|z|^6}{y^3} - \tiz^2_0 \right] \\
		\delta \tiz_1 &=&  -\frac{2}{\sqrt{3}} \left[3 \frac{e^{\phi}x|z|^4}{y^3} + \tiz_0 \tiz_1 \right] \\
		\delta \sigma &=& \frac{2}{\sqrt{3}} \left[\frac{e^{\phi}|z|^6}{y^3} \z^0 - \frac{3 e^{ \phi}x|z|^4}{y^3} \z^1  - \tiz_0 \sigma - \frac{1}{27} {\tiz_1}^3 \right] \ea \right.  \label{gen-G22-1} 
		\eea
and the $F_{q1}$ generator is: 
		\bea
 &&F_{q1}\left\{ \barcl \delta x &=& -\frac{2}{\sqrt{3}} [2(x^2-y^2) \z^1 - \tiz_0 +\frac{1}{3} x \tiz_1]\\ 
		\delta y &=& - \frac{2}{\sqrt{3}}(4 x y \z^1+ \frac{1}{3} y \tiz_1) \\
		\delta \phi &=& -\frac{2}{\sqrt{3}} \tiz_1 \\
		\delta \z^0 &=& -2 \sqrt{3} (\frac{ e^{\phi} x^2}{y^3} - {\z^1}^2) \\
		\delta \z^1 &=& -\frac{2}{\sqrt{3}} [\frac{e^{\phi} (3 x^3 +2 x y^2)}{y^3}+  \s - \z^0 \tiz_0 + \frac{1}{3} \z^1 \tiz_1 ]\\
		\delta \tiz_0 &=& -\frac{2}{\sqrt{3}} [ \frac{3 e^{\phi}x|z|^4}{y^3} + \tiz_0 \tiz_1] \\
		\delta \tiz_1 &=&  2 \sqrt{3} [\frac{ e^{\phi}(3 x^4+ 4 x^2 y^2+y^4)}{y^3} - 2 \z^1 \tiz_0 - \frac{1}{9} {\tiz_1}^2  \\
		\delta \sigma &=& - 2 \sqrt{3} [ \frac{ e^{\phi}x|z|^4}{y^3} \z^0 - \frac{e^{ \phi}(3x^4+4 x^2 y^2+y^4)}{y^3} \z^1  + \frac{1}{3} \tiz_1 \sigma + {\z^1}^2 \tiz_0] \ea \right.  
		\eea
Finally, the last generator is:
		\bea 
		&& F \; \left\{ \barcl \delta x &=& 2 \sqrt{3} [(x^2-y^2) ({\z^1}^2 + \frac{1}{3} \z^0 \tiz_1 )+ \frac{1}{9} {\tiz_1}^2 -  \z^1 \tiz_0 + x( \z^0 \tiz_0 + \frac{1}{3} \z^1 \tiz_1 )] \\ 
		\delta y &=& 2 \sqrt{3} (2 x y {\z^1}^2 + \frac{1}{3} y \z^1 \tiz_1 + y \z^0 \tiz_0 + \frac{2}{3} xy \z^0 \tiz_1) \\
		\delta \phi &=& -4 \sqrt{3}  (\half \z^0 \tiz_0 + \half \z^1 \tiz_1 - \s)\\
		\delta \z^0 &=& 2 \sqrt{3} \left[e^{\phi} (M \z)_1 -{\z^1}^3 -{\z^0}^2\tiz_0 -\z^0 \z^1 \tiz_1 + \z^0 \s \right] \\
		\delta \z^1 &=& 2 \sqrt{3} \left[e^{\phi}(M  \z)_2 + \z^0 {\tiz_1}^2 -\z^0 \z^1 \tiz_0 -{\z^1}^2 \tiz_1 + \z^1 \s \right] \\
		\delta \tiz_0 &=& 2 \sqrt{3}\left[-e^{\phi}(M \z )_3 -\frac{1}{27}{\tiz_1}^3 + \tiz_0 \s \right] \\
		\delta \tiz_1 &=& 2 \sqrt{3} \left[-e^{\phi}(M \z)_4 + 3 {\z^1}^2 \tiz_0 -\frac{2}{3}\z^1 \tiz_1^2  + \tiz_1 \s \right] \\
		\delta \s &=& 2 \sqrt{3}  \left\{-e^{ \phi} [ (M \z)_3 \z^0 + (M \z)_4 \z^1] -e^{2 \phi} +  2 {\z^1}^3 \tiz_0 - \frac{2 {\z^1}^2 \tiz_1^2}{6} + \s^2 \right\} \ea \right.    \label{gen-G22-2}\eea
here $(M\zeta)_{i}$ is the i-th component of the vector $M \zeta$ with $M$ defined in (\ref{def-M}) and the   the vector $\zeta$ is defined in (\ref{def-zeta}).
		The generators (\ref{gen-G22-3}-\ref{gen-G22-2}) form a $G_{2,2}$ group of global symmetries. Some of the relevant nonvanishing commutation relations are:
\be \begin{array}{rcl} \left[ E_{p_I},E_{q_J} \right] & = & -2 \delta^{IJ} E  \\ 
\left[ Y_-,Y_+ \right] &= & Y_0 \\
\left[ E_{p_0},F_{p_0} \right] & = & H + 2 Y_0  \\
 \left[ E_{p_1},F_{p_1} \right] &= &  H + \frac{2}{3} Y_0 \\ 
\left[E_{p_1} , F_{q_1} \right] & = & - \frac{4 \sqrt{2}}{3} Y_+ \\
\left[ Y_+ , E_{p_1} \right] &=&  \sqrt{ \frac{3}{2}} E_{p_0} \\
\left[ Y_+ , E_{q_1} \right] &=& - \sqrt{2} E_{p_1} \\
\left[ Y_+ , E_{q_0} \right] &=& - \sqrt{ \frac{3}{2}} E_{q_1} \\
\left[ Y_+ , F_{p_0} \right] &=&  -\sqrt{\frac{3}{2}} F_{p_1} \\
\left[ Y_+ , F_{p_1} \right] &=&  \sqrt{ 2} F_{q_1} \\
\left[ Y_+ , F_{q_1} \right] &=&  \sqrt{\frac{3}{2}} F_{q_0}  \\
\left[E , F_{q_I} \right] & = & - E_{p_I} \ea \qquad \brcl 
\left[ F_{p_I},F_{q_J} \right] & = & 2 \delta^{IJ} F  \\ 
\left[ E , F \right] &=& H \\  
\left[ E_{q_0},F_{q_0} \right] & = & H - 2 Y_0 \\ 
\left[ E_{q_1},F_{q_1} \right] & = & H - \frac{2}{3} Y_0  \\ 
\left[E_{q_1} , F_{p_1} \right] & = & \frac{4 \sqrt{2}}{3} Y_- \\
\left[ Y_- , E_{p_0} \right] &=& - \sqrt{ \frac{3}{2}} E_{p_1} \\
\left[ Y_- , E_{p_1} \right] &=&  \sqrt{2} E_{q_1} \\
\left[ Y_- , E_{q_1} \right] &=&  \sqrt{ \frac{3}{2}} E_{q_0} \\
\left[ Y_- , F_{p_1} \right] &=&  \sqrt{\frac{3}{2}} F_{p_0} \\
\left[ Y_- , F_{q_0} \right] &=&  -\sqrt{ \frac{3}{2}} F_{q_1} \\
\left[ Y_- , F_{q_1} \right] &=&  -\sqrt{2} F_{p_1}  \\ 
\left[E , F_{p_I} \right] & = &  E_{q_I}  \ea \ee
Finally, we get the following expressions for the Noether current associated with the symmetries $H
$:
\be j_H^{\mu}  =   3 \z^0 j^{\mu}_{E_{p0}} + 3 \z^1 j^{\mu}_{E_{p1}} + \frac{ \tiz_0 }{ \sqrt{3}} j^{\mu}_{E_{q0}} + \frac{ \tiz_1}{ \sqrt{3}} j^{\mu}_{E_{q1}} -2 \sqrt{3} 
(2\sigma - \z^I \tiz_I)j^{\mu}_E +2 \partial^{\mu} \phi \label{eq-qH} 
\ee
and the current associated with $Y_{0}$:
\be
 j^{\mu}_{Y_0}  =  \frac{ 3}{ 2} \big( 3 \z^0 j^{\mu}_{E_{p0}} + \z^1 j^{\mu}_{E_{p1}} - \frac{ \tiz_0 j^{\mu}_{E_{q0}}}{ \sqrt{3}} -\frac{ \tiz_1 j^{\mu}_{E_{q1}} }{ 3 \sqrt{3}}\big)  + \sqrt{3} (3 \z^0 \tiz_0 + \z^1 \tiz_1) j^{\mu}_E  -3 \frac{   \partial^\mu |z|^2 }{ 2 y^2} \label{eq-qY0}
 \ee  
and the current associated with $Y_{-}$:
 \be
 j^{\mu}_{Y_-}   =   \frac{  \tiz_1}{ \sqrt{2}}  j^{\mu}_{E_{q0}} - \frac{ 3 \sqrt{3}}{\sqrt{2}} \z^0 j^{\mu}_{E_{p1}} + 3 \sqrt{2} \z^1 j^{\mu}_{E_{q1}} - 3 \sqrt{2} (3 {\z^1}^2 + \z^0 \tiz_1)j^{\mu}_E - \frac{ 3 \sqrt{3}}{ \sqrt{2}} \frac{ \partial^{\mu} x}{y^2} \label{eq-qY-}
 \ee  
and the current associated with $Y_{+}$:
 \be
 j^{\mu}_{Y_+}  = \frac{\sqrt{3}}{\sqrt{2}}  \big( 3  \z^1  j^{\mu}_{E_{p0}} - \frac{ \tiz_0}{ \sqrt{3}} j^{\mu}_{E_{q1}} -  \frac{2}{ 3} \tiz_1 j^{\mu}_{E_{p1}} - \frac{ x^{2}-y^{2}}{ y^2} \partial^\mu x -2 \frac{ x\partial^{\mu}y}{ y} \big) + \sqrt{2} ( 3 \z^1 \tiz_0 - \frac{ \tiz_1^2}{ 3})j^{\mu}_E  \label{eq-qY+} \ee
The current associated with the $F_{p_{I}}$ and $F_{q_{I}}$ are more complicated. $F_{p0}$ gives the current:
\bea \nn \frac{ j_{F_{p0}}^{\mu}}{ 18} &=&   \left( \frac{ e^{ \phi}}{ y^3} -  {\z^0}^2   \right) j^{\mu}_{E_{p0}} +  \left(\frac{ e^{\phi}x}{ y^3} -  \z^0 \z^1 \right) j^{\mu}_{E_{p1}} + \frac{1}{ 3 \sqrt{3}}  \left(\frac{ e^{\phi}x^3}{ y^3} + \s  \right) j^{\mu}_{E_{q0}}\\
&& \nn  - \frac{ 1} {\sqrt{3}}  \left( \frac{ e^{\phi}x^2}{ y^3} - {\z^1}^2  \right) j^{\mu}_{E_{q1}} + \frac{ 2}{ \sqrt{3}} \frac{ e^{\phi}}{ y^3} \left( 3 x^2 \z^1 - x^3 \z^0 + \tiz_0 + x \tiz_1 \right) j^\mu_E \\
&&   - \left( 2{\z^1}^3 + {\z^0}^2 \tiz_0 + \z^0 \z^1 \tiz_1 \right)  j^{\mu}_E -   \frac{ 1}{ 3}  \z^0  \; \partial^{\mu} \phi +   \frac{ x \z^0 - \z^1}{ y^2} \partial^{\mu}x+   \frac{ \partial^\mu y}{ y} \qquad  \label{eq-qFp0} 
\eea
The current associated to $F_{p1}$ is:
\bea
\nn  \frac{ j_{F_{p1}}^{\mu}}{ 18} & =&  \left( \frac{ e^{\phi}x}{ y^3} - \z^0 \z^1   \right) j^{\mu}_{E_{p0}} - \left( \frac{ e^{\phi}(x^3+ \frac{2}{3} x y^2)}{ y^3} - \frac{ \s}{ 3} + \frac{ 4 \z^1 \tiz_1 }{ 9} \right) \frac{ j^{\mu}_{E_{q1}}}{ \sqrt{3}} + \\
&&  \nn \left( \frac{ e^{\phi}(x^2 +\frac{y^2}{3})}{ y^3} + \frac{ 2 \z^0 \tiz_1}{ 9}- \frac{ {\z^1}^2}{ 3} \right) j^{\mu}_{E_{p1}}+ \left(\frac{ e^{ \phi}x^2 |z|^2}{ 3 y^3} - \frac{ \tiz_1^2}{ 27}  \right) \frac{ j^{\mu}_{E_{q0}}}{ \sqrt{3}} -  \\ 
&&  \nn  \frac{   \z^1}{3} \partial^{\mu} \phi+ 2 \frac{ e^{\phi}}{ y^3} \left( (3 x^3 +2 x y^2 )\z^1- x^2|z|^2 \z^0  + x \tiz_0 + (x^2 + \frac{ y^2}{ 3}) \tiz_1 \right) \frac{ j^\mu_E}{ \sqrt{3}} + \\ &&   2 \frac{  {\z^1}^2 \tiz_1   + {2 \over 3 }  \z^0 \tiz_1^2 -3 \z^0 \z^1 \tiz_0}{3 \sqrt{3}} j^{\mu}_E  + \frac{ (x^{2}-y^{2}) \z^0 + x \z^1 +  \frac{2}{3} \tiz_1}{ 3  y^2} \partial^{\mu}x +   \frac{ \z^1 + 2 x \z^0 }{ 3 y} \partial^\mu y   \quad \qquad  \label{eq-qFp1}
\eea
and the current associated to $F_{q0}$ is:
\bea
   \frac{  j_{F_{q0}}^{\mu}}{ 2 \sqrt{3}} & =&   \left( \frac{ e^{ \phi} x^3 }{ y^3} +  \z^0 \tiz_0 + \z^1 \tiz_1 -\s  \right) j^{\mu}_{E_{p0}} +  \left( \frac{  e^{ \phi}x^2|z|^2}{ y^3} - \frac{ \tiz_1^2}{ 9} \right) j^{\mu}_{E_{p1}} \nn \\
&&  +  \left(\frac{ e^{ \phi}|z|^6}{ y^3} - \tiz_0^2  \right) \frac{ j^{\mu}_{E_{q0}}}{ 3 \sqrt{3}} -  \left( \frac{ e^{ \phi} x |z|^4}{ y^3} + \frac{ \tiz_0 \tiz_1 }{ 3} \right) \frac{ j^{\mu}_{E_{q1}}}{ \sqrt{3}} \nn-   \frac{ \tiz_0}{3}  \partial^{\mu} \phi + \\ &&   \frac{ 2  e^{\phi} }{\sqrt{3} y^3} \left( 3 x |z|^4 \z^1 - |z|^6 \z^0 + x^3 \tiz_0 + x^2|z|^2  \tiz_1 \right) j^\mu_E \nn \\
&&  + \frac{ 2}{ \sqrt{3}} \left( \z^0 \tiz_0^2 + \z^1 \tiz_0 \tiz_1 - \frac{ 2}{ 27} \tiz_1^3 \right)  j^{\mu}_E  -   \frac{ 3 x \tiz_0 + (x^2-y^2) \tiz_1}{ 3 y^2} \partial^{\mu}x  +   \frac{ \tiz_0 + \frac{2}{3} x \tiz_1}{ y} \partial^\mu y \quad \qquad  \label{eq-qFq0} 
\eea
Finally, the current associated to $F_{q1}$ is: 
\bea
\frac{j_{F_{q1}}^{\mu}}{ 2 \sqrt{3}} & =&  3 \left( {\z^1}^2 - \frac{ e^{\phi}x^2}{ y^3} \right) j^{\mu}_{E_{p0}} - \left( \frac{ e^{\phi}(3 x^3+ 2 x y^2)}{ y^3} + \s -\z^0 \tiz_0 +  \frac{ \z^1 \tiz_1 }{ 3} \right)  j^{\mu}_{E_{p1}} - \frac{ \tiz_1}{3}  \partial^{\mu} \phi \nn \\
&& - \left(\frac{ e^{\phi}x|z|^4 }{ y^3} + \frac{ \tiz_0 \tiz_1 }{ 3} \right) \frac{ j^{\mu}_{E_{q0}}}{ \sqrt{3}} + \left(\frac{ e^{ \phi}(3 x^4 + 4 x^2 y^2+ y^4)}{ y^3} - 2 \z^1 \tiz_0  -\frac{ \tiz_1^2}{ 9}  \right) \frac{ j^{\mu}_{E_{q1}}}{ \sqrt{3}} \nn \\ 
&&   \frac{  2 \sqrt{3} e^{\phi}}{  y^3} \left(  3 x |z|^4 \z^0 -(3 x^4 +4 x^2 y^2+y^4 )\z^1  - x^2 \tiz_0 \nn - \big( x^3 + \frac{ 2 x y^2}{ 3} \big) \tiz_1 \right)  j^\mu_E + \\
&&  \frac{  2 \z^0 \tiz_0 \tiz_1 + 12 {\z^1}^2 \tiz_0 - \frac{ 2}{ 3} \z^1  \tiz^2_1 }{\sqrt{3}}  j^{\mu}_E  - \frac{ 2 (x^{2}-y^{2}) \z^1 -  \tiz_0 + \frac{x}{3} \tiz_1}{ y^2} \partial^{\mu}x -  \frac{ 4 x \z^1 + \frac{\tiz_1}{3} }{ y} \partial^\mu y \quad \qquad  \label{eq-qFq1} 
\eea 
These expressions greatly simplify in the truncations considered in section 5.


\newpage


\begin{thebibliography}{99}

\small{




\bibitem{Hawking:1988ae}
  S.~W.~Hawking,
  ``Wormholes in Space-Time,''
  Phys.\ Rev.\  D {\bf 37} (1988) 904.


\bibitem{Lavrelashvili:1988jj}
  G.~V.~Lavrelashvili, V.~A.~Rubakov and P.~G.~Tinyakov,
  ``PARTICLE CREATION AND DESTRUCTION OF QUANTUM COHERENCE BY TOPOLOGICAL
  CHANGE,''
  Nucl.\ Phys.\  B {\bf 299} (1988) 757.


\bibitem{Giddings:1987cg}
  S.~B.~Giddings and A.~Strominger,
  ``Axion Induced Topology Change In Quantum Gravity And String Theory,''
  Nucl.\ Phys.\  B {\bf 306} (1988) 890.



\bibitem{Coleman:1989zu}
  S.~R.~Coleman and K.~M.~Lee,
  ``WORMHOLES MADE WITHOUT MASSLESS MATTER FIELDS,''
  Nucl.\ Phys.\  B {\bf 329}, 387 (1990).
  
  
  
\bibitem{Giddings:1989bq}
  S.~B.~Giddings and A.~Strominger,
  ``String Wormholes,''
  Phys.\ Lett.\  B {\bf 230} (1989) 46.


\bibitem{Strominger:1983ns}
  A.~Strominger,
  ``Vacuum Topology And Incoherence In Quantum Gravity,''
  Phys.\ Rev.\ Lett.\  {\bf 52} (1984) 1733.

\bibitem{Hawking:1987mz}
  S.~W.~Hawking,
  ``Quantum Coherence Down the Wormhole,''
  Phys.\ Lett.\  B {\bf 195} (1987) 337.



\bibitem{Coleman:1988cy}
  S.~R.~Coleman,
  ``Black Holes as Red Herrings: Topological Fluctuations and the Loss of
  Quantum Coherence,''
  Nucl.\ Phys.\  B {\bf 307} (1988) 867.

\bibitem{Rey:1998yx}
  S.~J.~Rey,
  ``Holographic principle and topology change in string theory,''
  Class.\ Quant.\ Grav.\  {\bf 16} (1999) L37
  [arXiv:hep-th/9807241].




\bibitem{Gutperle:2002km}
  M.~Gutperle and W.~Sabra,
  ``Instantons and wormholes in Minkowski and (A)dS spaces,''
  Nucl.\ Phys.\  B {\bf 647} (2002) 344
  [arXiv:hep-th/0206153].


\bibitem{Maldacena:2004rf}
  J.~M.~Maldacena and L.~Maoz,
  ``Wormholes in AdS,''
  JHEP {\bf 0402} (2004) 053
  [arXiv:hep-th/0401024].


\bibitem{ArkaniHamed:2007js}
  N.~Arkani-Hamed, J.~Orgera and J.~Polchinski,
  ``Euclidean Wormholes in String Theory,''
  arXiv:0705.2768 [hep-th].


\bibitem{Gibbons:1995vg}
  G.~W.~Gibbons, M.~B.~Green and M.~J.~Perry,
  ``Instantons and Seven-Branes in Type IIB Superstring Theory,''
  Phys.\ Lett.\  B {\bf 370} (1996) 37
  [arXiv:hep-th/9511080].


\bibitem{Green:1997tv}
  M.~B.~Green and M.~Gutperle,
  ``Effects of D-instantons,''
  Nucl.\ Phys.\  B {\bf 498} (1997) 195
  [arXiv:hep-th/9701093].

 
\bibitem{Bergshoeff:2004fq}
  E.~Bergshoeff, A.~Collinucci, U.~Gran, D.~Roest and S.~Vandoren,
  ``Non-extremal D-instantons,''
  JHEP {\bf 0410} (2004) 031
  [arXiv:hep-th/0406038].

  
\bibitem{Einhorn:2002sj}
  M.~B.~Einhorn,
  ``Instantons and SL(2,R) symmetry in type IIB supergravity,''
  Phys.\ Rev.\  D {\bf 68} (2003) 067701
  [arXiv:hep-th/0212322].
  

  

\bibitem{Gutperle:2000sb}
  M.~Gutperle and M.~Spalinski,
  ``Supergravity instantons and the universal hypermultiplet,''
  JHEP {\bf 0006} (2000) 037
  [arXiv:hep-th/0005068].

\bibitem{Davidse:2004gg}
  M.~Davidse, U.~Theis and S.~Vandoren,
  ``Fivebrane instanton corrections to the universal hypermultiplet,''
  Nucl.\ Phys.\  B {\bf 697} (2004) 48
  [Erratum-ibid.\  B {\bf 750} (2006) 108]
  [arXiv:hep-th/0404147].
  
  
\bibitem{Theis:2002er}
  U.~Theis and S.~Vandoren,
  ``Instantons in the double-tensor multiplet,''
  JHEP {\bf 0209} (2002) 059
  [arXiv:hep-th/0208145].


  
\bibitem{Gutperle:2000ve}
  M.~Gutperle and M.~Spalinski,
  ``Supergravity instantons for N = 2 hypermultiplets,''
  Nucl.\ Phys.\  B {\bf 598} (2001) 509
  [arXiv:hep-th/0010192].


\bibitem{deVroome:2006xu}
  M.~de Vroome and S.~Vandoren,
  ``Supergravity description of spacetime instantons,''
  Class.\ Quant.\ Grav.\  {\bf 24} (2007) 509
  [arXiv:hep-th/0607055].
  
    
\bibitem{Ferrara:1995ih}
  S.~Ferrara, R.~Kallosh and A.~Strominger,
  ``N=2 extremal black holes,''
  Phys.\ Rev.\  D {\bf 52} (1995) 5412
  [arXiv:hep-th/9508072].
  
  
\bibitem{Ferrara:1996dd}
  S.~Ferrara and R.~Kallosh,
  ``Supersymmetry and Attractors,''
  Phys.\ Rev.\  D {\bf 54} (1996) 1514
  [arXiv:hep-th/9602136].
  
  
\bibitem{LopesCardoso:1998wt}
  G.~Lopes Cardoso, B.~de Wit and T.~Mohaupt,
  ``Corrections to macroscopic supersymmetric black-hole entropy,''
  Phys.\ Lett.\  B {\bf 451} (1999) 309
  [arXiv:hep-th/9812082].


\bibitem{Ooguri:2004zv}
  H.~Ooguri, A.~Strominger and C.~Vafa,
  ``Black hole attractors and the topological string,''
  Phys.\ Rev.\  D {\bf 70} (2004) 106007
  [arXiv:hep-th/0405146].


 

\bibitem{deWit:1984px}
  B.~de Wit, P.~G.~Lauwers and A.~Van Proeyen,
  ``Lagrangians Of N=2 Supergravity - Matter Systems,''
  Nucl.\ Phys.\  B {\bf 255} (1985) 569.

\bibitem{Ferrara:1989ik}
  S.~Ferrara and S.~Sabharwal,
  ``Quaternionic Manifolds for Type II Superstring Vacua of Calabi-Yau
  Spaces,''
  Nucl.\ Phys.\  B {\bf 332}, 317 (1990).
  
  
\bibitem{Cecotti:1988qn}
  S.~Cecotti, S.~Ferrara and L.~Girardello,
  ``Geometry of Type II Superstrings and the Moduli of Superconformal Field
  Theories,''
  Int.\ J.\ Mod.\ Phys.\  A {\bf 4} (1989) 2475.


\bibitem{Dai:1989ua}
  J.~Dai, R.~G.~Leigh and J.~Polchinski,
  ``New Connections Between String Theories,''
  Mod.\ Phys.\ Lett.\  A {\bf 4} (1989) 2073.
  
    
\bibitem{Bohm:1999uk}
  R.~Bohm, H.~Gunther, C.~Herrmann and J.~Louis,
  ``Compactification of type IIB string theory on Calabi-Yau threefolds,''
  Nucl.\ Phys.\  B {\bf 569} (2000) 229
  [arXiv:hep-th/9908007].
  
  
\bibitem{deWit:1990na}
  B.~de Wit and A.~Van Proeyen,
  ``Symmetries of dual quaternionic manifolds,''
  Phys.\ Lett.\  B {\bf 252} (1990) 221.
  }
  
  
\bibitem{deWit:1992wf}
  B.~de Wit, F.~Vanderseypen and A.~Van Proeyen,
  ``Symmetry structure of special geometries,''
  Nucl.\ Phys.\  B {\bf 400} (1993) 463
  [arXiv:hep-th/9210068].
  
  
  
   
\bibitem{Schwarz:1983qr}
  J.~H.~Schwarz,
  ``Covariant Field Equations Of Chiral N=2 D = 10 Supergravity,''
  Nucl.\ Phys.\ B {\bf 226} (1983) 269.

\bibitem{Howe:1983sr}
  P.~S.~Howe and P.~C.~West,
  ``The Complete N=2, D = 10 Supergravity,''
  Nucl.\ Phys.\ B {\bf 238} (1984) 181.


  \bibitem{wolfa}
  J. A. Wolf, ``Complex homogeneous contact manifolds and quaternionic symmetric
spaces'', J. of Math. Mech., 14 (1965), 1033.
  
  \bibitem{alexa}
  D.V. Alekseevskii, ``Riemannian manifolds with exceptional holonomy groups'',  Funct.
Anal. Appl. 2 (1968), 97; `` Compact quaternion spaces'', Funct. Anal. Appl. 2 (1968),
106; ``Classification of quaternionic spaces with transitive solvable group of motions,''
Math. USSR v. 9 (1975), 297.
  
  
\bibitem{Cortes:2005uq}
  V.~Cortes, C.~Mayer, T.~Mohaupt and F.~Saueressig,
  ``Special geometry of Euclidean supersymmetry. II: Hypermultiplets and  the
  c-map,''
  JHEP {\bf 0506} (2005) 025
  [arXiv:hep-th/0503094].


\bibitem{Chiodaroli:2008rj}
  M.~Chiodaroli and M.~Gutperle,
  ``Instantons and Wormholes for the Universal Hypermultiplet,''
  arXiv:0807.3409 [hep-th].
  
  
  
\bibitem{Behrndt:1997ch}
  K.~Behrndt, I.~Gaida, D.~Lust, S.~Mahapatra and T.~Mohaupt,
  ``From type IIA black holes to T-dual type IIB D-instantons in N = 2,  D = 4
  supergravity,''
  Nucl.\ Phys.\  B {\bf 508} (1997) 659
  [arXiv:hep-th/9706096].
 
\bibitem{Gunaydin:2005mx}
  M.~Gunaydin, A.~Neitzke, B.~Pioline and A.~Waldron,
  ``BPS black holes, quantum attractor flows and automorphic forms,''
  Phys.\ Rev.\  D {\bf 73} (2006) 084019
  [arXiv:hep-th/0512296].
  
   
 
\bibitem{Gunaydin:2007bg}
  M.~Gunaydin, A.~Neitzke, B.~Pioline and A.~Waldron,
  ``Quantum Attractor Flows,''
  JHEP {\bf 0709} (2007) 056
  [arXiv:0707.0267 [hep-th]].
 
  
\bibitem{Goldstein:2005hq}
  K.~Goldstein, N.~Iizuka, R.~P.~Jena and S.~P.~Trivedi,
  ``Non-supersymmetric attractors,''
  Phys.\ Rev.\  D {\bf 72} (2005) 124021
  [arXiv:hep-th/0507096].
  
  
\bibitem{Kallosh:2006bt}
  R.~Kallosh, N.~Sivanandam and M.~Soroush,
  ``The non-BPS black hole attractor equation,''
  JHEP {\bf 0603} (2006) 060
  [arXiv:hep-th/0602005].
  
  
\bibitem{Gaiotto:2007ag}
  D.~Gaiotto, W.~W.~Li and M.~Padi,
  ``Non-Supersymmetric Attractor Flow in Symmetric Spaces,''
  JHEP {\bf 0712} (2007) 093
  [arXiv:0710.1638 [hep-th]].
  
  
\bibitem{Saraikin:2007jc}
  K.~Saraikin and C.~Vafa,
  ``Non-supersymmetric Black Ho les and Topological Strings,''
  Class.\ Quant.\ Grav.\  {\bf 25} (2008) 095007
  [arXiv:hep-th/0703214].

\bibitem{Bergshoeff:2008be}
  E.~Bergshoeff, W.~Chemissany, A.~Ploegh, M.~Trigiante and T.~Van Riet,
  ``Generating Geodesic Flows and Supergravity Solutions,''
  arXiv:0806.2310 [hep-th].
  
  
\bibitem{Shmakova:1996nz}
  M.~Shmakova,
  ``Calabi-Yau black holes,''
  Phys.\ Rev.\  D {\bf 56} (1997) 540
  [arXiv:hep-th/9612076].
  
\bibitem{Behrndt:1997ny}
  K.~Behrndt, D.~Lust and W.~A.~Sabra,
  ``Stationary solutions of N = 2 supergravity,''
  Nucl.\ Phys.\  B {\bf 510} (1998) 264
  [arXiv:hep-th/9705169].
  
 
  
\bibitem{Denef:2000nb}
  F.~Denef,
  ``Supergravity flows and D-brane stability,''
  JHEP {\bf 0008} (2000) 050
  [arXiv:hep-th/0005049].
  

\bibitem{LopesCardoso:2000qm}
  G.~Lopes Cardoso, B.~de Wit, J.~Kappeli and T.~Mohaupt,
  ``Stationary BPS solutions in N = 2 supergravity with R**2 interactions,''
  JHEP {\bf 0012} (2000) 019
  [arXiv:hep-th/0009234].
  


 
\bibitem{Becker:1995kb}
  K.~Becker, M.~Becker and A.~Strominger,
  ``Five-Branes, Membranes And Nonperturbative String Theory,''
  Nucl.\ Phys.\  B {\bf 456}, 130 (1995)
  [arXiv:hep-th/9507158].
  
    
  
\bibitem{Alexandrov:2006hx}
  S.~Alexandrov, F.~Saueressig and S.~Vandoren,
  ``Membrane and fivebrane instantons from quaternionic geometry,''
  JHEP {\bf 0609} (2006) 040
  [arXiv:hep-th/0606259].
   
\bibitem{Alexandrov:2008ds}
  S.~Alexandrov, B.~Pioline, F.~Saueressig and S.~Vandoren,
  ``Linear perturbations of quaternionic metrics - I. The Hyperkahler case,''
  arXiv:0806.4620 [hep-th].
  
\bibitem{Alexandrov:2008gh}
  S.~Alexandrov, B.~Pioline, F.~Saueressig and S.~Vandoren,
  ``D-instantons and twistors,''
  arXiv:0812.4219 [hep-th].

  
\bibitem{Gunaydin:2007qq}
  M.~Gunaydin, A.~Neitzke, O.~Pavlyk and B.~Pioline,
  ``Quasi-conformal actions, quaternionic discrete series and twistors: SU(2,1)
  and $G_2(2)$,''
  Commun.\ Math.\ Phys.\  {\bf 283} (2008) 169
  [arXiv:0707.1669 [hep-th]].
 
\bibitem{Berkooz:2008rj}
  M.~Berkooz and B.~Pioline,
  ``5D Black Holes and Non-linear Sigma Models,''
  JHEP {\bf 0805} (2008) 045
  [arXiv:0802.1659 [hep-th]].

 
\bibitem{Ehlers} Ehlers, J.: Konstruktion und Charakterisierungen von L\"osungen der Einsteinschen Gravitationsgleichungen. Dissertation, Hamburg 1957

 
\bibitem{Geroch:1970nt}
  R.~P.~Geroch,
  ``A Method for generating solutions of Einstein's equations,''
  J.\ Math.\ Phys.\  {\bf 12} (1971) 918.
  
  
\bibitem{Kinnersley:1977pg}
  W.~Kinnersley,
  ``Symmetries Of The Stationary Einstein-Maxwell Field Equations. 1,''
  J.\ Math.\ Phys.\  {\bf 18} (1977) 1529.
  
\bibitem{Breitenlohner:1987dg}
  P.~Breitenlohner, D.~Maison and G.~W.~Gibbons,
  ``Four-Dimensional Black Holes from Kaluza-Klein Theories,''
  Commun.\ Math.\ Phys.\  {\bf 120} (1988) 295.



\bibitem{Andrianopoli:2001gm}
  L.~Andrianopoli, R.~D'Auria and S.~Ferrara,
  ``Consistent reduction of $ N = 2 \to N = 1$ four-dimensional supergravity
  coupled to matter,''
  Nucl.\ Phys.\  B {\bf 628} (2002) 387
  [arXiv:hep-th/0112192].

\bibitem{DAuria:2004kx}
  R.~D'Auria, S.~Ferrara and M.~Trigiante,
  ``c-map,very special quaternionic geometry and dual K\" ahler spaces,''
  Phys.\ Lett.\  B {\bf 587}, 138 (2004)
  [arXiv:hep-th/0401161].
  
  
\bibitem{Grimm:2004uq}
  T.~W.~Grimm and J.~Louis,
  ``The effective action of N = 1 Calabi-Yau orientifolds,''
  Nucl.\ Phys.\  B {\bf 699}, 387 (2004)
  [arXiv:hep-th/0403067].
  
\bibitem{Grimm:2004ua}
  T.~W.~Grimm and J.~Louis,
  ``The effective action of type IIA Calabi-Yau orientifolds,''
  Nucl.\ Phys.\  B {\bf 718}, 153 (2005)
  [arXiv:hep-th/0412277].
  
   
 
\bibitem{D'Auria:2005yg}
  R.~D'Auria, S.~Ferrara, M.~Trigiante and S.~Vaula,
  ``N = 1 reductions of N = 2 supergravity in the presence of tensor
  multiplets,''
  JHEP {\bf 0503} (2005) 052
  [arXiv:hep-th/0502219].
   
  
\bibitem{Brunner:2003zm}
  I.~Brunner and K.~Hori,
  ``Orientifolds and mirror symmetry,''
  JHEP {\bf 0411} (2004) 005
  [arXiv:hep-th/0303135].
  
\bibitem{Jockers:2005pn}
  H.~Jockers,
  ``The effective action of D-branes in Calabi-Yau orientifold
  compactifications,''
  Fortsch.\ Phys.\  {\bf 53}, 1087 (2005)
  [arXiv:hep-th/0507042].

   
\bibitem{Blumenhagen:2006xt}
  R.~Blumenhagen, M.~Cvetic and T.~Weigand,
  ``Spacetime instanton corrections in 4D string vacua - the seesaw mechanism
  for D-brane models,''
  Nucl.\ Phys.\  B {\bf 771} (2007) 113
  [arXiv:hep-th/0609191].

\bibitem{Akerblom:2007uc}
  N.~Akerblom, R.~Blumenhagen, D.~Lust and M.~Schmidt-Sommerfeld,
  ``Instantons and Holomorphic Couplings in Intersecting D-brane Models,''
  JHEP {\bf 0708} (2007) 044
  [arXiv:0705.2366 [hep-th]].
   
   
\bibitem{Cvetic:2007sj}
  M.~Cvetic, R.~Richter and T.~Weigand,
  ``D-brane instanton effects in Type II orientifolds: local and global
  issues,''
  arXiv:0712.2845 [hep-th].
  
\bibitem{Gimon:2007mh}
  E.~G.~Gimon, F.~Larsen and J.~Simon,
  ``Black Holes in Supergravity: the non-BPS Branch,''
  JHEP {\bf 0801} (2008) 040
  [arXiv:0710.4967 [hep-th]].
\end{thebibliography}
\end{document}